\documentclass[
 prd,
superscriptaddress,
showkeys,
preprintnumbers,
nofootinbib,
 amsmath,amssymb,
 aps,
 twocolumn,
]{revtex4-2}

\usepackage{graphicx}
\usepackage{dcolumn}
\usepackage{bm}
\usepackage{hyperref}
\usepackage[mathlines]{lineno}
\usepackage{mathtools}
\usepackage{fontawesome5}
\usepackage{listings}

\usepackage[
]{geometry}

\usepackage{xcolor}
\usepackage{physics}
\usepackage{fancyref}
\usepackage{subcaption}
\usepackage{stfloats} 
\usepackage{placeins}
\usepackage{float}
\usepackage{caption}

\usepackage{slashed}
\usepackage{ragged2e}

\definecolor{nicered}{rgb}{0.6,0,0}
\definecolor{nicegreen}{rgb}{0.1,0.5,0.1}
\definecolor{niceblue}{rgb}{0,0.4,0.8}
\definecolor{newgreen}{rgb}{0,0.667,0}
\hypersetup{
	colorlinks=true,      
	linkcolor=black,      
	citecolor=blue,       
	filecolor=magenta,     
	urlcolor=blue
}

\newcommand{\gdm}{g_\text{DM}}
\newcommand{\mdm}{m_\text{DM}}

\usepackage{orcidlink}

\usepackage{framed}
\definecolor{shadecolor}{rgb}{0.95,0.95,0.95}
\newenvironment{claim}{\begin{shaded}\noindent\ignorespaces}{\end{shaded}}

\newcommand{\CodeURL}{https://github.com/Nape93-max/SE-BSF4DM}

\lstdefinelanguage{Cpp}{
    language=C++,
    morekeywords={bool, class, public, private, protected, template, namespace, using},
    sensitive=true,
    morecomment=[l]{//},
    morecomment=[s]{/*}{*/},
    morestring=[b]",
    morestring=[b]'
}

\begin{document}
\preprint{TUM-HEP-1571/25} 
\preprint{MITP-25-058} 

\title{Sommerfeld Effect and Bound State Formation for Dark Matter Models with Colored Mediators with \texttt{SE+BSF4DM}}

\author{Mathias Becker \orcidlink{0000-0002-0203-8411}}
\email{mathias.becker@unipd.it}
\affiliation{Dipartimento di Fisica e Astronomia, Universit\`{a} degli Studi di Padova, Via Marzolo 8, 35131 Padova, Italy}
\affiliation{INFN, Sezione di Padova, Via Marzolo 8, 35131 Padova, Italy}

\author{Emanuele Copello \orcidlink{0000-0002-8404-8479}}
\email{ecopello@uni-mainz.de}
\affiliation{\text{PRISMA}$^+$ Cluster of Excellence \& Mainz Institute for Theoretical Physics, Johannes Gutenberg-Universität Mainz, 55099 Mainz, Germany}

\author{Julia Harz \orcidlink{0000-0002-8362-4083}}
\email{julia.harz@uni-mainz.de}
\affiliation{\text{PRISMA}$^+$ Cluster of Excellence \& Mainz Institute for Theoretical Physics, Johannes Gutenberg-Universität Mainz, 55099 Mainz, Germany}

\author{Martin Napetschnig \orcidlink{0000-0001-9161-9340}}
\email{martin.napetschnig@tum.de}
\affiliation{\text{PRISMA}$^+$ Cluster of Excellence \& Mainz Institute for Theoretical Physics, Johannes Gutenberg-Universität Mainz, 55099 Mainz, Germany}
\affiliation{Technical University of Munich, TUM School of Natural Sciences, Physics Department T70, 85748 Garching, Germany}

\date{\today}

\begin{abstract}
    In the universal framework of simplified $t$-channel dark matter models, the calculation of the relic abundance can be dominated by mediator annihilation when the dark matter and mediator masses are almost degenerate. We analyze four representative models with scalar and fermionic mediators, confront them with direct detection limits and highlight the differences and common features between them. The mediator annihilations are considerably enhanced by the Sommerfeld effect and bound state formation. Albeit their effect is subdominant in the coannihilation regime, excited bound state levels are included as well. 
    We find that Sommerfeld and bound-state effects can lead to $\mathcal{O}(1)$ corrections to the constraints on the DM mass in the coannihilating regime, with the precise magnitude depending on the specific model realization. 
    In addition we provide \texttt{SE+BSF4DM}, an intuitive and easy to use add-on to \texttt{micrOMEGAs}, allowing for an automated inclusion of these effects for a generic $t$-channel Dark Matter Model, which is publicly available on \href{\CodeURL}{
  \faGithub~\textcolor{blue}{Github}}.
\end{abstract}


\maketitle

\section{Introduction}\label{sec:intro}
The nature of dark matter (DM) remains one of the most pressing open questions in astroparticle physics.
Despite extensive efforts to constrain weakly interacting massive particles (WIMPs) experimental searches have severely cornered the parameter space of classical WIMPs in the 10\,-\,few\,100\,GeV mass range \cite{LZ:2022lsv, PICO:2023uff, XENON:2024wpa, Arcadi:2017kky, LHCDarkMatterWorkingGroup:2018ufk, CMS:2019ykj, ATLAS:2023rvb, PerezAdan:2023rsl}.
This has spurred interest in rich dark sectors \cite{Harris:2022vnx}, where DM emerges as the lightest stable particle (LSP).  
Not only do such scenarios allow for much higher DM masses in the multi-TeV range due to coannihilation \cite{Griest:1990kh, Gondolo:1997km, Harz:2012fz, Baker:2015qna}, but they also open up the possibility that the coannihilating partners exhibit (gauge) interactions with the standard model (SM), which are prohibited for the DM candidate itself.
Such a scenario has attracted a lot of attention in recent years \cite{Ellis:2014ipa, Ibarra:2015nca, ElHedri:2017nny, ElHedri:2018atj, Biondini:2019int}. 

Owing to the fact that the dark sector particles are non-relativistic in the standard freeze-out paradigm, non-perturbative effects can have a sizeable impact on the effective annihilation cross section and need to be taken into account for an accurate prediction of the DM relic density in case the dark sector particles are subject to a long-range force. 
The \textit{Sommerfeld effect} \cite{Sommerfeld:1931qaf, Sakharov:1948plh, Beneke:2014hja, Beneke:2019qaa, Hisano:2006nn,  Branahl:2019yot} can enhance or suppress the cross section depending on the long-range potential being attractive or repulsive (see Sec. \ref{sec:colorDec_Sommerfeld}). 
Radiative formation and subsequent decay of dark sector \textit{bound states} constitute an additional annihilation channel that can even be the dominant contribution to the total cross section, especially if colored coannihilations take place due to the strong QCD coupling\,\cite{vonHarling:2014kha, Ellis:2015vaa, An:2016gad, Asadi:2016ybp, Petraki:2016cnz, Liew:2016hqo}. \\
In order to constrain the available parameter space using experimental data, it is essential to minimize the number of free parameters in a given model. 
This leads to the concept of \textit{minimal}, or \textit{simplified} dark matter models, which serve as benchmarks for experimental searches \cite{Cirelli:2005uq, Cirelli:2007xd, Cirelli:2009uv, DiFranzo:2013vra, Abdallah:2015ter, Morgante:2018tiq, Belanger:2018sti, Belyaev:2018xpf, Belyaev:2022qnf, Ghosh:2022zef, arcadi2025veilchartingwimpterritories}. 

Simplified $t$-channel DM models (\textit{DMSimpt}) are a widely studied class of models, in which the mediator couples directly to DM and a SM quark \cite{Garny:2015wea, Giacchino:2015hvk, Mohan:2019zrk, Arina:2020udz, Arina:2020tuw, Becker:2022iso, Arina:2023msd, Jueid:2024cge, Arina:2025zpi, Biondini:2025gpg, olgoso2025darkterazfactory}. 
In these setups, the mediator shares the same quantum numbers as the quark it couples to, while the DM candidate is a SM singlet. 
This contrasts with simplified \textit{s-channel} DM models (\textit{DMSimp}) \cite{Backovic:2015soa}, where both the mediator and the DM candidate are mostly SM singlets.  

In this work, we analyze four realizations of simplified $t$-channel dark matter models, comparing scalar and fermionic mediators coupled to either the lightest (up) or the heaviest (top) SM quarks (see Tab.~\ref{tab:Models}). 
We quantify the impact of the aforementioned non-perturbative effects - Sommerfeld enhancement and bound-state formation (BSF) - on the dark matter relic abundance and associated mass constraints, with particular emphasis on coannihilation regimes where long-range QCD effects can enhance cross sections by up to 300\%~\cite{Harz:2018csl}. 
We derive updated direct detection limits and show that these effects can significantly shift the allowed dark matter mass range. 
To enable efficient and consistent relic density calculations, we introduce \texttt{SE+BSF4DM}, an automated C++ package that incorporates Sommerfeld and bound-state effects to the DM abundance calculation of \texttt{micrOMEGAs}. 
While several tools exist for generic relic density computations \cite{Belanger:2006is, Bringmann:2018lay, Ambrogi:2018jqj, Harz:2023llw, Palmiotto:2022rvw}, the inclusion of long-range effects remains largely unautomated.
So far the only software package that provided next-to-leading order corrections including the Sommerfeld effect interfaced to \texttt{micrOMEGAs} has been \texttt{DM@NLO}~\cite{Harz:2012fz,Harz:2014tma,Harz:2014gaa,Harz:2016dql,Schmiemann:2019czm,Branahl:2019yot,Harz:2022ipe,Harz:2023llw}, which is so far limited to the Minimal Supersymmetric Model (MSSM) only.
The recent release of \texttt{BSFfast}~\cite{binder2025bsffastrapidcomputationboundstate} provides highly efficient tabulations of bound-state formation cross sections for extensive phenomenological scans.
The \texttt{SE+BSF4DM} package eliminates the need for manual implementations for any model within, or mappable onto, the \textit{DMSimpt} class by performing a direct, automated calculation of both Sommerfeld effect and bound state formation within the \texttt{micrOMEGAs} environment, offering a complementary approach focused on integration and ease of use for this specific framework.

This paper is structured as follows: section \ref{sec:model} briefly presents the \textit{DMSimpt} class of models and revisits the basics of DM freeze-out. 
Section \ref{sec:long_range_effects} is devoted to a concise discussion of the Sommerfeld effect and bound state formation (BSF) in non-Abelian gauge theories. 
We apply our code to four representative models to highlight the impact of the non-perturbative effects, present our results and direct detection limits in section \ref{sec:Results} and summarize in Sec. \ref{sec:conclusions}. 
Appendix \ref{app:color_decomp} gives further details about the color decomposition of the QCD potential.  
The numerical implementation and illustrative examples on how to use our code is provided in Appendix \ref{sec:Numerics}. 

\onecolumngrid
\begin{center}
\begin{minipage}{\textwidth}
\centering
\captionof{table}{\label{tab:Models}%
Summary of $t$–channel models considered in this work.}
\vspace{0.5em}
\begin{tabular}{c|c|c}
\hline
\hline
 & up-quark & top-quark \\
\hline
Scalar     & \texttt{S3MuR}: & \texttt{S3MtR}: \\
mediator   & $\mathcal{L}_\text{int} = \gdm \left(\overline{\chi} \: X^\dagger \: u_R + \overline{u}_R \: X \: \chi \right)$ & $\mathcal{L}_\text{int} = \gdm \left(\overline{\chi} \: X^\dagger \: t_R + \overline{t}_R \: X \: \chi \right)$ \\ 
\hline 
Fermionic &  \texttt{F3SuR}: & \texttt{F3StR}: \\ 
mediator  & $\mathcal{L}_\text{int} = \gdm \left(\overline{X} \: \chi \: u_R + \overline{u}_R \: \chi \: X  \right)$ & $\mathcal{L}_\text{int} = \gdm \left(\overline{X} \: \chi \: t_R + \overline{t}_R \: \chi \: X \right)$ \\
\hline
\hline
\end{tabular}
\end{minipage}
\end{center}
\vspace{1em}
\twocolumngrid

\section{Simplified $t$-channel Models and Dark Matter Cosmology}\label{sec:model}

In the following, we briefly review the \textit{DMSimpt} class of simplified $t$-channel DM models, for which our code is applicable \cite{Arina:2020udz}. 
In its most general form, the DM candidate in these models, denoted generically with $\chi$, is a SM singlet and can either be a real or complex spin-0, spin-$1/2$ or spin-1 field. 
A $\mathbb{Z}_2$ symmetry ensures the stability of the LSP. 
The mediator, generically denoted by $X$, couples the DM candidate to either the left-handed quark doublet, right-handed down- or up-type quark field, which are denoted by $Q,d$ and $u$, respectively. 
The mediator may either be a spin-0 or spin-$1/2$ particle (real or complex)\footnote{Vector mediators are not included in the \textit{DMSimpt} class. Apart from few exceptions (e.g.~\cite{Saez:2018off}), vector fields in the fundamental representation of $SU(3)$ are rarely studied, because vector fields usually serve as gauge bosons and not as $t$-channel mediators.}. 
The three quark types times the three possible mediator spins allow for nine different Yukawa couplings \small {$\gdm$} at the renormalizable level:
\begin{align}
    \mathcal{L}_\text{int} = g_\text{DM} \: \chi\: \widetilde{X}\:\widetilde{q} + \text{ h.c.}, \label{eq:t_chann_coupling_generic}
\end{align}
where $\widetilde{q}={Q,u,d}$ and $\widetilde{X} = {X^\dagger, \overline{X}}$ in the case of a scalar or fermionic mediator, respectively. \\ 
For definiteness, we focus on four different realizations of the aforementioned $t$-channel interactions with a single Yukawa coupling $\gdm$ between one mediator species and one single quark flavor. 
However, our code is applicable to more complicated setups as well.
Adopting a similar notation to Ref.~\cite{Arina:2020udz}, our models are summarized in Tab.~\ref{tab:Models}. 

In the \texttt{S3M} (\texttt{F3S}) models, the DM candidate is a Majorana fermion (real scalar) while the mediator is a complex scalar (Dirac fermion). 
For both models we consider the option of couplings to the right-handed up-quark (\texttt{uR}) and the right-handed top-quark (\texttt{tR}). 
The coupling to down-type quark fields gives similar results \cite{Giacchino:2015hvk, Becker:2022iso}. 
The impact of Sommerfeld effect and bound state formation on the relic density for one representative model coupling to left-handed quarks was shown in \texttt{S3Muni} for universal couplings in Ref.\,\cite{Becker:2022iso}.

At the mass scale of the DM mediators, we set the couplings of the dark scalars (which we, for the moment, denote by $\Phi$) to the SM Higgs to zero: $\lambda_{|\Phi|^2 H^2}(\mu = M_\Phi) = 0$, such that they have no impact on the relic abundance.\footnote{We also assume quartic self interactions of $\Phi$ to be negligible for our analysis.} 
However, as it has been pointed out in \cite{Biondini:2019int}, at lower scales such couplings can grow sizeable due to their Renormalization Group (RG) running.
We therefore include this running in the direct detection amplitudes.  

The total dark sector number density $\mathrm{n_{tot}}$ divided by the entropy density $s$ defines the total DM \textit{yield} $\widetilde{Y} = \frac{\mathrm{n_{tot}}}{s}$, which, in our setup, is given by:
\begin{equation}
    \widetilde{Y}=Y_\chi +  \left(  Y_{X}+Y_{X^{\dagger}} \right) = Y_\chi+2 \: Y_{X}.
    \label{eq:totYield}
\end{equation}
The Boltzmann equation describing the evolution of the yield is expressed in terms of the dimensionless variable $x = \mdm/T$, where $\mdm$ is the mass of the DM candidate (i.e. the LSP $\chi$) and $T$ is the temperature of the SM thermal bath. 
The time evolution of $\widetilde{Y}$ is then given by \cite{Griest:1990kh}
\begin{equation}
    \dfrac{\dd \widetilde{Y}}{\dd x}=-c\,g_{*,\text{eff}}^{1/2}\dfrac{\langle\sigma_{\text{eff}} v_{\text{rel}}\rangle}{x^2}\left(\widetilde{Y}^2-\widetilde{Y}^2_{\text{eq}}\right),
    \label{eq:BoltzmannEq}
\end{equation}
where $c = \sqrt{\pi/45}\, M_{\text{Pl}}\, m_{\text{DM}}$ with $M_{\text{Pl}}$ being the Planck mass.
The factor $g_{*,\text{eff}}^{1/2} = \dfrac{g_{*S}}{\sqrt{g_*}}\left(1 + \dfrac{T}{3g_{*S}}\dfrac{\dd g_{S}}{\dd T}\right)$ encodes the \textit{effective} degrees of freedom of the thermal bath, where $g_*$ and $g_{*S}$ represent, respectively, the energy and entropy degrees of freedom. The equilibrium yields for the DM candidate $\chi$ and mediator $X$ are \cite{KolbTurner1994}
\begin{align}
Y_{\chi}^{\text{eq}} &\equiv \frac{n_\chi}{s} \simeq \dfrac{90}{(2\pi)^{7/2}}\dfrac{g_\chi}{g_{*S}}x^{3/2}e^{-x}, \\
Y_{X}^{\text{eq}} &= Y_{X^{\dagger}}^{\text{eq}} \simeq \dfrac{90}{(2\pi)^{7/2}}\dfrac{g_X}{g_{*S}}[(1+\delta)x]^{3/2} e^{-(1+\delta) x}, \label{eq:mediator_yield}
\end{align}
with $g_\chi$ and $g_X$ being the internal degrees of freedom of $\chi$ and $X$, respectively and we introduced the relative mass splitting between the mediator and DM,
\begin{equation}
\delta \equiv \dfrac{m_X - m_\chi}{m_\chi} = \dfrac{\Delta m}{m_\chi}. \label{eq:mass_differences}
\end{equation}
The effective annihilation cross section is then given by \cite{Griest:1990kh}:
\begin{equation}
    \langle\sigma_{\text{eff}} v_{\text{rel}}\rangle=\sum_{ij}\langle\sigma_{ij}v_{ij}\rangle \dfrac{Y_i^{\text{eq}}}{\widetilde{Y}^{\text{eq}}}\dfrac{Y_j^{\text{eq}}}{\widetilde{Y}^{\text{eq}}},
    \label{eq:effective_sigmav_coannih}
\end{equation} 
The validity of Eq.~\eqref{eq:effective_sigmav_coannih} requires: $i)$ kinetic equilibrium between the dark sector and SM bath and $ii)$ chemical equilibrium within the dark sector.
Both conditions are satisfied when the $t$-channel coupling $\gdm$ is sufficiently large.
While condition $i)$ is generally maintained through QCD interactions of the mediators, $\gdm$ must still be large enough to ensure efficient momentum transfer between the mediators and the DM.
Condition $ii)$ imposes a stronger requirement - the conversion rate $\Gamma^{\chi \rightarrow X}$ must exceed the Hubble rate, which occurs at larger $\gdm$ values than needed for kinetic equilibrium. When both conditions are met, the system can be described by the single Boltzmann equation, Eq.~\eqref{eq:BoltzmannEq}, and -- in case of small mass differences $\delta$ -- is the so-called \textbf{coannihilation regime}, on which we focus in this work. 
If $\Gamma^{\chi \rightarrow X}$ is comparable to the Hubble rate during mediator freeze-out, $H(T=m_X)$, the conversion rates have to be explicitly included in the Boltzmann equations and one can no longer define a single effective annihilation cross section as in Eq.~\eqref{eq:effective_sigmav_coannih}. 
This scenario is dubbed \textbf{conversion-driven freeze-out} \cite{Garny:2017rxs} or \textbf{coscattering} \cite{DAgnolo:2017dbv}.
Lastly, for even lower values of $\gdm$, the DM candidate $\chi$ is never in thermal contact with the thermal bath and the mediator $X$ is extremely long lived, such that the majority of DM is produced due to  mediator decays before or after its thermal freeze-out. 
This is called the \textbf{freeze-in} \cite{Hall:2009bx} or \textbf{super-WIMP} scenario \cite{Covi:1999ty, Feng:2003uy, Garny:2018ali}, respectively. \\
The effect of Sommerfeld enhancement and bound state formation in the coannihilation regime has been assessed in Refs.~\cite{Ellis:2015vaa, An:2016gad, Asadi:2016ybp, Petraki:2016cnz, Liew:2016hqo, Harz:2017dlj, Harz:2018csl, Harz:2019rro, Becker:2022iso}. In the coscattering and super-WIMP regime, Refs.~\cite{Garny:2021qsr, Binder:2023ckj, Beneke:2024nxh} have also included the effects of excited bound states.

\section{Long-range effects in simplified $t$-channel models}
\label{sec:long_range_effects}

In the following, we summarize the key expressions for the Sommerfeld effect and bound state formation. 
Before, some general definitions for the non-relativistic central-force problem for two colored particles are given, which are used throughout this paper.

The annihilating mediators are generically denoted with $X_1$ and $X_2$, have masses $m_{X_1}$ and $m_{X_2}$ and are in the representations $\textbf{R}_1$ and $\textbf{R}_2$ of $SU(3)_c$, respectively.
In the simplified $t$-channel DM models, the representations are restricted to be the fundamental or the anti-fundamental representation: $\textbf{R}_i = \mathbf{3}, \: \overline{\mathbf{3}},\:i=1,2$. 
The reduced mass of the two-body system is given by $\mu = \frac{m_{X_1} m_{X_2}}{m_{X_1} + m_{X_2}}$ and the possible color representations of the intial and final states are obtained by considering the decompositions into irreducible representations: 
\begin{align}
\mathbf{3} \otimes \overline{\mathbf{3}} &= \mathbf{1} \oplus \mathbf{8}\label{eq:color_decomp_33bar} \\
\mathbf{3} \otimes \mathbf{3} &= \overline{\mathbf{3}} \oplus \mathbf{6}\label{eq:color_decomp_33} \\
\overline{\mathbf{3}} \otimes \overline{\mathbf{3}} &= \mathbf{3} \oplus \overline{\mathbf{6}}\label{eq:color_decomp_3bar3bar}. 
\end{align}
The strong gauge coupling $\alpha_s$ depends on the energy scale $Q$ and the Coulombic potential depends on the irreducible representation of the final or initial state:
\begin{equation}
    V_{[\hat{\textbf{R}}]}(r)=-\frac{\alpha_g^{[\hat{\textbf{R}}]}\left( Q \right)}{r},
    \label{eq:Vr}
\end{equation}
where we introduced the effective coupling constant $\alpha^{[\hat{\textbf{R}}]}_g \left( Q \right)$ given by
\begin{align}
    \alpha^{[\hat{\textbf{R}}]}_g \left( Q \right) &= \alpha_s \left( Q \right)\times\frac{1}{2}[C_2 (\textbf{R}_\mathbf{1})+C_2 (\textbf{R}_\mathbf{1})-C_2 (\hat{\textbf{R}})] \\ 
    &\equiv\alpha_s \left( Q \right) \times k_{[\hat{\textbf{R}}]},
    \label{eq:alpha_g}
\end{align}
where $C_2(\textbf{R})$ are the quadratic Casimirs.

Depending on the sign of $k_{[\hat{\textbf{R}}]}$, the potential is either attractive or repulsive 

\begin{align}
  V(r)_{\mathbf{3}\otimes\Bar{\mathbf{3}}} &=
    \begin{cases}
    -\dfrac{4}{3}\dfrac{\alpha_s}{r}\quad[\mathbf{1}]\\[8pt]
    +\dfrac{1}{6}\dfrac{\alpha_s}{r}\quad[\mathbf{8}]
    \end{cases}, \label{eq:ColorPotential}\\
    V(r)_{\mathbf{3}\otimes\mathbf{3}} &=
    \begin{cases}
    -\dfrac{2}{3}\dfrac{\alpha_s}{r}\quad[\mathbf{\bar{3}}]\\[8pt]
    +\dfrac{1}{3}\dfrac{\alpha_s}{r}\quad[\mathbf{6}]
    \end{cases}. \label{eq:ColorPotential2}
\end{align}
The combination of $\Bar{\mathbf{3}}\otimes\Bar{\mathbf{3}}$ results in the same potential as $\mathbf{3}\otimes\mathbf{3}$ since $C_2 (\textbf{R})=C_2 (\Bar{\textbf{R}})$.
The typical momentum scale for scattering states is given by $k = \mu v_\text{rel}$, leading to an energy $E_\mathbf{k} = \frac{\mu}{2}v^2_\text{rel}$ of the scattering state. 

The potentials introduced above modify annihilation processes through the Sommerfeld effect and, if attractive, allow for the formation of bound states. Both effects are reviewed in the following.

\subsection{Color decomposition and Sommerfeld effect}\label{sec:colorDec_Sommerfeld}
The resummed exchange of gluons between the colored particles prior to annihilation can significantly modify the cross section, an effect known as the Sommerfeld effect \cite{Sommerfeld:1931qaf, Sakharov:1948plh, Beneke:2014hja, Beneke:2019qaa, Hisano:2006nn,  Branahl:2019yot}.
At low relative velocities, multiple gluon exchanges effectively distort the wavefunction of the incoming two-particle system.
This distortion either enhances or suppresses the annihilation probability compared to the perturbative calculation, depending on whether the QCD potential is attractive or repulsive in a given color configuration.
The initial two-particle state exists in a superposition of the irreducible color representations obtained from the decompositions in Eqs.~\,\eqref{eq:color_decomp_33bar}\,-\,\eqref{eq:color_decomp_3bar3bar}. Each such representation evolves independently under its own distinct Coulomb-like potential, given by Eq.~\eqref{eq:ColorPotential}. Consequently, the non-perturbative correction differs for each color configuration.
Moreover, the short-distance annihilation process has contributions from different partial waves, characterized by orbital angular momentum $\ell$ and spin $s$.
The full effect is therefore captured by correcting each partial wave in each color configuration with a specific Sommerfeld factor $S_\ell^{[\mathbf{R}]}(\zeta)$, which depends on the effective coupling strength in that representation and the relative velocity.
Thus, the total cross section is obtained by summing the perturbative partial-wave cross sections $\sigma_{\ell s}$, each multiplied by its corresponding color projection coefficient $c^{[\mathbf{R}]}_{\ell s}$ and its corresponding Sommerfeld factor $S_\ell^{[\mathbf{R}]}(\zeta)$. The parameter $\zeta = \frac{\alpha_g^{[\mathbf{R}]}}{v_\text{rel}}$ quantifies the relative strength of the attractive or repulsive potential compared to the kinetic energy of the scattering state.
The full Sommerfeld corrected cross section, decomposed into partial waves, reads \cite{Cassel:2009wt,ElHedri:2016onc}
\begin{equation}\label{eq:Sommerfeld_corrected_XS}
    \sigma_{\text{SE}}= \sum_{\ell,s} \sum_{[\mathbf{R}]} c^{[\textbf{R}]}_{\ell s} S_\ell^{[\textbf{R}]}(\zeta)\,\sigma_{\ell s},
\end{equation}
where $\sum_{[\mathbf{R}]} c^{[\textbf{R}]}_{ls} = 1$ for all $(\ell,s)$ contributions to the total perturbative cross section $\sigma_{\text{pert.}} = \sum_{\ell,s} \sigma_{\ell s}$. 
The second sum in Eq.\,\eqref{eq:Sommerfeld_corrected_XS} sums over all irreducible representations in the initial state of the process predicted by the color decomposition. 
The Sommerfeld factor for the $\ell$-th partial wave in Eq.\,\eqref{eq:Sommerfeld_corrected_XS} is given by
\begin{equation}
S_{\ell}^{[\textbf{R}]}=S_\ell\left(k_{[\textbf{R}]}\dfrac{\alpha_s}{v_{\text{rel}}}\right).
    \label{eq:somfact_color}
\end{equation}

When employing Eq.~\eqref{eq:somfact_color} for simplified $t$-channel models, the four types of Sommerfeld factors needed are:
\begin{align}
    S_{0,[\mathbf{1}]}&=S_0\left(\frac{4\alpha_s}{3v_\text{rel}}\right),\; S_{0,[\mathbf{8}]}=S_0\left(\frac{-\alpha_s}{6v_\text{rel}}\right),\; \\ S_{0,[\bar{\mathbf{3}}]}&=S_0\left(\frac{2\alpha_s}{3v_\text{rel}}\right),\; S_{0,[\mathbf{6}]}=S_0\left(\frac{-\alpha_s}{3v_\text{rel}}\right).
\end{align}
In case of a Coulomb potential, the function $S_0(\zeta_s)$ is given by \cite{Sommerfeld:1931qaf, Sakharov:1948plh} 
\begin{equation}
    S_0(\zeta_s)=\dfrac{2\pi\zeta_s}{1-e^{-2\pi\zeta_s}},
    \label{eq:S0_coulomb}
\end{equation}
where $\zeta_s=\alpha_{g,[\textbf{R}]}/v_{\text{rel}}=k_{[\textbf{R}]}\,\alpha_s/v_{\text{rel}}$.
Depending on the sign of $k_{[\textbf{R}]}$, the Sommerfeld effect results in either a significant enhancement ($S_{0,[\textbf{1}]}$ and $S_{0,[\bar{\mathbf{3}}]}$) or suppression ($S_{0,[\textbf{8}]}$ and $S_{0,[\textbf{6}]}$) of the perturbative cross-section, provided that $v_\text{rel}\lesssim \alpha_{g,[\textbf{R}]}$.
Processes which are dominated by higher powers of $v_\text{rel}$ require the Sommerfeld factor for higher partial waves, which is given by \cite{Cassel:2009wt,Iengo:2009ni,ElHedri:2016onc}
\begin{equation}
S^{[\textbf{R}]}_\ell(\zeta)=S^{[\textbf{R}]}_0(\zeta)\prod_{k=1}^\ell \left(1+\frac{\zeta^2}{k^2}\right).
\label{eq:SommerfeldPartialWave}
\end{equation}
It has to be emphasized that the expansion of $\sigma v_\text{rel}$ in powers of $v_\text{rel}$ \textit{does not} correspond to an expansion in partial waves $\ell$.
The leading order contributions to $\sigma_{ls} v_\text{rel}$ start with powers of the momentum squared $p^{2 \ell}$, but may contain contributions with powers $p^{2 \ell'}$ with $\ell'>\ell$ as well.
For example, the cross section to order $v_\text{rel}^4$ can be decomposed as follows
\begin{equation}\label{eq:XS_partial_wave_vrel_expansion}
    \sigma v_\text{rel} = \underset{\text{s-wave}}{\left(a + b \: v_\text{rel}^2 + \ldots \right)} + \underset{\text{p-wave}}{\left(c\:v_\text{rel}^2 + \ldots \right)} + \mathcal{O}(v_\text{rel}^4).
\end{equation}
From Eq.\,\eqref{eq:XS_partial_wave_vrel_expansion} it should be clear that a naive expansion in $v_\text{rel}$ mixes contributions from different partial waves. In \texttt{micrOMEGAs}, the cross section for numerical values of $v_\text{rel}$ as a function $\sigma v_\text{rel}(v_\text{rel})$ is accessible. This can be exploited to numerically decompose the cross section as $\sigma v_\text{rel}\approx a + (b+c)v_\text{rel}^2+\ldots$. Since it is not numerically possible to disentangle the $s$- and $p$-wave contributions in the $v_\text{rel}^2$ term, we take the pragmatic approach of employing the Sommerfeld effect only for the leading order $v_\text{rel}^0$ of the $s$-wave contribution, which we identify as the coefficient $a$ in the velocity expansion; this procedure introduces an error of order $v_\text{rel}^2$.
The precise implementation of the Sommerfeld effect for the colored annihilation channels considered in this work is detailed in Appendix~\ref{app:color_decomp}, see Eqs.~\eqref{eq:color_decomp_pure8}–\eqref{eq:color_decomp_qq}. 

\subsection{Bound-state formation, transitions, ionizations and decays}\label{sec:BSF_ion_dec}
If the potential between the particles is attractive, $X_1$ and $X_2$ can form hydrogen-like bound states $\mathcal{B}(X_1 X_2)$ upon radiation of a gluon: 
\begin{equation}\label{eq:BSF_radiative_process}
    \left(X_1 + X_2\right)_{[\hat{\mathbf{R}}]} \rightarrow \mathcal{B}(X_1 X_2)_{[\mathbf{R}_\mathcal{B}]} + g_{[\mathbf{8}]}.
\end{equation}
This process, which we refer to as bound state formation (BSF) in the following, is depicted in Fig.~\ref{fig:bsf}. 
\begin{center}
\begin{minipage}{\columnwidth} 
    \centering
    \includegraphics[scale=0.7]{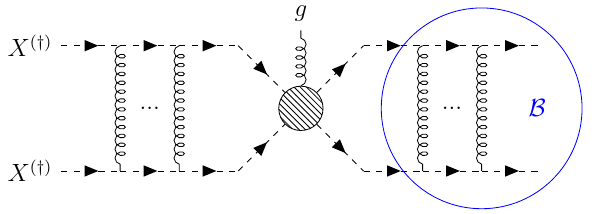}
    \captionof{figure}{Diagrammatic illustration of radiative bound state formation. 
    The labels $X^{(\dagger)}$ for the incoming legs indicate $X^\dagger-X^\dagger$, $X^\dagger-X$ and $X-X$ BSF. From Ref.\,\cite{Becker:2022iso}.}
    \label{fig:bsf}
\end{minipage}
\end{center}
We focus on mediators with equal masses, for which it has been shown~\cite{Becker:2022iso} that the rate for particle-particle BSF is strongly suppressed. 
While we take into account the Sommerfeld effect for all color configurations as listed in Eqs.~\eqref{eq:ColorPotential} and \eqref{eq:ColorPotential2}, we consider therefore only $\mathbf{3} \otimes \overline{\mathbf{3}}$ bound states and set $\hat{\mathbf{R}} = \mathbf{8}, \mathbf{R}_{\mathcal{B}} = \mathbf{1}$ in Eq.~\eqref{eq:BSF_radiative_process}.

The non-relativistic bound state $\mathcal{B}$ is characterized by its quantum numbers $\{n,l,m,s\}$, where $n$ denotes the principal quantum number, $l$ the orbital angular momentum, $m$ its projection along a chosen axis, and $s$ the total spin of the constituents.  
The typical momentum in a bound state with quantum number $n$ is given by the \textit{Bohr momentum} $p_\mathcal{B} = \mu \frac{\alpha_B}{n}$, where $\alpha_B$ is determined by solving
\begin{equation}\label{eq:alpha_B_self_con}
    \alpha_B = \alpha_s\left(Q = \mu \frac{\alpha_B}{n}\right)
\end{equation}
self-consistently.
From the Bohr momentum, the respective binding energy can be obtained:
\begin{equation}\label{eq:Ebin_def}
    E_\mathcal{B} = -\frac{p_\mathcal{B}^2}{2 \mu} = -\frac{1}{2} \mu \left( \frac{\alpha_B}{n} \right)^2. 
\end{equation}
Neglecting the kinetic energy of the bound states, momentum conservation fixes the energy $\omega$ of the radiated gluon:
\begin{equation}
    \omega \simeq E_\mathbf{k} - E_\mathcal{B} = \frac{\mu}{2}\left[ v_\text{rel}^2 + \frac{\alpha_B^2}{n^2} \right].\label{eq:omega_energy_gluon}
\end{equation}
The bound state formation process is denoted by the blob in Fig.~\ref{fig:bsf} and the gluon radiation is mediated by the coupling $\alpha_\text{BSF} = \alpha_s(Q = \omega)$.

The thermal average of the BSF cross section is defined as
\begin{align}\label{eq:def_sigma_v_averaged}
    &\expval{\sigma_{\text{BSF}, i} v_\text{rel}} = \left(\dfrac{\mu}{2\pi T}\right)^{3/2} \int \dd^3 v_\text{rel} \times \nonumber \\
    & \exp(-\mu v_\text{rel}^2/2T) \left[1+f_g(\omega)\right]\sigma_{\text{BSF}, i}v_\text{rel} \, ,
\end{align}
with $\omega$ given in Eq.~\eqref{eq:omega_energy_gluon} and~$f_g(\omega)=(\exp(\omega/T)-1)^{-1}$ is the gluon distribution function. The BSF cross section itself is given by~\cite{Yao_2019}
\begin{align}
    \sigma_{\text{BSF}, i=\{n,l\}}v_\text{rel} &= \frac{4 \omega^3}{3 N_c^2} \alpha_g^{[\mathbf{1}]}(\mu v_\text{rel}) \times \notag \\
   &(2 \ell + 1) \left|\expval{\psi^{[\mathbf{1}]}_{nl} | \mathbf{r} |\psi^{[\mathbf{8}]}_{\mathbf{p}_\text{rel}}} \right|^2.\label{eq:BSF_XS_gen_form}
\end{align}
Here, $\psi^{[\mathbf{1}]}_{nl}$ denotes the non-relativistic bound-state wavefunction with principal quantum number $n$ and orbital angular momentum $l$, while $\psi^{[\mathbf{8}]}_{\mathbf{p}_\text{rel}}$ is the corresponding scattering-state wavefunction in the color-octet channel with relative momentum $\mathbf{p}_\text{rel}$. 
In a similar manner, transition rates between bound state levels can be calculated. As all these states are color singlets, the gauge boson mediating the transition cannot be a gluon. However, photon mediated transitions are allowed in our class of models. The transition rate from a higher ($nl$) to a lower lying state ($n'l'$) is given by \cite{bethe1957quantum}
\begin{align}
    \Gamma_\text{trans}^{nl \rightarrow n' l'} &= \frac{4}{3} \alpha_\text{QED}\: Q_X^2 \: \omega_{nn'}^3 \times \notag \\ 
    &(2 \ell + 1) \left|\expval{\psi_{nl}|\mathbf{r}|\psi_{n'l'}}\right|^2, \label{eq:transition_rate_bethe}
\end{align}
where $\omega_{nn'}>0$ is the difference in binding energy between the two levels ($E_n > E_{n'}$), and $\psi_{nl}$ and $\psi_{n'l'}$ are bound state wave functions. The rate for the inverse process can be obtained via detailed balance:

\begin{align}
    \Gamma_\text{trans}^{n'l' \rightarrow n l} &= \Gamma_\text{trans}^{nl \rightarrow n' l'} \frac{Y^\text{eq}_{\mathcal{B}_{nl}}}{Y^\text{eq}_{\mathcal{B}_{n'l'}}}.\label{eq:Gamma_trans_detailed_balance}
\end{align}

Expressions for the squared matrix elements appearing in Eqs.~\eqref{eq:BSF_XS_gen_form} and \eqref{eq:transition_rate_bethe} can be found in Refs.~\cite{Binder:2023ckj, Beneke:2024nxh, Biondini:2023zcz}. 

Bound states can be ionized by the thermal gluon plasma. The corresponding rate is given by
\begin{subequations}
\begin{align}
    \expval{\Gamma^i_{\text{ion}}}&= g_g \int_{\omega_\text{min}}^\infty \dd\omega \frac{\omega^2}{2\pi^2}f_g(\omega)\sigma^i_\text{ion},\label{eq:gamma_ion}\\
    \text{with }\sigma^i_\text{ion}&=\dfrac{g_X^2}{g_g g_{\mathcal{B}, i}}\left(\frac{\mu^2 v_\text{rel}^2}{\omega^2}\right)\sigma_{\text{BSF}, i} \, .
\label{eq:sigma_ion}
\end{align}
\end{subequations}
Here, $g_X$, $g_g=8$ and $g_{\mathcal{B}, i}$ are the internal degrees of freedom of the mediator triplets, the gluons and the bound states $i$, respectively. 

The decay rate $\Gamma^{n}_\text{dec,[\textbf{R}]}$ of a $\ell=0$ bound state in a given representation \textbf{R} into gluons is computed by taking the $s$-wave perturbative annihilation cross-section times relative velocity of the corresponding scattering states and multiplying it with the bound state wavefunction evaluated at the origin
\begin{equation}
    \Gamma^{n \ell}_\text{dec,[\textbf{R}]}=(\sigma^{s\text{-wave}}_{\text{ann},[\textbf{R}]}\,v_\text{rel})|\psi_{n00}^{[\textbf{R}]}(0)|^2.
    \label{eq:decayrate_general}
\end{equation}
Considering the color-singlet ground state decaying into a pair of gluons, one obtains~\cite{Harz:2018csl, Garny:2021qsr}
\newline
\begin{equation}
    \Gamma^{n}_\text{dec,[\textbf{1}]} = \frac{\mu}{4 n^3} C_F \left(\alpha_s^\text{ann} \right)^2 \left(\alpha_{s, [\textbf{1}]}^B \right)^3,  
\end{equation}
where $|\psi^{[\textbf{1}]}_{n00}(0)|^2 = \mu^3 \frac{\left(\alpha_{s, [\textbf{1}]}^B \right)^3}{\pi n^3}$ has been used. The decay rate for $\ell \neq 0$ is strongly suppressed and taken to be zero.

\subsection{Including (excited) bound states in the Boltzmann equations}\label{subsec:excited_states_formalism}
While simplified $t$-channel DM models feature gauge-singlet DM candidates, the color-charged mediators can undergo BSF and their annihilation are modified by the Sommerfeld effect.
Thus, both the Sommerfeld effect and BSF affect  can contribute efficiently to the effective annihilation cross section of DM via coannihilations.   
The treatment of mediator coannihilations via an effective Boltzmann equation for the dark sector is well established~\cite{Griest:1990kh}.
To incorporate BSF effects, we treat each bound state $\mathcal{B}_i$ as an independent dynamical component having its own Boltzmann equation, analogous to the DM and mediator species.
Each $\mathcal{B}_i$ has $g_{\mathcal{B}_i} = (2s+1)(2 \ell+1)$ configurations with equal energy (degeneracy).

This gives a set of coupled Boltzmann equations between the DM candidate $\chi$, the mediators $X$ and the bound states $\mathcal{B}_i$ ~\cite{Binder:2021vfo, Garny:2021qsr}:
\begin{align}
    \frac{d Y_X}{dx} &= \left( \frac{d Y_X}{dx} \right)_\text{ann} \notag \\ 
-\dfrac{c\,g_{*,\text{eff}}^{1/2}}{x^2} &\left[ \sum_i \expval{\sigma_{\text{BSF},i} v_\text{rel}} \left( Y_X^2 - Y_X^{\text{eq}\,2} \frac{Y_{\mathcal{B}_i}}{Y^\text{eq}_{\mathcal{B}_i}} \right) \right]. \label{eq:BEq_mediator_BSF} \\ 
    \frac{d Y_{\mathcal{B}_i}}{dx} &= -\dfrac{c\,g_{*,\text{eff}}^{1/2}}{x^2} \Biggl[ 
      \expval{\Gamma^i_\text{ion}} \left( Y_{\mathcal{B}_i} - Y^\text{eq}_{\mathcal{B}_i} \frac{Y_X^2}{Y_X^{\text{eq}\,2}} \right) \notag \\ 
      \quad &+ \expval{\Gamma_\text{dec}^i} \left( Y_{\mathcal{B}_i} - Y^\text{eq}_{\mathcal{B}_i} \right) \notag \\
    \quad &- \sum_{j \neq i} \expval{\Gamma_\text{trans}^{j\rightarrow i}} \left( Y_{\mathcal{B}_j} - Y_{\mathcal{B}_i} \frac{Y^\text{eq}_{\mathcal{B}_j}}{Y^\text{eq}_{\mathcal{B}_i}} \right) \Biggr]. \label{eq:BEq_BS_BSF} 
\end{align}
The equilibrium yield for the bound state species $i$ is defined as
\begin{equation}\label{eq:def_YBi}
    Y_{\mathcal{B}_i}^\text{eq} = \frac{n_{\mathcal{B}_i}^\text{eq}}{s} \simeq \frac{90}{(2 \pi)^{7/2}} \frac{g_{\mathcal{B}_i}}{g_{*S}} \left( \frac{m_{\mathcal{B}_i}}{T}\right)^{3/2} e^{-\frac{m_{\mathcal{B}_i}}{T}} \, ,
\end{equation}
where $m_{\mathcal{B}_i} = m_{X_1} + m_{X_2} - E_{\mathcal{B}_i}$ is the mass of the bound state and $\expval{\Gamma_\text{ion}}$, $\expval{\Gamma_\text{dec}}$ and $\expval{\Gamma_\text{trans}}$ are the ionization, decay and transition rates, respectively, and are given in Sec.~\ref{sec:BSF_ion_dec}. 

Since at least one of the averaged rates on the right-hand-side of the Boltzmann equation is much larger than the Hubble rate $H$~\cite{Binder:2021vfo}, the bound state yields quickly adjust to a steady solution $\frac{d Y_{\mathcal{B}_i}}{dx} = 0$, which turns the bound state Boltzmann equations into algebraic equations. The same applies to a single bound state, as it was first shown in Ref.~\cite{Ellis:2015vaa}. 

All bound state effects including arbitrary many excited states can be effectively incorporated into the evolution equation for $X$ (see Eq.~\eqref{eq:BEq_mediator_BSF}) in the following form~\cite{Binder:2021vfo, Garny:2021qsr}
\begin{align}
    \frac{Y_{\mathcal{B}_i}}{Y^\text{eq}_{\mathcal{B}_i}} &= 1 + \sum_j \left(M^{-1} \right)_{ij} \frac{\langle \Gamma^j_\text{ion} \rangle}{\langle \Gamma^j \rangle}\left(\frac{Y_X^2}{Y_X^{\text{eq}\,2}} -1\right),\label{eq:BS_yields_formal_solution}\\
    \langle \Gamma^i \rangle &= \langle \Gamma^i_\text{ion} \rangle + \langle \Gamma^i_\text{dec} \rangle + \sum_{j \neq i} \langle \Gamma^{i\rightarrow j}_\text{trans} \rangle,\label{eq:total_width} \\
    M_{ij} &\equiv \delta_{ij} - \frac{\langle \Gamma^{i \rightarrow j}_\text{trans} \rangle}{\langle \Gamma^i \rangle}\label{eq:def_MMatrix}.
\end{align}
The indices $i,j$ run over all labels of bound state quantum numbers $(n,l,m,s)$. 
With these manipulations, one obtains the following effective thermally averaged annihilation cross section for the mediators~\cite{Binder:2021vfo, Garny:2021qsr}

\begin{widetext}
\begin{claim}
\begin{equation}
\expval{\sigma_{XX} v_\text{rel}}_\text{eff} = \expval{\mathcal{S} \left( \sigma_{XX} v_\text{rel} \right) }_\text{ann} + \sum_i \expval{\sigma_{\text{BSF}, i} v_\text{rel}} \left(1 - \sum_j \left( M^{-1}\right)_{ij} \frac{\expval{\Gamma^j_\text{ion}}}{\expval{\Gamma^j}} \right)\label{eq:BSF_master_equation}
\end{equation}
\end{claim}
\end{widetext}

The first term in Eq.~\eqref{eq:BSF_master_equation} contains the annihilation cross section of the mediator $X$, where the Sommerfeld enhancement/suppression is indicated with $\mathcal{S} \left( \dots \right)$. 
The second term captures the contribution of the bound states with the short-hand label $i=(n,\ell,m,s)$. Eq.~\eqref{eq:BSF_master_equation} forms the backbone of our \texttt{micrOMEGAs} package.
Limiting cases allow for simpler expressions for the effective BSF cross section in Eq.~\eqref{eq:BSF_master_equation}.
These are:  

\begin{itemize}
    \item \textbf{No transition limit}:\vspace{2mm} \newline 
    $\expval{\Gamma^i_\text{dec}}, \expval{\Gamma^i_\text{ion}} \gg \expval{\Gamma_\text{trans}^{i\rightarrow j}} \: \forall \: i,j$. 
    In this limit, the matrix $M$ in Eq.~\eqref{eq:def_MMatrix}  becomes the trivial identity matrix.
    One obtains a weighted sum of bound state contributions with weight factors analogous to the ones used for the ground state~\cite{Becker:2022iso, Harz:2018csl, Petraki:2015hla, Mitridate:2017izz}
    \begin{align}
        \langle \sigma_\text{BSF} v_\text{rel} \rangle_\text{eff} &= \sum_i \langle \sigma_{\text{BSF}, i} v_\text{rel} \rangle w_i \label{eq:no_tran_limit} \\
        w_i &= \frac{\langle \Gamma^i_\text{dec} \rangle}{\langle \Gamma^i_\text{dec} \rangle + \langle \Gamma^i_\text{ion} \rangle}.
    \end{align}
    \item \textbf{Efficient transition limit}:\vspace{2mm} \newline $\expval{\Gamma_\text{trans}^{i\rightarrow j}} \gg \expval{\Gamma^i_\text{ion}} ,\expval{\Gamma^i_\text{dec}}  \: \forall \: i,j$. In this regime, transitions are so efficient that \textit{all} bound state levels (including those with zero decay rate at leading order in $\alpha_s$) contribute to the annihilation cross section.
    In this limit, the effective BSF cross section does not depend explicitly on the transition rates $\expval{\Gamma_\text{trans}^{i\rightarrow j}}$. Instead, \textit{effective decay and ionization rates} enter in the effective bound state contribution to $\expval{ \sigma_\text{BSF} v_\text{rel}}_\text{eff}$.
    \begin{align}
        \qquad \expval{\sigma_\text{BSF} v_\text{rel}}_\text{eff} &= \sum_i \expval{\sigma_{\text{BSF}, i} v_\text{rel}} w\label{eq:sigmaBSF_eff_trans_limit}, \\
        w &= \frac{\Gamma^\text{eff}_\text{dec} }{ \Gamma^\text{eff}_\text{dec} + \Gamma^\text{eff}_\text{ion}},\\
        \Gamma^\text{eff}_\text{ion/dec} &\equiv \frac{\sum_i \expval{\Gamma^i_\text{ion/dec}} Y^\text{eq}_{\mathcal{B}_i}}{\sum_j Y^\text{eq}_{\mathcal{B}_j}}. \label{eq:def_Gamma_eff}
    \end{align}
    In the efficient transition limit the bound states are in chemical equilibrium among themselves. 
    \item \textbf{Ionization equilibrium}: \vspace{2mm} \newline
    $\expval{\Gamma^i_\text{ion}} \gg \expval{\Gamma^i_\text{dec}}, \expval{\Gamma_\text{trans}^{i\rightarrow j}} \: \forall \: i,j$. This hierarchy of rates applies to high temperatures/early times, when ionization and recombination processes are in equilibrium. The effective BSF cross section is then given by:
    \begin{align}
        &\expval{\sigma_\text{BSF} v_\text{rel}}_\text{eff} = \frac{1}{g_{X_1}g_{X_2}} \times \nonumber \\ 
        & \left(\frac{2 \pi}{T \mu} \right)^{3/2} \sum_i g_{\mathcal{B}_i}\, e^{E_{\mathcal{B}_i}/T}\, \Gamma^i_\text{dec} \label{eq:sigmaBSF_IonEq}
    \end{align}
    This limit is rarely of practical interest, as the hierarchy $\expval{\Gamma^i_\text{ion}} \gg \expval{\Gamma^i_\text{dec}}$ does not hold any more at low temperatures. 
\end{itemize}

The user may choose between one of these three limiting scenarios with the \texttt{bsf\_scenario=1,2,3} option, while the default is the full solution to Eq.~\eqref{eq:BSF_master_equation} (\texttt{bsf\_scenario=4}).

\section{Results and discussion}\label{sec:Results}
\subsection{Impact of long-range effects on the relic density}\label{subsec:results_relic}
We run parameter scans of the four $t$-channel models presented in Sec.~\ref{sec:model} for values of $\mdm \in [200, 10^4]$\,GeV, $\delta \in [10^{-4}, 10]$ and $\gdm \in [10^{-4}, \sqrt{4 \pi}]$. 
For this range of $\gdm$, the dark sector is in chemical equilibrium. 
Moreover, colored gauge interactions ensure that the mediators $X$ are in kinetic equilibrium with the SM.
This choice of parameter space is motivated by requiring dark matter to annihilate in the non-relativistic regime, which enhances the impact of long-range effects. 
To ensure this, we focus on dark-matter masses above 200~GeV, so that annihilations always proceed into lighter SM states (including top quarks). 
We restrict our analysis to the parameter space where the assumptions of freeze-out and coannihilation remain valid, and where perturbative unitarity is maintained. 
In addition, in all of our scans we ensure that $\Delta m > 1$\,GeV, as otherwise post-confinement annihilations of the mediators are shown to have a drastic impact on the relic density \cite{Gross:2018zha}, because annihilations of confined hadrons with geometric cross sections boost the annihilation cross section.
Moreover, we also demand that $E_\mathcal{B_i}\gtrsim \Lambda_\text{QCD}$, to ensure that $\alpha_s$ is only evaluated in its perturbative regime when determining bound-state formation cross section~\cite{Biondini:2023zcz}. \\  
To emphasize the impact of long-range effects, we compare the regions of parameter space in the $(\Delta m - \mdm)$-plane predicting the correct relic abundance with four different setups: $i)$ The conventional perturbative calculation, $ii)$ including the Sommerfeld effect, $iii)$ adding bound state formation of the $n = 1$ state and, finally, $iv)$ the full tower of excited states up to $n=6$ according to Eq.~\eqref{eq:BSF_master_equation}.
We find that the effect of excited states with $n>6$ changes the relic density less than 1\,\% in the parameter space we consider.
This is because excited states give a significant contribution at high values of $x$, which are suppressed by the exponential factor $e^{-2 \delta x}$ in the effective coannihilation cross section Eq.\,\eqref{eq:effective_sigmav_coannih}. Our comparison for the four $t$-channel models considered in this work is illustrated in Fig.\,\ref{fig:BandScans}. 

\onecolumngrid
\begin{center}
\begin{minipage}{\textwidth}  
    \centering
    \begin{minipage}{0.49\textwidth}
        \centering
        \includegraphics[scale=0.35]{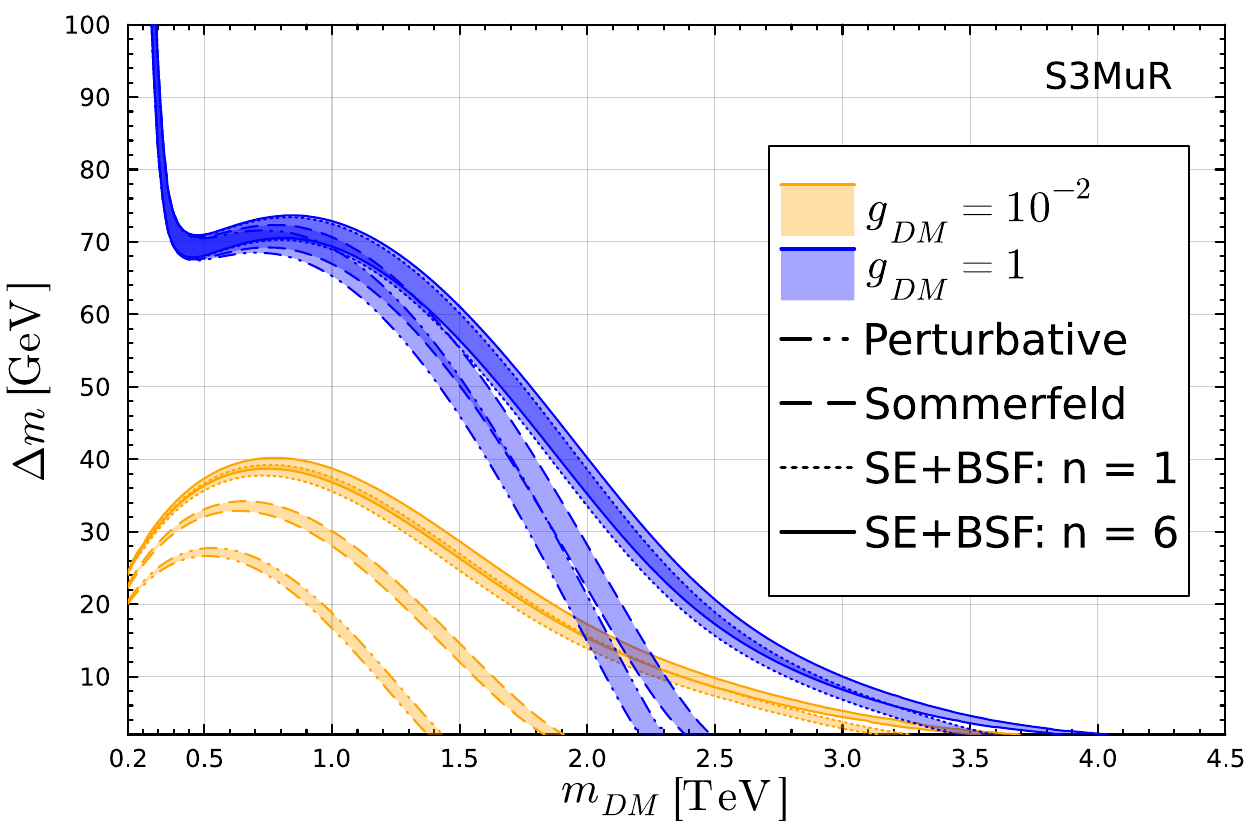}
    \end{minipage}%
    \hfill
    \begin{minipage}{0.49\textwidth}
        \centering
        \includegraphics[scale=0.35]{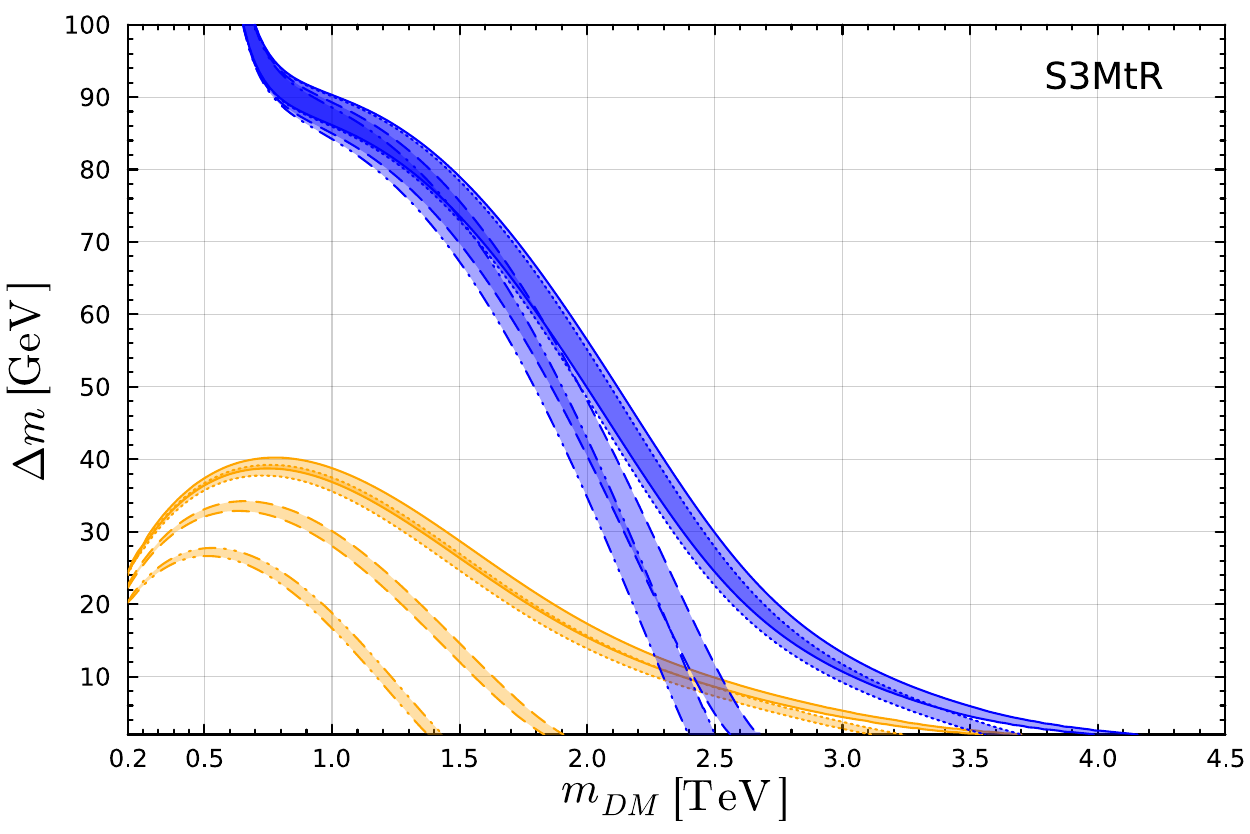}
    \end{minipage}
    \begin{minipage}{0.49\textwidth}
        \centering
        \includegraphics[scale=0.35]{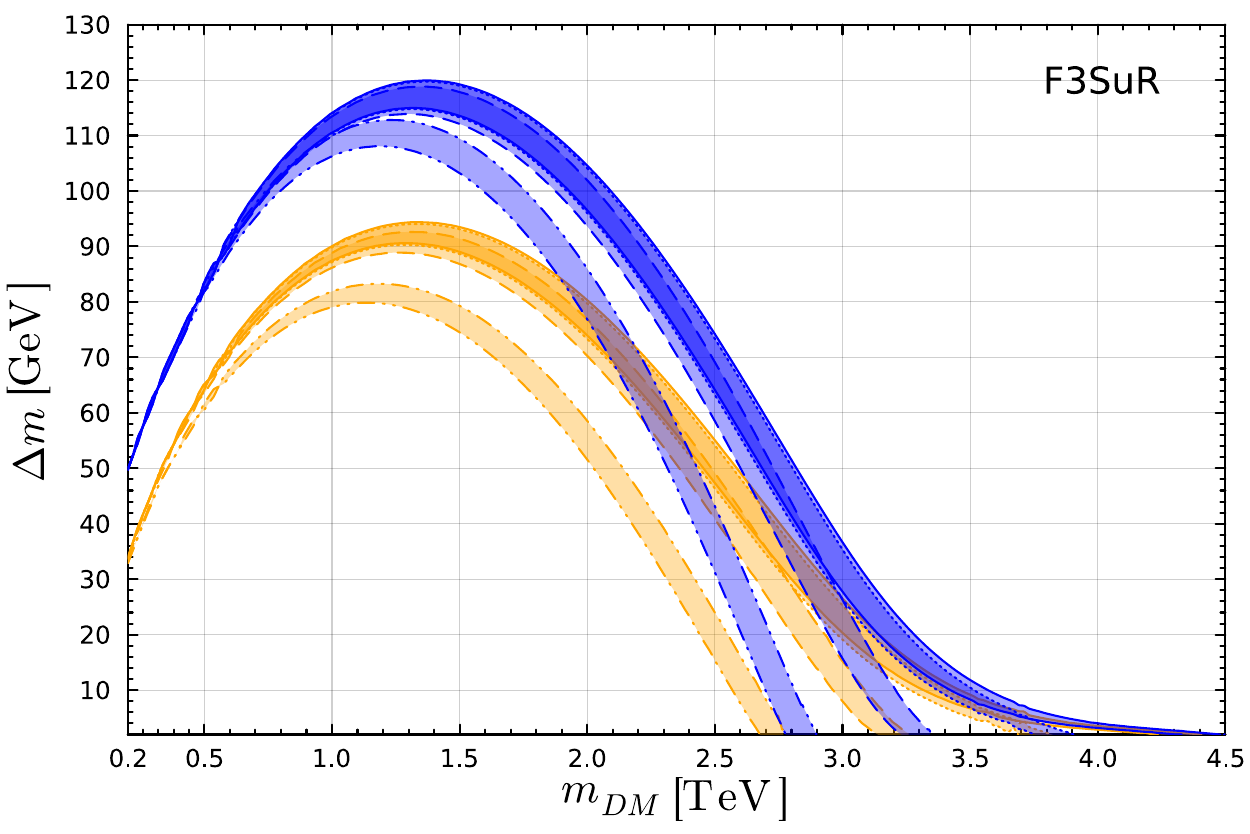}
    \end{minipage}%
    \hfill
    \begin{minipage}{0.49\textwidth}
        \centering
        \includegraphics[scale=0.35]{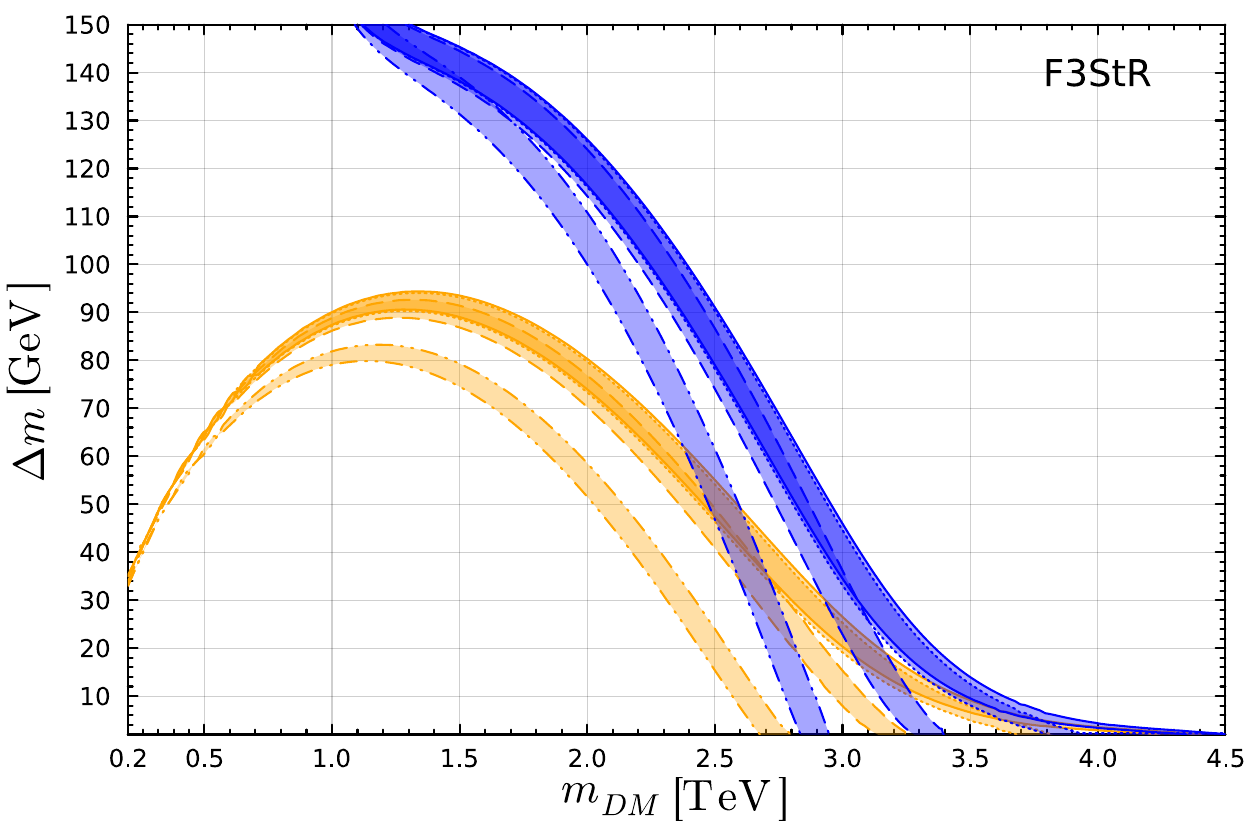}
    \end{minipage}

    \vspace{0.5em}  
    \captionof{figure}{\justifying Mass splitting $\Delta m$ vs. DM mass $\mdm$ for the four models considered. Blue (yellow) bands correspond to a $5 \sigma$ uncertainty of $\Omega h^2$ \cite{Planck:2018vyg} for $\gdm = 1$ ($10^{-2}$). Dash-dotted bands correspond to the conventional perturbative calculation, dashed bands include the Sommerfeld effect. Bound state formation for the ground state ($n = 1$) is indicated by the dotted lines, which in most of parameter space overlay with the solid bands for $n = 6$.}
    \label{fig:BandScans}
\end{minipage}
\end{center}
\vspace{1em}
\twocolumngrid

\begin{figure}[!t]
\centering
\begin{minipage}{\textwidth}
\centering
\begin{subfigure}{0.3\textwidth}
\centering
\includegraphics[width=0.49\linewidth]{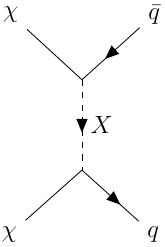}
\caption{DM Annihilation} \label{fig:processCC}
\end{subfigure}
\hfill
\begin{subfigure}{0.3\textwidth}
\centering
\includegraphics[width=0.7\linewidth]{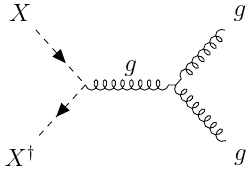}
\caption{Colored Annihilation}\label{fig:processXXd1} 
\end{subfigure}
\hfill
\begin{subfigure}{0.3\textwidth}
\centering
\includegraphics[width=0.55\linewidth]{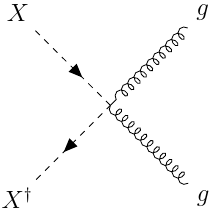}
\caption{Colored Annihilation}\label{fig:processXX4} 
\end{subfigure}

\vspace{4pt}

\begin{subfigure}{0.3\textwidth}
\centering
\includegraphics[width=0.49\linewidth]{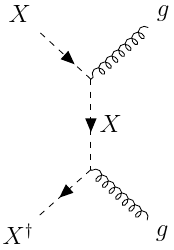}
\caption{Colored Annihilation}\label{fig:processXXd3}
\end{subfigure}
\hfill
\begin{subfigure}{0.3\textwidth}
\centering
\includegraphics[width=0.7\linewidth]{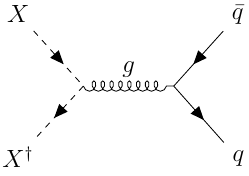}
\caption{Colored Annihilation}\label{fig:processXXd2}
\end{subfigure}
\hfill
\begin{subfigure}{0.3\textwidth}
\centering
\includegraphics[width=0.49\linewidth]{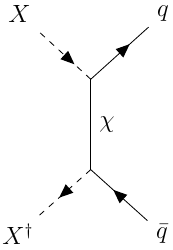}
\caption{Colored Annihilation}\label{fig:processXXdqq}
\end{subfigure}

\vspace{4pt}

\begin{subfigure}{0.3\textwidth}
\centering
\includegraphics[width=0.49\linewidth]{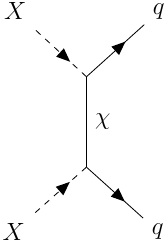}
\caption{Colored Annihilation}\label{fig:processXXqq}
\end{subfigure}
\hfill
\begin{subfigure}{0.3\textwidth}
\centering
\includegraphics[width=0.7\linewidth]{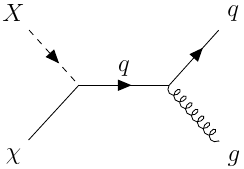}
\caption{Co-annihilation}\label{fig:processXcs}
\end{subfigure}
\hfill
\begin{subfigure}{0.3\textwidth}
\centering
\includegraphics[width=0.49\linewidth]{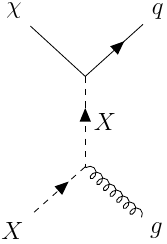}
\caption{Co-annihilation} \label{fig:processXct}
\end{subfigure}

\caption{\small Representative Feynman diagrams corresponding to the main processes contributing to (co-)annihilations. From Ref.\,\cite{Becker:2022iso}.}
\label{fig:feynm_diagr}
\end{minipage}
\end{figure}

We observe that long-range effects have a larger impact for lower $\gdm$, as in this case colored annihilations of the mediators dominate. 
Bound state formation has its biggest impact for low mass splittings, where it is not exponentially suppressed by the coannihilation weight. 
Since the cross sections for excited states peak at higher values of $x = \mdm/T$ than the ground state, the effect of excited states is only visible for large DM masses (see the difference between the solid and the dotted curves in Fig.~\ref{fig:BandScans}) and hence expected to be of minor importance in the coannihilation regime (see also Ref.~\cite{Garny:2021qsr}). \\
The dependence on $\gdm$ differs between fermionic and scalar mediators due to their distinct annihilation properties. The curves in Fig.~\ref{fig:BandScans} reflect the following features:
\begin{itemize}
    \item Perturbative annihilations: \texttt{F3S} models naturally achieve larger annihilation cross sections than \texttt{S3M} models because of the $\overline{X} X \rightarrow \overline{q} q$ channel (see Fig.\,\ref{fig:processXXd2} and \ref{fig:processXXdqq}), which is helicity suppressed for scalar mediators because they do not carry spin, resulting in higher predicted DM masses (and mass splittings) in perturbative calculations.

\onecolumngrid
\begin{center}
\begin{minipage}{\textwidth}
    \centering
    \begin{minipage}{0.49\textwidth}
        \centering
        \includegraphics[scale=0.35]{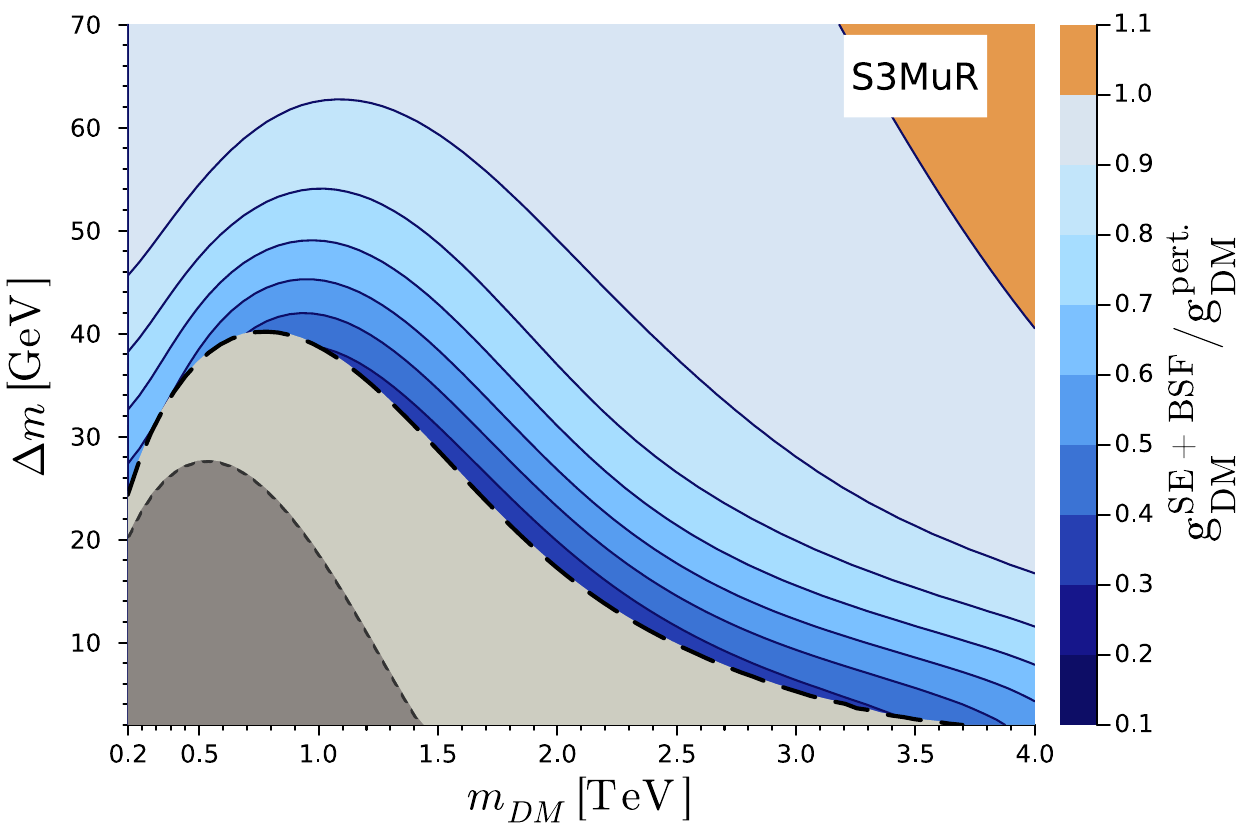}
    \end{minipage}%
    \hfill
    \begin{minipage}{0.49\textwidth}
        \centering
        \includegraphics[scale=0.35]{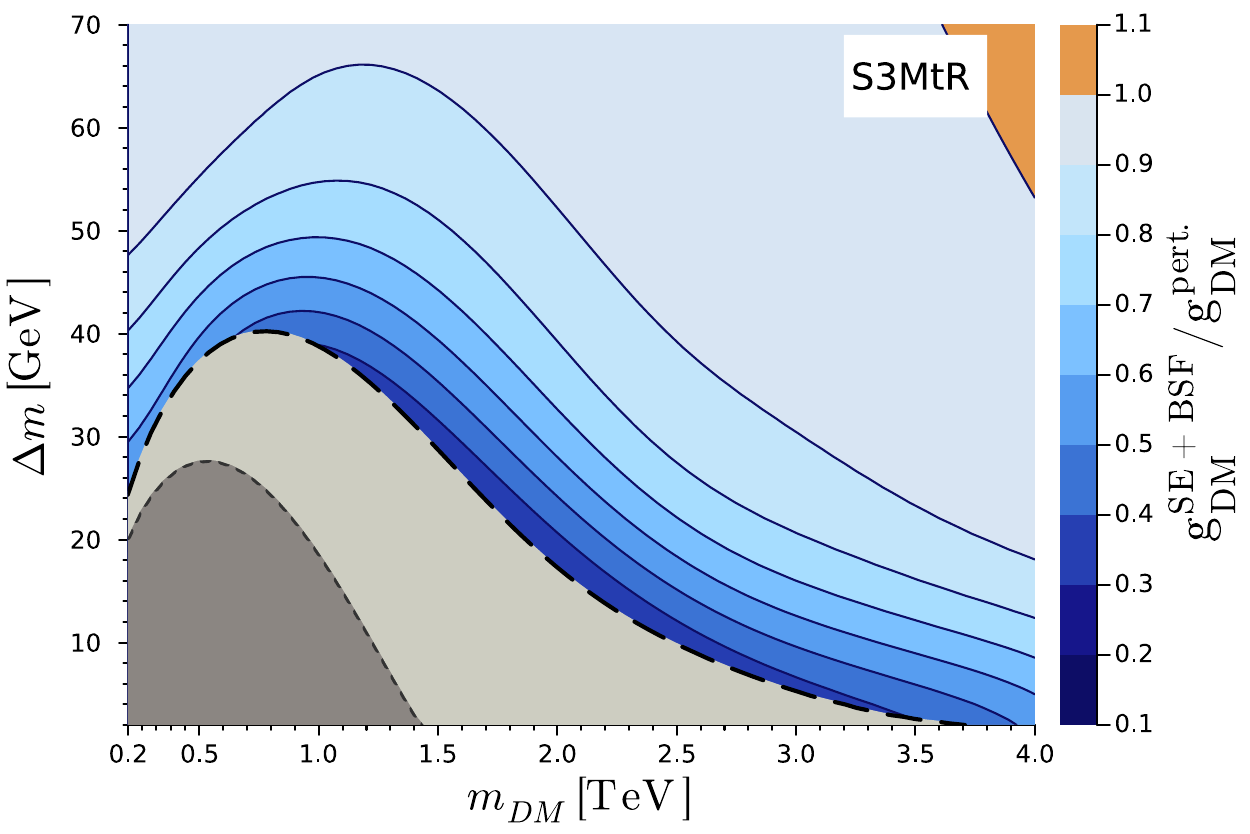}
    \end{minipage}

    \begin{minipage}{0.49\textwidth}
        \centering
        \includegraphics[scale=0.35]{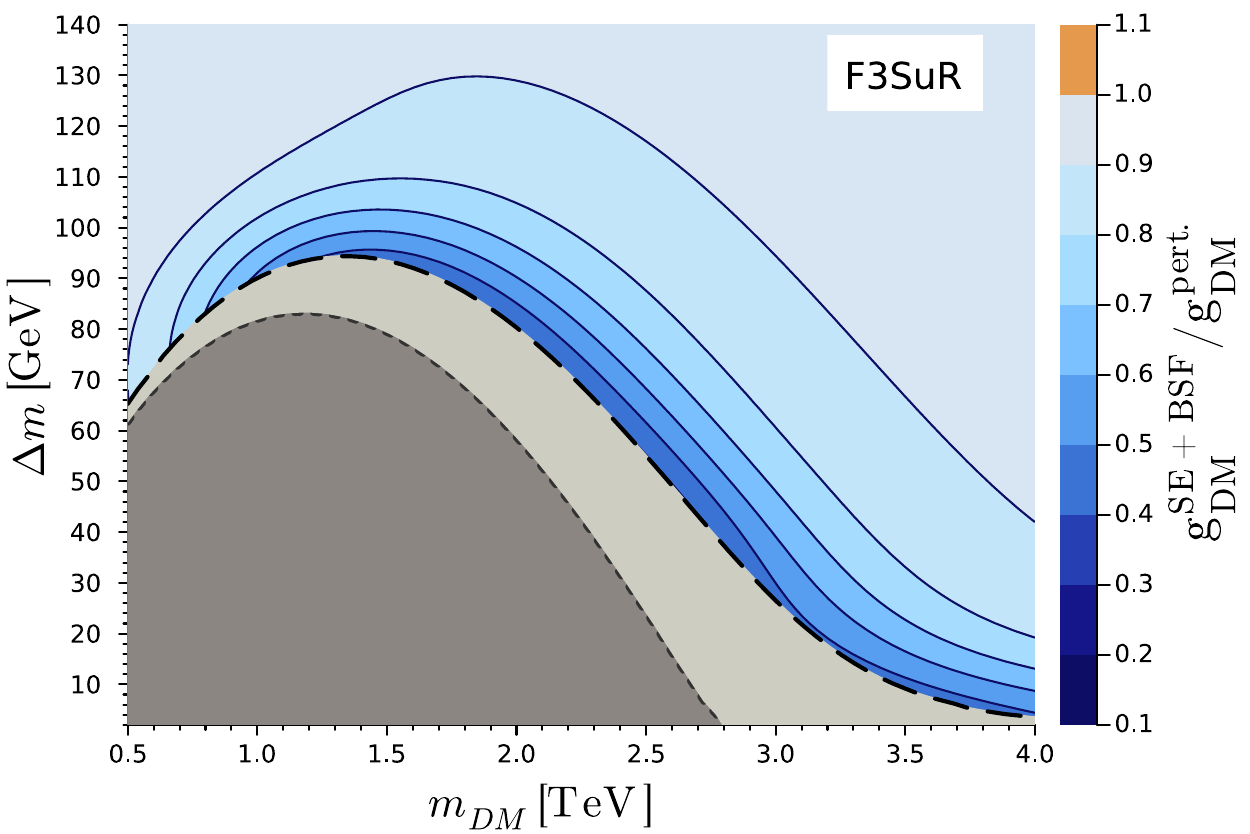}
    \end{minipage}%
    \hfill
    \begin{minipage}{0.49\textwidth}
        \centering
        \includegraphics[scale=0.35]{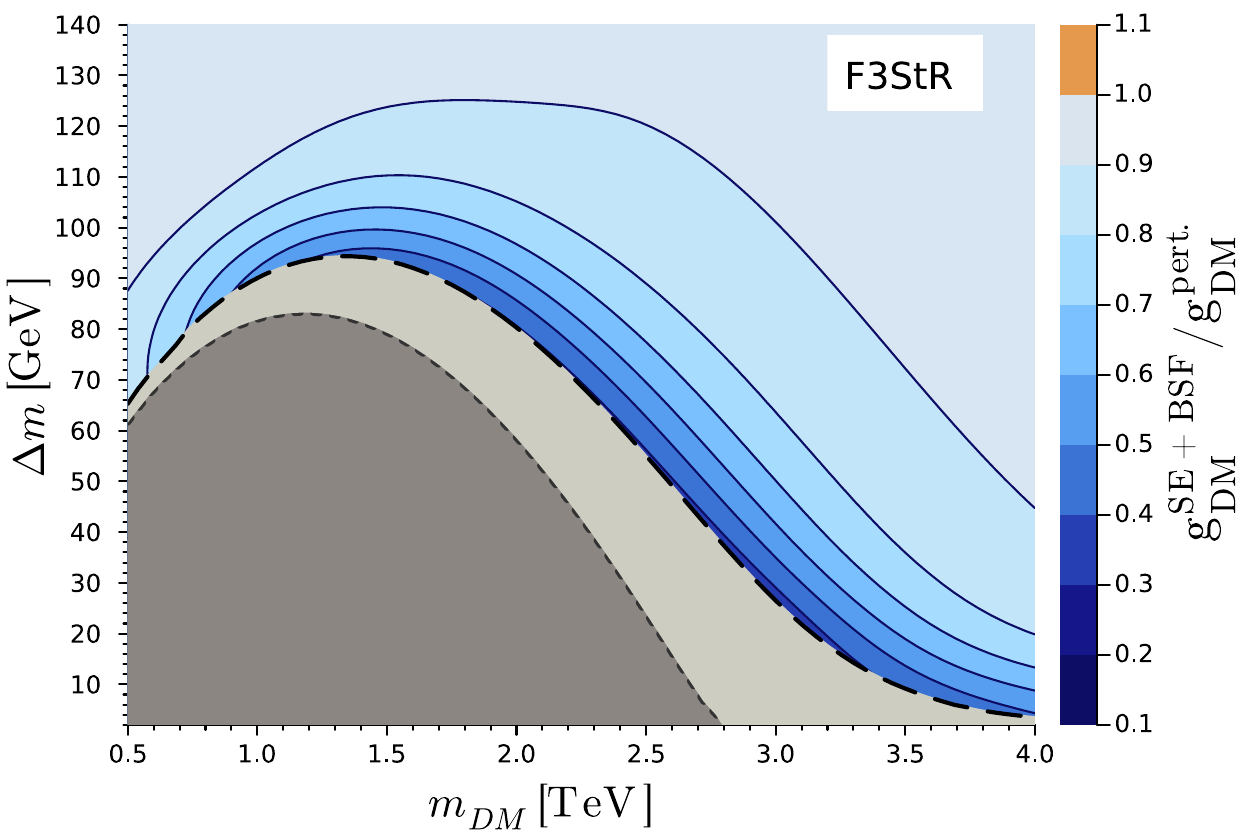}
    \end{minipage}

    \vspace{0.5em}  
    \captionof{figure}{\justifying Ratio of Yukawa couplings $\gdm$ for which $\Omega h^2 = 0.12$ with and without non-perturbative effects. Blue contours indicate a smaller value of $\gdm$ required to satisfy the relic density constraint when long-range effects are included, while orange contours indicate the opposite. The gray areas indicates the regime of non thermal DM. 
    In dark gray, the corresponding regime without including long-range effects is shown, while the light gray region indicates the same regime taking into account long-range effects.}
    \label{fig:gDM_ratio_plots}
\end{minipage}
\end{center}
\vspace{1em}
\twocolumngrid
    
    \item Sommerfeld effect: \texttt{F3S} models show a stronger impact of from the Sommerfeld enhancement at large $\gdm$ (compare the offset between the blue dash-dotted and blue dashed curves in the upper panel (\texttt{S3M}) and lower panel (\texttt{F3S}) in Fig.~\ref{fig:BandScans}), as now also the unsuppressed $\overline{X} X \rightarrow \overline{q} q$ channels (see Fig.\,\ref{fig:processXXd2} and \ref{fig:processXXdqq}) are enhanced by the Sommerfeld effect.
    \item Bound State Formation (BSF) contributes similarly for both scalar and fermionic mediators.
    However, for fermionic mediators, there is an additional consideration related to spin.
    The fermionic mediator pair can form both spin-singlet and spin-triplet bound states. 
    At late times, after chemical freeze-out, when bound states decay faster than they are ionized by thermal photons, the spin-triplet states contribute more strongly to the annihilation rate.
    However, such late times are not directly impacting the freeze-out abundance.
    Despite this potential enhancement in the late universe, the overall impact of BSF on the relic density is actually smaller for fermionic mediators than for scalar mediators. 
    This is because fermionic models already have larger perturbative annihilation cross sections (as explained earlier), so the relative enhancement from BSF is reduced.
    This comparative effect is visible in the offset between the dash-dotted (perturbative only) and solid (including BSF) curves in Fig.\,\ref{fig:BandScans}, where the offset is smaller for fermionic mediators (lower panel) than for scalar mediators (upper panel).
\end{itemize}

\onecolumngrid
\twocolumngrid

For the \texttt{S3MuR}, \texttt{S3MtR} and \texttt{F3StR} models one observes a steep increase of the $\gdm = 1$ bands (see Fig.~\ref{fig:BandScans}). 
This is because, for low $\mdm$, $\chi$ annihilations mediated by $\gdm$ dominate $\langle \sigma v_\text{rel} \rangle$ and mediator (co-)annihilation becomes almost irrelevant (diagram \ref{fig:processCC} dominates) such that the relic density is practically independent of $m_X$ for $\Delta m \lesssim \mdm$. This is the case for all the plots we show in this paper.
For the \texttt{F3SuR} model, this feature appears for masses $\mdm < 200$\,GeV or for higher values of $\gdm$, because the $\overline{X} X \rightarrow \overline{q} q$ annihilations (diagrams \ref{fig:processXXd2} and \ref{fig:processXXdqq}) neither experience a phase space suppression (like in the \texttt{F3StR} model due to the coupling to the massive top quark), nor are these annihilations helicity suppressed (as for the \texttt{S3} models), such that it takes higher values of $\gdm$ to become larger than the $\alpha_s$ mediated $\overline{X} X$ annihilations (diagrams \ref{fig:processXXd1} - \ref{fig:processXXd3}).

The impact of non-perturbative effects on the parameter space is summarized in Fig.\,\ref{fig:gDM_ratio_plots}, which shows the ratio of Yukawa couplings $g_\text{DM}$ required to obtain the observed relic density $\Omega h^2 = 0.12$ with and without long-range effects. 
When the inclusion of Sommerfeld enhancement and bound state formation lowers the required coupling, the corresponding parameter region is shown in blue, indicating an overall enhancement of the annihilation rate. Conversely, regions in orange correspond to a higher required $g_\text{DM}$, reflecting scenarios where non-perturbative effects suppress the effective cross section.
The gray shaded areas denote the regime where DM is not produced via thermal freeze-out, instead DM production must proceed via conversion-driven freeze-out or freeze-in. The darker gray shows the corresponding region without considering long-range effects, while the lighter gray extension illustrates how non-perturbative processes can enlarge the boundary of the non-freeze-out regime.

As a cross-check and validation of our work we compare our results with existing literature.
Ref.~\cite{Arina:2020tuw} has studied the \texttt{S3MuR} and \texttt{F3SuR} without including long-range effects and our perturbative results agree with their findings. The Sommerfeld effect has been included for \texttt{S3MuR} in Ref.~\cite{Arina:2023msd}, confirming the results presented in this work. 
The impact of the Sommerfeld effect for \texttt{F3SuR} has been assessed in Refs.~\cite{Giacchino:2015hvk, Arina:2023msd} and Ref.~\cite{Colucci:2018vxz} studied \texttt{F3StR}.

Our perturbative results show perfect agreement with previous studies, while comparisons of the Sommerfeld effect for the \texttt{F3S} models reveal modest deviations ($\leq$20\,\% for $\delta \simeq 10^{-2}$).
These differences could originate from our approximation of the color decomposition\footnote{The color decomposition that we employ is detailed in Appendix,\ref{app:color_decomp}.}, which may differ from that adopted in Refs.~\cite{Giacchino:2015hvk, Arina:2023msd, Colucci:2018vxz}, as well as from our omission of subleading contributions such as loop corrections and three-body decays.
However, importantly, these inaccuracies at small mass splitting are subleading compared to the inclusion of bound-state formation, which constitutes the dominant annihilation channel in this regime.

\subsection{Experimental constraints}\label{subsec:results_exclusions}
In the following, we aim to confront the parameter space with constraints from direct detection and collider searches as summarized in Figs.~\ref{fig:Money_plots_S3M} and \ref{fig:Money_plots_F3S}. 

In these plots, the dark matter coupling is fixed by imposing not to overproduce the relic density: the left panels show results from conventional perturbative calculations, while the right panels incorporate our refined treatment, which includes Sommerfeld enhancement and BSF. 

Crucially, our refined calculation predicts lower values of the $t$-channel coupling $\gdm$ compared to the perturbative approach, as long-range effects mediated by $\alpha_s$ dominate the annihilation cross-section for a sufficiently small mass splitting and $\gdm$ can thus be much lower whilst reproducing the correct relic abundance (see Fig.~\ref{fig:gDM_ratio_plots}). 
This reduction in $\gdm$ significantly enlarges the allowed parameter space, reopening regions that would otherwise be excluded from direct detection experiments. 

The hatched areas at high $\Delta m$ mark the estimated breakdown of perturbative unitarity~\cite{Cahill-Rowley:2015aea,Schuessler:2007av} ($\gdm^2 > 4 \pi$), while the dark gray region indicates when chemical equilibrium in the dark sector is lost ($\Gamma^{\chi \rightarrow X} \lesssim H$). 
The absence of chemical equilibrium is invalidating the treatment of using a single effective Boltzmann equation for the DM relic abundance, as discussed in Eq.~\eqref{eq:effective_sigmav_coannih} in Sec.~\ref{sec:model}. 
We label this are as the \textbf{No Freeze-Out} region in our plots.

Importantly, this region remains \textit{unconstrained} by direct detection experiments due to the extremely small values of $\gdm \lesssim 10^{-7}$ — insufficient to produce detectable nuclear recoils.
Instead, collider searches for long-lived particles become the primary probe in the gray region, as explored in Refs.~\cite{Garny:2021qsr, Becker:2022iso, Heisig_2024, Arina:2025zpi}. 
In the following we discuss the experimental limits arising from the search for bound state resonances as well as direct detection experiments shown in Figs.~\ref{fig:Money_plots_S3M} and \ref{fig:Money_plots_F3S} in more detail. We summarize how their interpretation is altered when the Sommerfeld effect and bound state formation are considered. 

\subsubsection{Bound State Searches at the LHC}\label{subsubsec:bsf_at_LHC}
Dark Matter mediator masses $m_X = \mdm + \Delta m$ can be probed at colliders through resonant diphoton searches. When produced at the LHC, mediator pairs $(\widetilde{X} X)$ may form bound states that decay predominantly into gluons — which is the dominant decay channel for the relic density calculations — but also into photons, albeit with a much smaller branching fraction, irrelevant for the effective BSF cross section ($\frac{\Gamma(\mathcal{B}(\widetilde{X} X) \rightarrow \gamma \gamma)}{\Gamma(\mathcal{B}(\widetilde{X} X) \rightarrow g g)} = \frac{9}{8}\frac{\alpha^2_{QED}}{\alpha^2_s} \sim \mathcal{O}(10^{-3})$, \cite{Martin:2008sv}). 
However, the diphoton channel offers superior sensitivity due to its cleaner experimental signature \cite{Younkin:2009zn, Becker:2022iso}. 
We derive constraints on $m_X$ by applying ATLAS diphoton search results \cite{ATLAS:2017ayi}, following the methodology of Refs.~\cite{Becker:2022iso, Martin:2008sv, Younkin:2009zn}.
The production cross section for a bound state $\mathcal{B}$ is given by \cite{Martin:2008sv}
\begin{align}
    \sigma(pp \rightarrow \mathcal{B}) = \frac{\pi^2}{8 m^2_\mathcal{B}} \Gamma(\mathcal{B} \rightarrow gg) P_{gg}\left( \frac{m_\mathcal{B}}{\text{13 TeV}} \right) \label{eq:BSF_LHC_prod_XS}
\end{align}
where $P_{gg}$ is the two gluon luminosity. This production cross section holds for one bound state made out of scalars. For couplings to $n_f$ flavors of quarks, the limits are $n_f$ times stronger because of the flavor universal QCD coupling. On the other hand, bound states made out of fermions have a production cross section suppressed by $1/4$, as only the spin singlet state decays into two gauge bosons. For these two reasons, we only obtain limits for the \texttt{S3MuR} and \texttt{S3MtR} models for $200$\:GeV$\lesssim m_\mathcal{B} \lesssim 450$\:GeV (see the blue areas in the lower left corners in Fig.~\ref{fig:Money_plots_S3M}). 
It is important to emphasize that these limits are independent of the $t$-channel coupling $\gdm$, provided the condition $\Gamma_X \lesssim E_B$ is satisfied. 
Thus, they also apply to any dark matter production mechanism within the gray \textbf{No Freeze-Out} region. 
In this context, we stress that conventional resonant searches can still yield valuable insights for this class of models, potentially bridging the gap between prompt and long-lived particle searches. For more details and the demonstration of their impact in the parameter space we refer the reader to Ref.~\cite{Becker:2022iso}.

\subsubsection{Direct Dark Matter Detection}\label{subsubsec:direct_detection}

To obtain limits from direct dark matter detection experiments, we calculate the DM-nucleus cross section using the standard effective field theory (EFT) for Majorana and real scalar DM found in the literature \cite{Drees:1992rr, Drees:1993bu, Hisano:2010ct, Gondolo:2013wwa, Hisano:2015bma, Mohan:2019zrk, Arcadi:2023imv}. The effective Lagrangian for fermionic DM relevant to our discussion is given by\footnote{The corresponding terms for scalar DM look the same but all coefficients for operators involving $\gamma$ matrices vanish.}

\begin{align}
    \mathcal{L}_\text{eff} &= \sum_{q = u,d}\widetilde{c}_q (\overline{\chi} \gamma^\mu \gamma_5 \chi) (\overline{q} \gamma_\mu \gamma_5 q) \label{eq:MajoranaDM_eff_Lag} \\
     &+ \sum_{q =u,d,s} f_q m_q \overline{\chi} \chi \overline{q}q + f_G \: \frac{\alpha_s}{\pi} G_{\mu \nu}^a G^{a \, \mu \nu}  \notag \\ 
     + \sum_{q = u,d,s} &\left( g_1^q \: \frac{\overline{\chi} i \partial^\mu \gamma^\nu \chi}{\mdm} + g_2^q \: \frac{\overline{\chi} (i \partial^\mu)(i \partial^\nu)\chi}{\mdm^2}\right) \mathcal{O}^q_{\mu \nu} \notag.
\end{align}
The first term in Eq.~\eqref{eq:MajoranaDM_eff_Lag} induces a spin-dependent (SD) coupling for fermionic DM\footnote{Since scalar DM carries no spin, SD DM-nucleon interactions can only arise with a velocity suppression.} coupled to first generation quarks.
Thus, of the four models discussed in this paper, this operator is only active for \texttt{S3MuR}, excluding the purple area in the upper row of Fig.~\ref{fig:Money_plots_S3M} and Fig.~\ref{fig:Money_plots_S3M_linear}. 
The first term in the second line is the quark matching coefficient, while the second term constitutes the gluon matching coefficient.
Finally, the operator $\mathcal{O}^q_{\mu \nu}$ in the third line is the \textit{quark twist-2 operator}. For the \texttt{F3S} models (scalar DM), the first of the two quark twist-2 coefficients and the spin-dependent tree level interaction are absent: $g_1^q = \widetilde{c}_q = 0$.

We adopt the conventions of Ref.~\cite{Hisano:2010ct} and split the gluonic matching coefficient $f_G$ into long distance (LD) - and short distance (SD) contributions\footnote{The acronym SD is standard at the same time for \textit{spin-dependent} and \textit{short distance} in the direct detection literature. However, in a given context there is little potential for confusion.}, 
\begin{align}
    f_G &= \sum_{q = u,d,s,c,b,t}\left. f^{(q)}_G \right|_\text{SD} + \sum_{Q = c,b,t}\left. f^{(Q)}_G \right|_\text{LD}  \label{eq:gluon_matching_SD_LD},
\end{align}
thereby correctly disentangling contributions stemming from light quarks ($u,d,s$) into perturbative ones (accounted for in the matching coefficients) and non-perturbative ones (captured already at the level of the tabulated nuclear matrix elements \cite{Del_Nobile_2022}). 

As pointed out in Ref.~\cite{Arcadi:2023imv}, Higgs-penguin contributions to the gluon matching coefficient $f_G$ can be relevant for models coupling to top quarks and we include them in our analysis,\footnote{Readers should be aware that some expressions in Ref.~\cite{Arcadi:2023imv} display inconsistent mass dimensionality. Though often correctable via dimensional analysis, we encountered multiple instances requiring explicit correction during our implementation.} using the relation given in Ref.~\cite{Hisano:2015bma}. 

We do not, however, include contributions from the anapole moment \cite{Ibarra:2022nzm} to the \texttt{S3M} models or employ an RG improvement on the Wilson coefficients of the DM-nucleon EFT \cite{Mohan:2019zrk}, as these refinements are beyond the scope of this paper and we deem their impact to be small for the following reasons:
\begin{itemize}
    \item For simplified $t$-channel DM  models, the anapole moment is typically velocity suppressed and thus subleading~\cite{Arina:2025zpi}. 
    \item For real scalar DM, the RG evolution down to the nucleon scale has been discussed in \cite{Brod:2017bsw} (without applications to direct direction limits, as the focus of this work was the EFT operator basis up to dimension 7). 
    The RG improvement in the direct detection of Majorana DM has been worked out in Ref.~\cite{Mohan:2019zrk} and it has been found that in particular $f_G$ receives large corrections of order $\mathcal{O}(\alpha_s)$.
    However, our work differs with respect to Ref.~\cite{Mohan:2019zrk} in the way we define our effective operator to gluons.
    Crucially, following the convention of Ref.~\cite{Hisano:2015bma}, we explicitly include a factor of $\alpha_s$ in the \textit{definition} of the operator coupling to gluons (whereas $\alpha_s$ is not part of the definition of the corresponding operator in Ref.~\cite{Mohan:2019zrk}), which renders the operator defined in our convention manifestly RG invariant: $\mu\frac{d}{d \mu} \frac{\alpha_s}{\pi} G^a_{\mu \nu} G^{a\, \mu \nu} = 0$. 
    This convention automatically includes the running of the operator in its definition, such that an additional RG improvement is not needed in our setup. 
\end{itemize}
Moreover we neglect gluon twist-2 operators, as these operators are suppressed with one order of $\alpha_s$ compared to the DM-gluon effective operator mediated by $f_G$ \cite{Hisano:2010ct}. \\ 
We obtain direct detection bounds on our models by comparing the SD and SI cross sections to the recent limits of the LUX-ZEPLIN (LZ) experiment~\cite{LZ:2024zvo}: 
\begin{itemize}
    \item For \texttt{S3MuR}, the most constraining limits (shown in the upper rows of Figs.~\ref{fig:Money_plots_S3M} and \ref{fig:Money_plots_S3M_linear}) stem from the SD scattering, while the SI limits are driven by the up-quark twist-2 contributions $g_1^q + g_2^q$ \cite{Hisano:2010ct, Mohan:2019zrk, Arcadi:2023imv}. 
    \item The constraints for the \texttt{S3MtR} model (shown in the lower rows of Figs.~\ref{fig:Money_plots_S3M} and \ref{fig:Money_plots_S3M_linear}) have a different origin: The Higgs-penguin contribution is essentially independent of $\Delta m$~\cite{Arcadi:2023imv}, while the DM-gluon effective interaction strongly decreases with larger $\Delta m$~\cite{Hisano:2010ct}. 
    Thus, the DM-gluon effective interaction causes the limits for $\Delta m \lesssim 250$\,GeV, where the upper excluded region stems from the aforementioned loop-induced Higgs penguin contribution. 
    \item The \texttt{F3SuR} model is almost completely excluded by direct detection (as can be seen in the upper rows of Figs.~\ref{fig:Money_plots_F3S} and \ref{fig:Money_plots_F3S_linear}), where - as in the case of \texttt{S3MuR} - the biggest contribution stems from the quark twist-2 coefficient $g_2^q$.
    Apart from the unconstrained \textbf{No Freeze-Out} region, there is only a small area in the upper right corner that is not excluded, as for high masses both the experimental sensitivity and the quark twist-2 coefficient $g_2^q$ is smaller.
    \item Finally, \texttt{F3StR}, displayed in the lower rows of Figs.~\ref{fig:Money_plots_F3S} and \ref{fig:Money_plots_F3S_linear}, is constrained by the Higgs penguin contribution to both the $f_q$ and $f_G$ Wilson coefficients and by gluon loops contributing to $f_G$~\cite{Arcadi:2023imv}, leading to a connected excluded area. 
\end{itemize}
Direct detection severely constrains the viable parameter space across all four models. However, including long-range effects (the right panels of Figs.~\ref{fig:Money_plots_S3M}-\ref{fig:Money_plots_F3S_linear}) enlarges the gray, unconstrained \textbf{No Freeze-Out} region significantly. This enlargement of the theoretically allowed parameter space at weak coupling represents the key impact of long-range effects in our analysis. \\ 

When including Sommerfeld effect and BSF, parameter space is reopened, particularly in the strongly coannihilating (small $\Delta m$) regime.
Most strikingly, the upper mass limits from direct detection (the largest allowed $\mdm$ values for $\Delta m \rightarrow 0$ in Figs.~\ref{fig:Money_plots_S3M}-\ref{fig:Money_plots_F3S_linear}) shift by a significant amount:
\begin{itemize}
    \item \texttt{S3MuR}: 1.4\,TeV $\rightarrow$ 4.0\,TeV
    \item \texttt{S3MtR}: 2.8\,TeV $\rightarrow$ 4.6\,TeV
    \item \texttt{F3SuR}: 1.6\,TeV $\rightarrow$ 4.0\,TeV
    \item \texttt{F3StR}: 3.0\,TeV $\rightarrow$ 5.0\,TeV
\end{itemize}

\onecolumngrid
\begin{center}
\begin{minipage}{\textwidth}
    \centering

    \begin{minipage}{0.48\textwidth}
        \centering
        \includegraphics[scale=0.8]{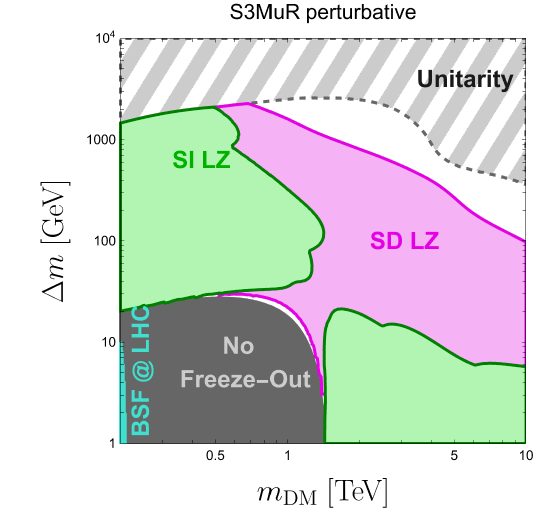}
        \\[0.3em]
    \end{minipage}\hfill
    \begin{minipage}{0.48\textwidth}
        \centering
        \includegraphics[scale=0.8]{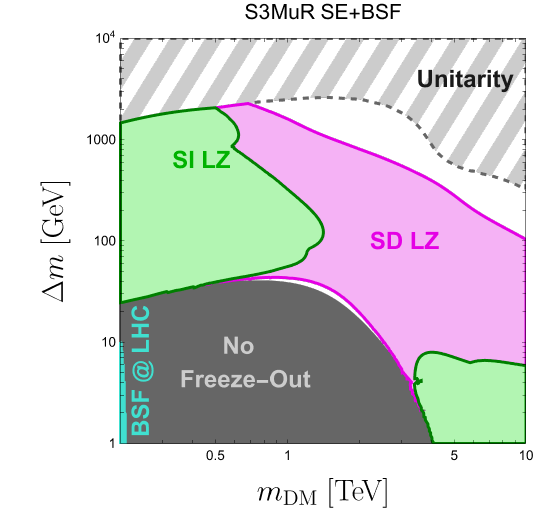}
        \\[0.3em]
    \end{minipage}

    \vspace{0.5em}

    \begin{minipage}{0.48\textwidth}
        \centering
        \includegraphics[scale=0.8]{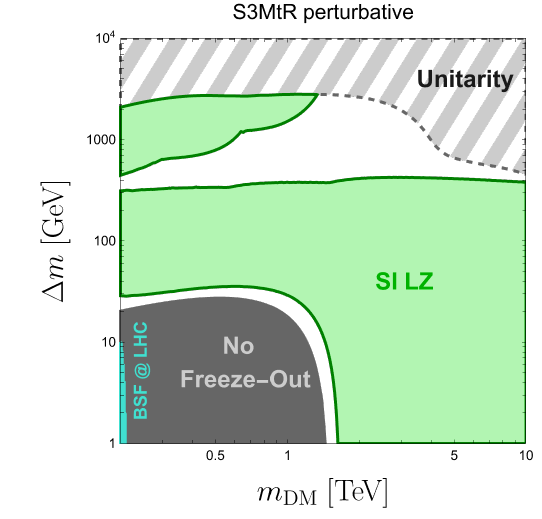}
        \\[0.3em]
    \end{minipage}\hfill
    \begin{minipage}{0.48\textwidth}
        \centering
        \includegraphics[scale=0.8]{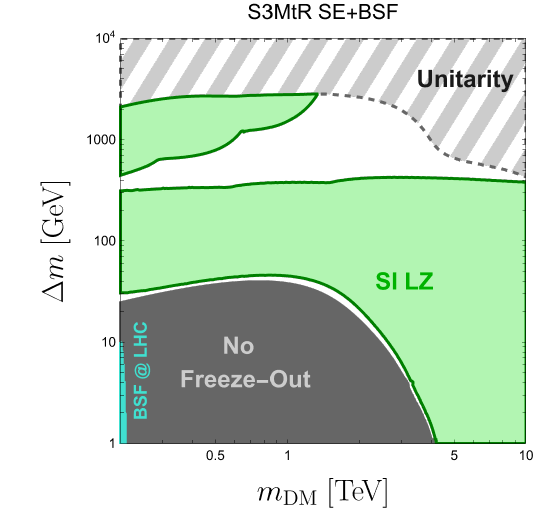}
        \\[0.3em]
    \end{minipage}
    \vspace{1em}
    \captionof{figure}{\justifying Experimental bounds on the \texttt{S3M} models with Majorana Dark Matter 
    and a complex scalar mediator. The upper row is for the models coupling to the up-quark, 
    the lower for couplings to the top. The left column shows the constraints without 
    long-range effects, while the right column includes them. }
    \label{fig:Money_plots_S3M}
\end{minipage}
\end{center}
\vspace{1em}
\twocolumngrid

\onecolumngrid
\begin{center}
\begin{minipage}{\textwidth}
    \centering

    \begin{minipage}{0.48\textwidth}
        \centering
        \includegraphics[scale=0.6]{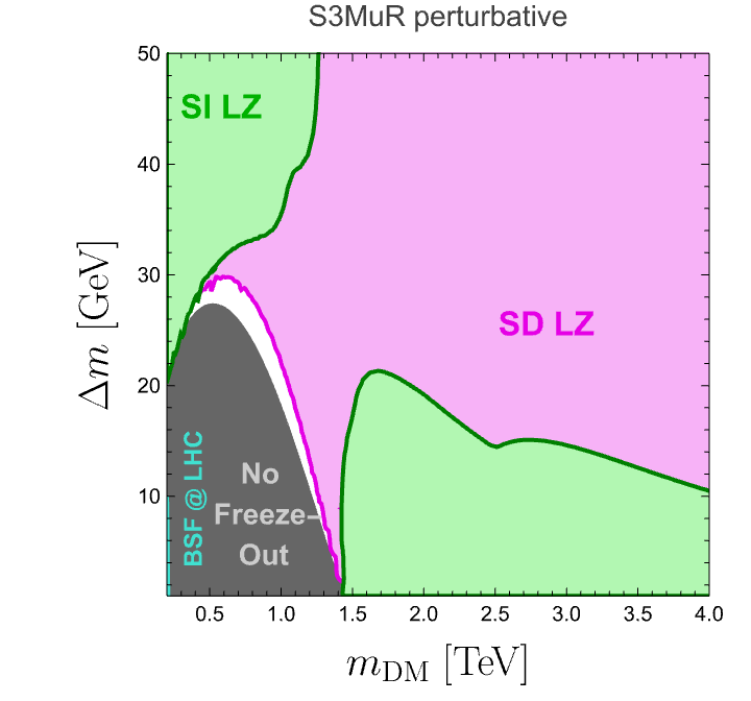}
        \\[0.3em]
    \end{minipage}\hfill
    \begin{minipage}{0.48\textwidth}
        \centering
        \includegraphics[scale=0.6]{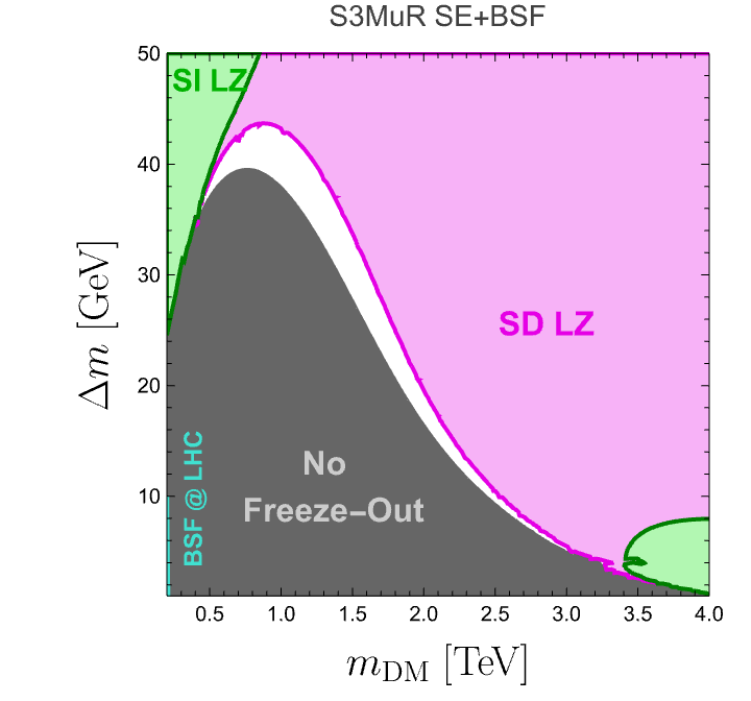}
        \\[0.3em]
    \end{minipage}

    \vspace{0.5em}

    \begin{minipage}{0.48\textwidth}
        \centering
        \includegraphics[scale=0.27]{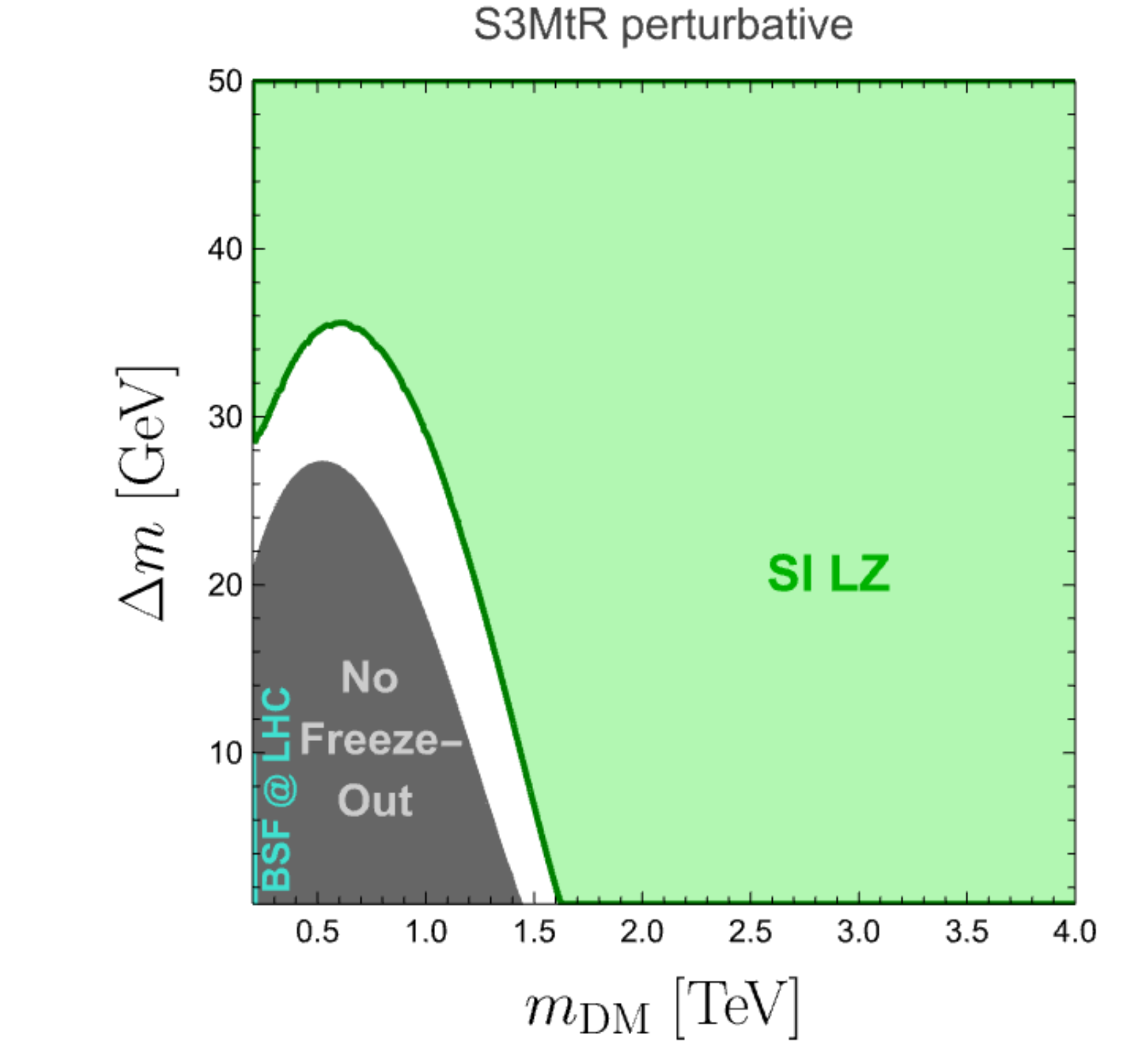}
        \\[0.3em]
    \end{minipage}\hfill
    \begin{minipage}{0.48\textwidth}
        \centering
        \includegraphics[scale=0.27]{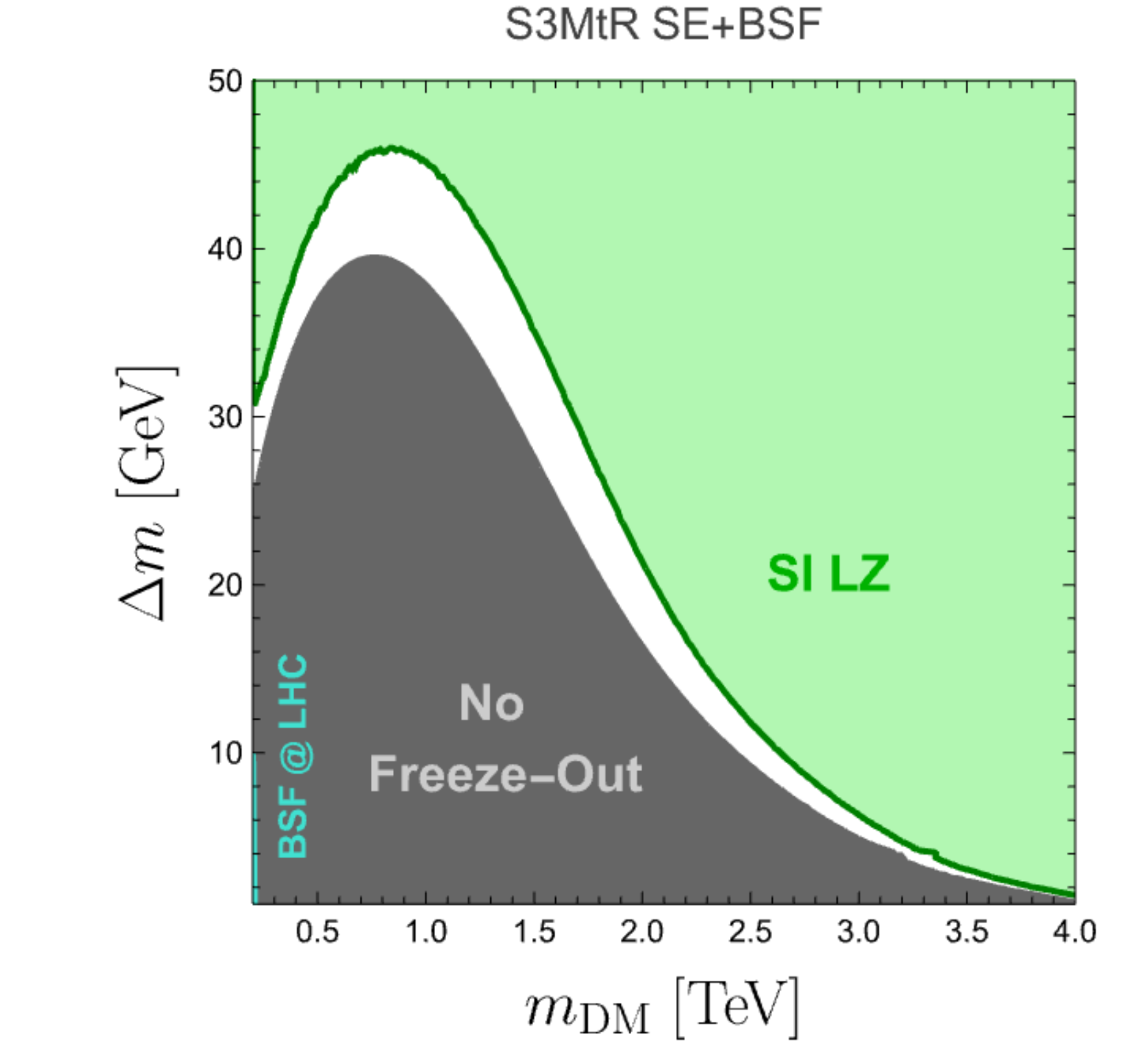}
        \\[0.3em]
    \end{minipage}
    \vspace{1em}
    \captionof{figure}{\justifying Same as Fig.\,\ref{fig:Money_plots_S3M}, but plotted on a linear Dark Matter mass axis. Experimental bounds on the S3M model with Majorana Dark Matter and a complex scalar mediator, showing couplings to up quarks (upper row) and top quarks (lower row), without (left) and with (right) long-range effects.}
    \label{fig:Money_plots_S3M_linear}
\end{minipage}
\end{center}
\vspace{1em}
\twocolumngrid

These shifts - up to nearly 3 TeV in mass - highlight the importance of including long-range interactions in the phenomenology of DM.
Consequently, we find an expanded viable parameter space for the S3MuR, S3MtR, and F3StR models.
This underscores that earlier constraints, derived from perturbative treatments or from including only the Sommerfeld effect, were overly restrictive for these scenarios.
In contrast, the impact on the already strongly constrained F3SuR model is negligible.


\onecolumngrid
\begin{center}
\begin{minipage}{\textwidth}
    \centering

    \begin{minipage}{0.48\textwidth}
        \centering
        \includegraphics[scale=0.85]{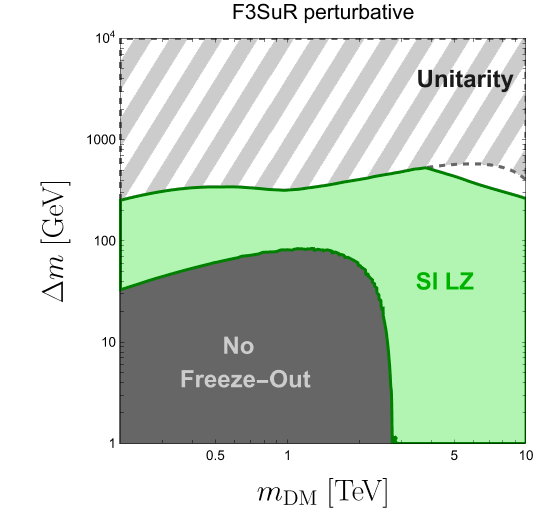}
        \\[0.3em]
    \end{minipage}\hfill
    \begin{minipage}{0.48\textwidth}
        \centering
        \includegraphics[scale=0.85]{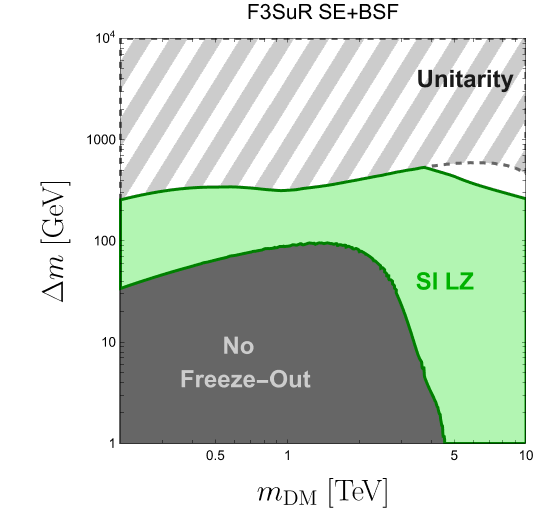}
        \\[0.3em]
    \end{minipage}

    \vspace{0.5em}

    \begin{minipage}{0.48\textwidth}
        \centering
        \includegraphics[scale=0.85]{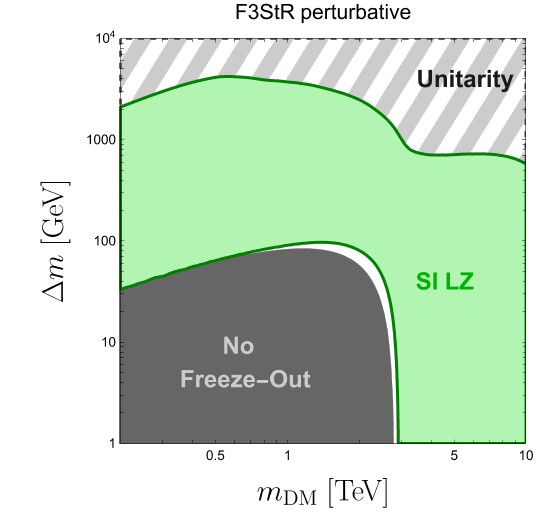}
        \\[0.3em]
    \end{minipage}\hfill
    \begin{minipage}{0.48\textwidth}
        \centering
        \includegraphics[scale=0.85]{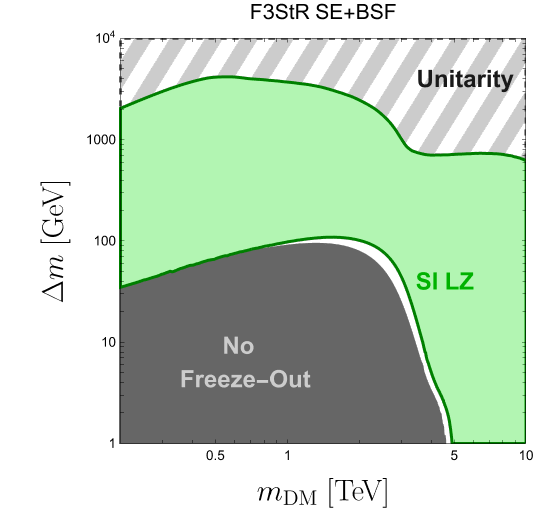}
        \\[0.3em]
    \end{minipage}

    \vspace{1em}

    \captionof{figure}{\justifying Experimental bounds on the \texttt{F3S} models with real scalar Dark Matter and a Dirac fermionic mediator. The results are presented as in Fig.~\ref{fig:Money_plots_S3M}.}
    \label{fig:Money_plots_F3S}
\end{minipage}
\end{center}
\vspace{1em}
\twocolumngrid

\onecolumngrid
\begin{center}
\begin{minipage}{\textwidth}
    \centering

    \begin{minipage}{0.48\textwidth}
        \centering
        \includegraphics[scale=0.25]{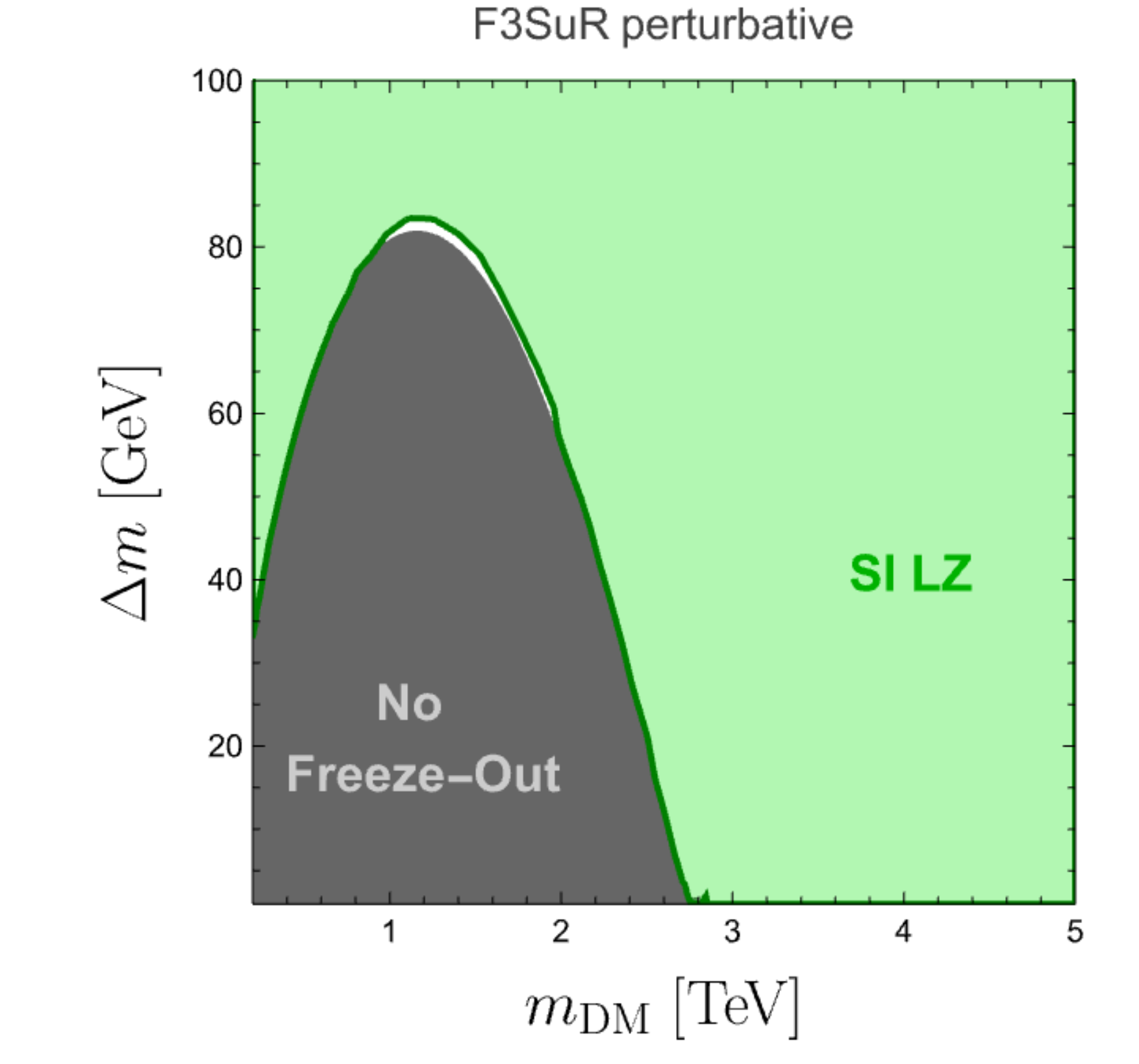}
        \\[0.3em]
    \end{minipage}\hfill
    \begin{minipage}{0.48\textwidth}
        \centering
        \includegraphics[scale=0.25]{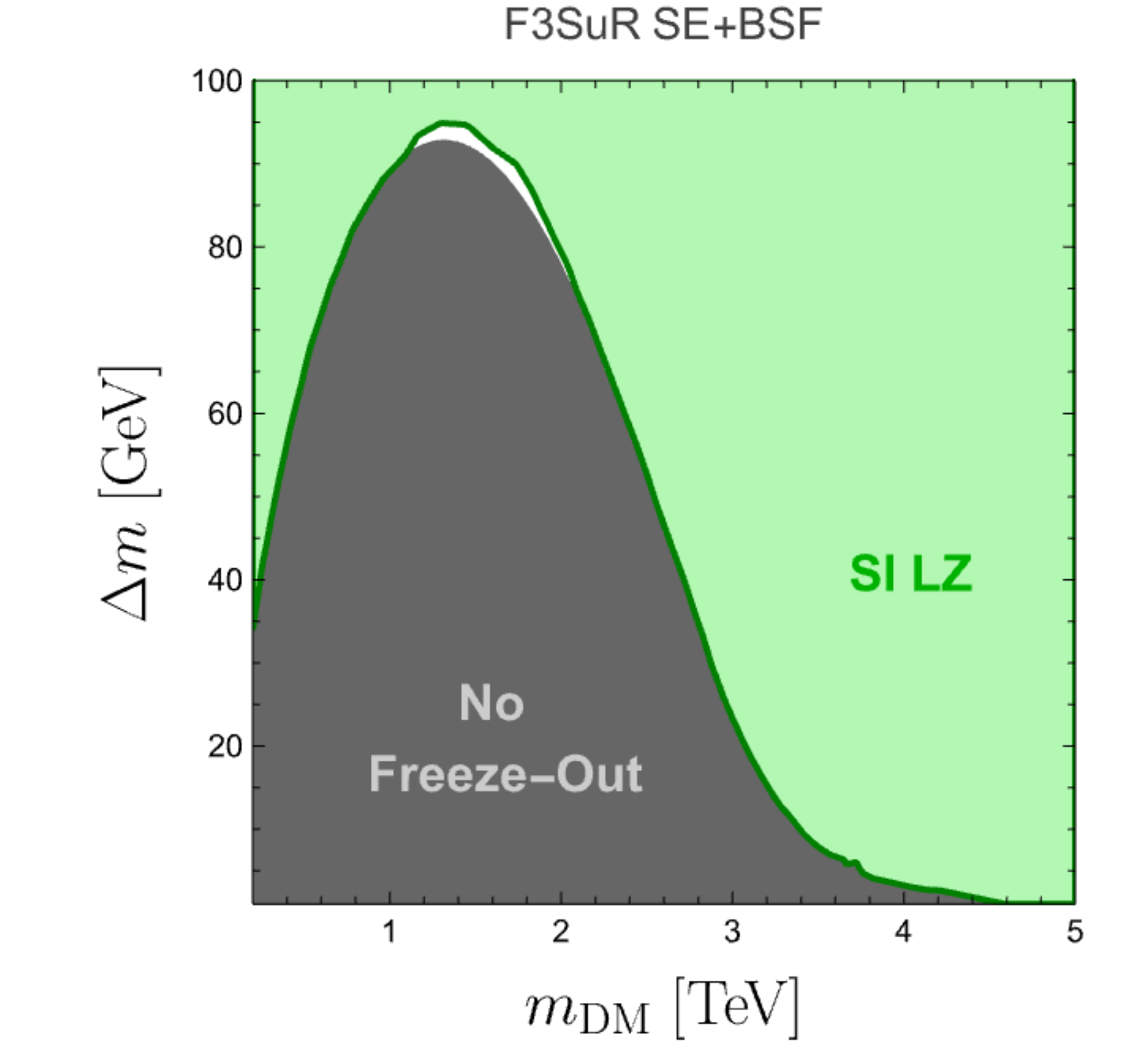}
        \\[0.3em]
    \end{minipage}

    \vspace{0.5em}

    \begin{minipage}{0.48\textwidth}
        \centering
        \includegraphics[scale=0.25]{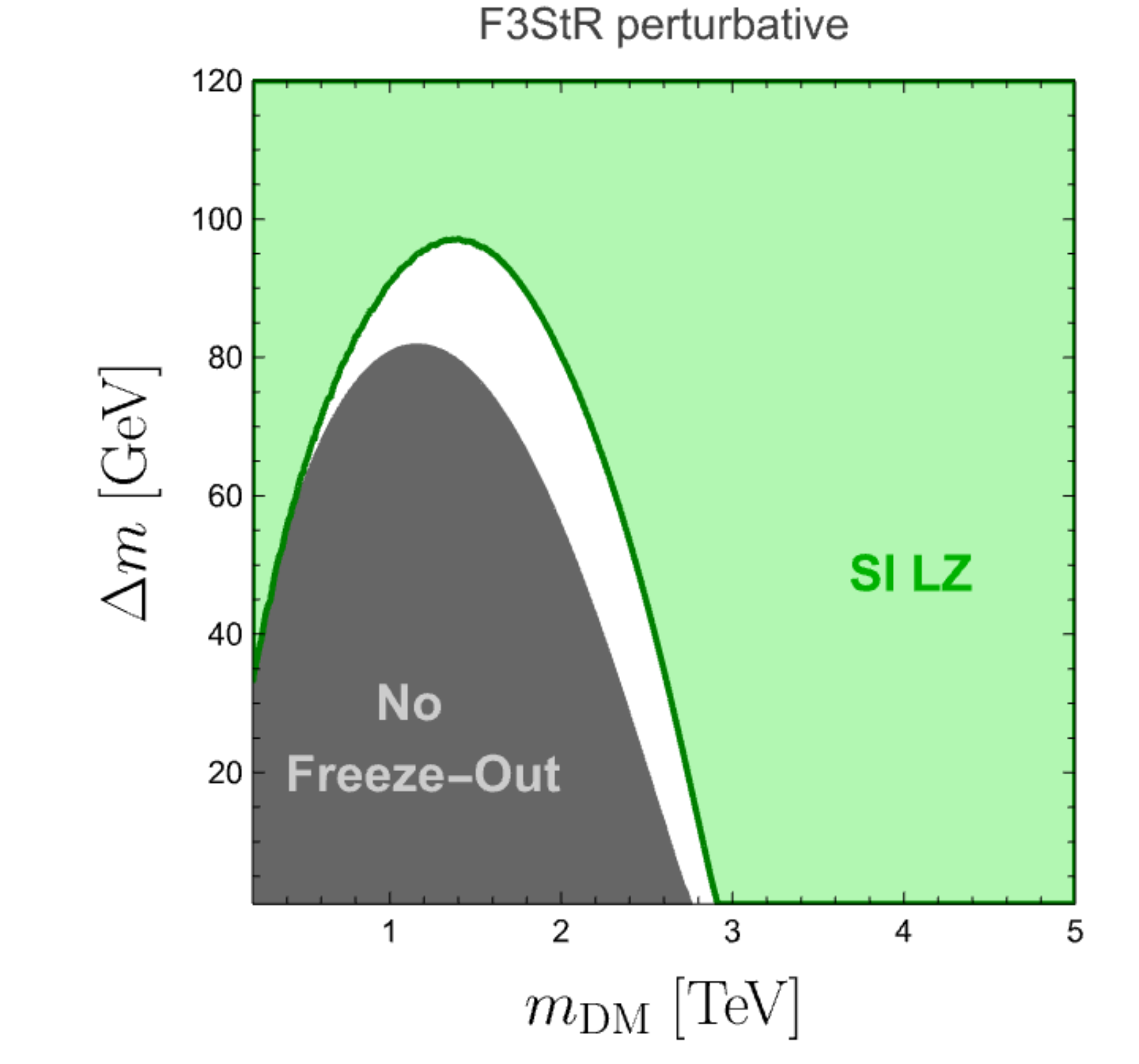}
        \\[0.3em]
    \end{minipage}\hfill
    \begin{minipage}{0.48\textwidth}
        \centering
        \includegraphics[scale=0.25]{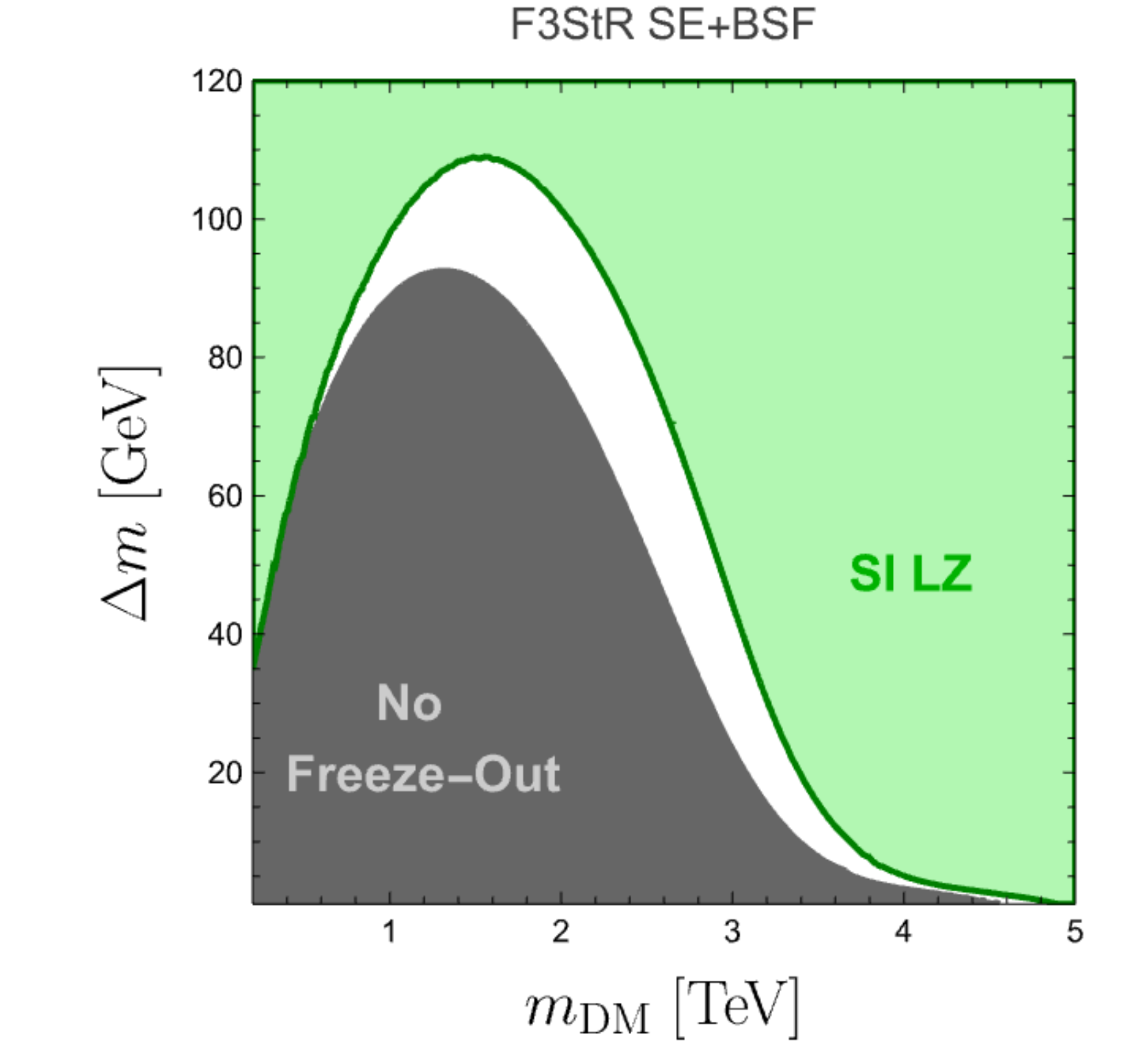}
        \\[0.3em]
    \end{minipage}

    \vspace{1em}

    \captionof{figure}{\justifying Fig.\,\ref{fig:Money_plots_F3S}, but plotted on a linear Dark Matter mass axis. 
    Experimental bounds on the F3S model with real scalar Dark Matter and a Dirac fermion mediator, showing couplings to up quarks (upper row) and top quarks (lower row), without (left) and with (right) long-range effects.}
    \label{fig:Money_plots_F3S_linear}
\end{minipage}
\end{center}
\vspace{1em}
\twocolumngrid

Our results demonstrate that a consistent treatment of non-perturbative effects is essential for accurate predictions in this class of models.

\section{Summary}\label{sec:conclusions}
Dark matter models with QCD-charged coannihilation partners remain among the most compelling and testable thermal relic scenarios. 
In such models, long-range effects, such as the Sommerfeld enhancement and bound state formation, can significantly modify the annihilation dynamics in the early universe, particularly in regions with small mass splittings where coannihilation is efficient. These effects alter the relic abundance and thereby also impact the interpretation of collider and direct detection limits with respect to the viable parameter space.
To quantify these effects precisely, we analyzed four representative benchmark models involving scalar or fermionic mediators coupled to either light or heavy quarks. 
In all cases, we find that including long-range effects leads to substantial shifts in the predicted dark matter–Standard Model $t$-channel coupling, when the dark matter and mediator masses are nearly degenerate. 
Consequently, the resulting experimental limits---particularly from direct detection---can change significantly, and upper bounds on the dark matter mass in the coannihilating regime are increased by factors of order unity.
To facilitate broader application of these insights, we developed \texttt{SE+BSF4DM}, a package linked to \texttt{micrOMEGAs}, which enables an intuitive and easy-to-use precision calculaton of the relic density including the Sommerfeld effect and bound state formation for QCD charged dark sector models.
Thanks to its simple structure, the tool integrates seamlessly into standard dark matter phenomenology pipelines and is available via \href{\CodeURL}{
  \faGithub~\textcolor{blue}{Github}}.

\section*{Acknowledgements}
We thank Alexander Pukhov for his support concerning the implementation in \texttt{micrOMEGAs} and Tanushree Bhattacharya for collaboration in initial stages of this project during her internship at the University of Mainz. M.~N thanks Merlin Reichard for useful input on the code and Dipan Sengupta, Sreemanti Chakraborti, Stefan Lederer and David Cabo Almeida for fruitful discussions.\\
The authors would like to express special thanks to the Mainz Institute for Theoretical Physics (MITP) of the Cluster of Excellence PRISMA$^+$ (Project ID 390831469), for its hospitality and support. 
M.~B., E.~C. and J.~H. knowledge support from the Emmy Noether grant “Baryogenesis, Dark Matter and Neutrinos: Comprehensive analyses and accurate methods in particle cosmology” (HA 8555/1-1, Project No. 400234416) funded by the
Deutsche Forschungsgemeinschaft (DFG, German Research Foundation). M.~N. acknowledges support from the DFG Collaborative Research Centre “Neutrinos, Dark Matter and Messengers” (SFB 1258). M.~B., E.~C. and J.~H. acknowledge support by the Cluster of Excellence “Precision Physics, Fundamental Interactions, and Structure of Matter” ($\text{PRISMA}^+$ EXC 2118/1) funded by the DFG within the German Excellence Strategy (project No. 390831469).
This work was supported by Istituto Nazionale di Fisica Nucleare (INFN) through the Theoretical Astroparticle Physics (TAsP) project, and in part by the Italian MUR Departments of Excellence grant 2023-2027 “Quantum Frontiers”. The work of M.B. was supported in part by the Italian Ministry of University and Research (MUR) through the PRIN 2022 project n. 20228WHTYC (CUP:I53C24002320006 and C53C24000760006).

\appendix

\setcounter{equation}{0}

\section{Color decomposition}\label{app:color_decomp}
This appendix contains further details on the employed color decomposition for scalar and fermionic mediators, which constitute the 'default' settings implemented in our package.
The coefficients $c^{[\textbf{R}]}_{\ell s}$ of Eq.~\eqref{eq:Sommerfeld_corrected_XS} for an arbitrary $2 \rightarrow 2$ process are given by:
\begin{align}
    &c^{[\textbf{R}]}_{\ell s} = \notag \\ 
    &\frac{1}{\overline{|\mathcal{M}_\text{full}|^2}} \left\vert \frac{2 \ell + 1}{2} \int_{-1}^1 d(\cos \theta) P_\ell (\cos \theta) P^{[\textbf{R}]} \mathcal{M}^s_\text{full} \right\vert^2,\label{eq:color_decom_coef_general}
\end{align}
where $P_\ell (\cos \theta)$ are the Legendre polynomials of the first kind that project out the $\ell$-th partial wave, $P^{[\textbf{R}]}$ is the projector onto the $SU(3)_c$ representation $\textbf{R}$ and $\mathcal{M}^s_\text{full}$ is the perturbative matrix element in a total spin $s$ configuration (to arbitrary loop order).
As elaborated in the main text, we will simplify the analysis by keeping only the $v^0_\text{rel}$ contribution, thereby automatically singling out solely contributions to the s-wave.
Thus one obtains
\begin{align}
    c^{[\textbf{R}]}_{0 s} &= \frac{\sigma^{[\textbf{R}]}_{0s}}{\sigma_0},\label{eq:coef_kQfac} \\
    \mathcal{S} \left( \sigma \right) &\simeq \sum_{[\mathbf{R}]} S_0^{[\mathbf{R}]}  \sum_s c^{[\textbf{R}]}_{0 s} \sigma_{0s}.\label{eq:Sommerfeld_numerical_sigma}
\end{align}

Explicitly, this implies that we only keep the $l=0$ and $v_\text{rel}\rightarrow 0$ terms in Eq.~\eqref{eq:Sommerfeld_corrected_XS}.
Thus, the following coefficients are employed in our code:
\begin{align}
    \mathcal{S} \left( \sigma_{X X^\dagger \rightarrow g \mathcal{A}} \right) &= S^{[\textbf{8}]}_0 \sigma_{X X^\dagger \rightarrow g \mathcal{A}},\label{eq:color_decomp_pure8} \\
    \mathcal{S} \left( \sigma_{X X^\dagger \rightarrow \mathcal{B} \mathcal{C}} \right) &= S^{[\textbf{1}]}_0 \sigma_{X X^\dagger \rightarrow \mathcal{B} \mathcal{C}},\label{eq:color_decomp_pure1} \\
    \mathcal{S}\left(\sigma_{X X^\dagger \rightarrow g g} \right) &\simeq \left(\frac{2}{N_c^2 - 2} S^{[\textbf{1}]}_0 \right. \notag \\ 
    &\left. + \frac{N_c^2 - 4}{N_c^2 - 2} S^{[\textbf{8}]}_0 \right) \sigma_{X X^\dagger \rightarrow g g},\label{eq:color_decomp_gg} \\
    \mathcal{S}\left( \sigma_{X_i^\dagger X_j \rightarrow \overline{q}_i q_j} \right) &\simeq \notag \\
    \sigma_{X_i^\dagger X_j \rightarrow \overline{q}_i q_j} &\times  
    \begin{cases}
        S^{[\textbf{8}]}_0, \text{$X$ fermionic \& } i=j  \\ 
        \frac{1}{N_c^2} S^{[\textbf{1}]}_0 + \frac{N_c^2 - 1}{N_c^2} S^{[\textbf{8}]}_0, \text{ else}
    \end{cases} ,\label{eq:color_decomp_qqbar} \\
    \mathcal{S}\left( \sigma_{X_i X_j \rightarrow q_i q_j} \right) &\simeq \notag \\
    \sigma_{X_i X_j \rightarrow q_i q_j} &\times
    \begin{cases}
        S^{[\textbf{6}]}_0, \text{$X$ scalar \& } i=j  \\ 
        \frac{N_c - 1}{2 N_c} S^{[\overline{\textbf{3}}]}_0 + \frac{N_c + 1}{2 N_c} S^{[\textbf{6}]}_0, \text{ else}
    \end{cases}  .\label{eq:color_decomp_qq}
\end{align}

A few comments to these equations are in order:
\begin{itemize}
    \item The particle $\mathcal{A}$ in Eq.~\eqref{eq:color_decomp_pure8} is a color-neutral SM vector boson $\mathcal{A} = \{A, W^\pm, Z \}$.
    \item In Eq.~\eqref{eq:color_decomp_pure1}, the particles $\mathcal{B}$ and $\mathcal{C}$ represent any color neutral particle of the SM: $\mathcal{B},\mathcal{C} = \{A, W^\pm, Z, h, \ell \}$.
    \item The relation Eq.~\eqref{eq:color_decomp_gg} is exact for the s-wave part. 
    One might expect different coefficients for fermionic $X$ particles, as they can be in spin singlet or triplet configurations. 
    That the spin triplet does not appear in Eq.~\eqref{eq:color_decomp_gg} is a manifestation of the Landau-Yang theorem, according to which a (color-neutral) massive spin 1 state cannot decay into two massless spin 1 particles.   
    \item For the process described in Eq.~\eqref{eq:color_decomp_qqbar}, there are two competing diagrams: one is a $t$-channel DM exchange mediated by $\gdm$, the other are s-channel SM gauge boson exchanges, in particular gluons. 
    Due to interferences between these diagrams, the coefficients formally depend on all the masses and couplings of the particles involved in the process. 
    The decomposition Eq.~\eqref{eq:color_decomp_qqbar} holds for the s-wave part and in the limit $\gdm^2/\alpha_s \rightarrow 0$, introducing a theoretical error of $\mathcal{O}\left(\frac{\gdm^4}{4 \pi^2 (N_c^2 - 1))\alpha_s^2}\right)$.
    We assume this limit of this particular Sommerfeld coefficient as the default setting in our code since long-range QCD effects are mostly of interest when colored annihilations dominate the annihilation cross section. 
    \item In equation Eq.~\eqref{eq:color_decomp_qq}, the annihilation of scalars into two identical quarks represents a special situation where the symmetry of the wavefunction forbids a contribution of the (antisymmetric) $[\overline{\mathbf{3}}]$ channel for $\ell = 0$. 
\end{itemize}
Equations with an $=$ sign are exact, while those with an $\simeq$ sign only hold in certain limits, as indicated above.
We state the general results for an $SU(N_c)$ group here, while in the code we explicitly set $N_c = 3$.
Following the discussion of Appendix~\ref{sec:Numerics}, the user can change these coefficients manually in the code by modifying the \texttt{SommerfeldFactor\_BSMmodel} function.  

\section{\texttt{micrOMEGAs} package for Sommerfeld effect and bound state formation}\label{sec:Numerics}

For a full manual, we refer the reader to Ref.~\cite{becker2025sebsf4dmmicromegaspackage}.
In practice, the code will go through all possible coannihilating pairs of dark sector particles and calculates the contributions of Sommerfeld effect and BSF in case the pair belongs to a $\mathbf{3} \otimes \overline{\mathbf{3}}$ or $\mathbf{3} \otimes \mathbf{3}$ color configuration of QCD (see also App.~\ref{app:color_decomp}). \\

\onecolumngrid
\begin{center}
\begin{minipage}{\textwidth} 
    \centering
    \includegraphics[scale=0.4]{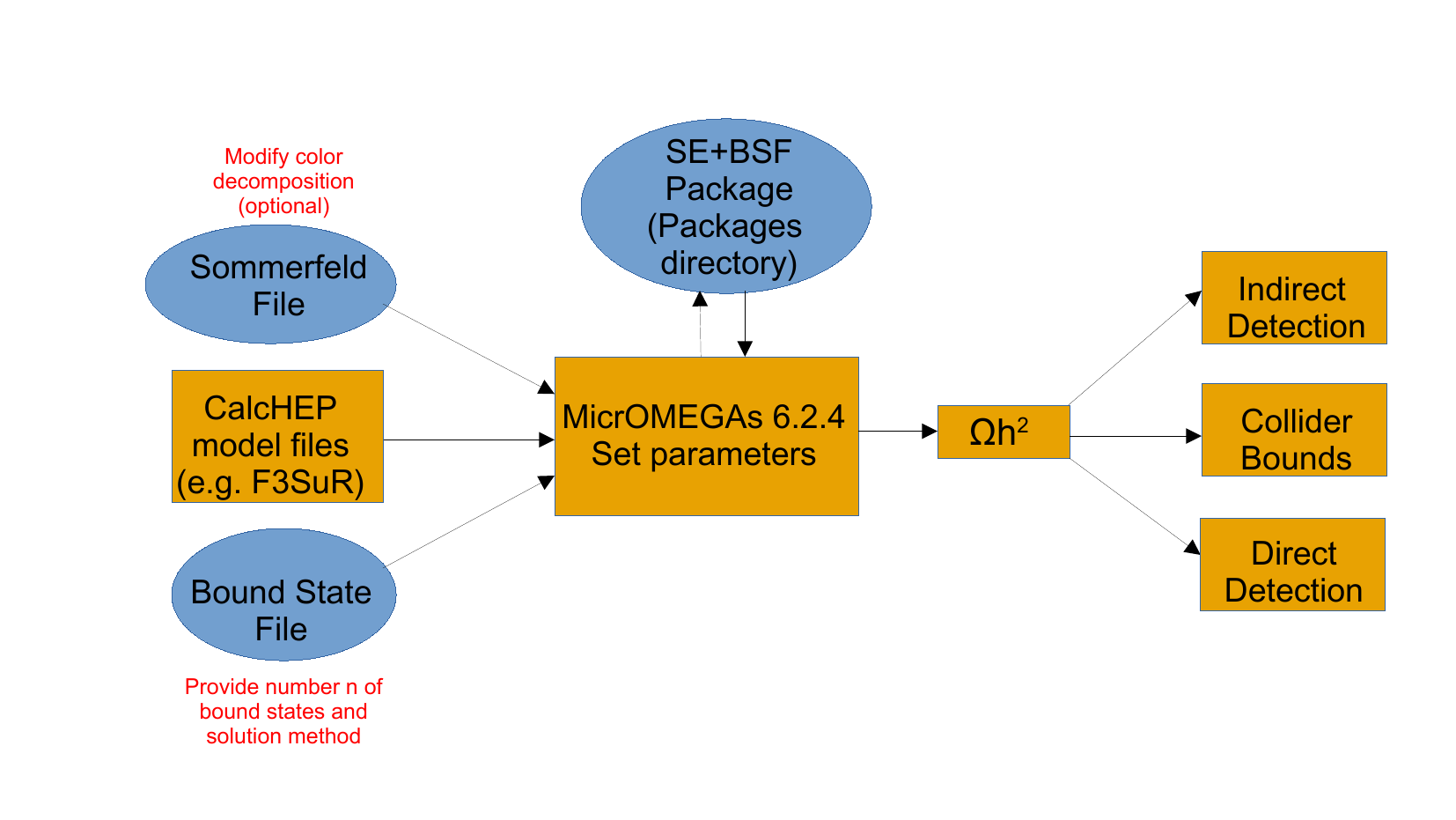}
    \captionof{figure}{\justifying Schematic workflow of the files needed to be included for the \texttt{SE+BSF4DM} package.}
    \label{fig:workflow_code}
\end{minipage}
\end{center}
\vspace{1em} 
\twocolumngrid

Within an ordinary \texttt{micrOMEGAs} installation, our package can be included by downloading the publicly available files from our \href{\CodeURL}{\faGithub~\textcolor{blue}{Github}} project page. The directory \texttt{SE\_BSF} containing the files \texttt{SE\_BSF\_header.h} and \texttt{SE\_BSF\_functions.cpp} has to be downloaded into the \texttt{Packages} directory of \texttt{micrOMEGAs}.
These files contain all the model-independent definitions and functions needed for evaluating the expressions presented in Sec.~\ref{sec:long_range_effects}. 
We stress that our package requires no modifications to the \texttt{micrOMEGAs} source code and integrates seamlessly into existing installations.
It extends \texttt{micrOMEGAs} capabilities to include long-range effects while maintaining all of \texttt{micrOMEGAs} original features. \\  

Once a new model is defined in \texttt{micrOMEGAs}, e.g.~\texttt{F3StR}, a new directory with the same name is created in the main \texttt{micrOMEGAs} folder.
To include long-range effects for the newly generated model, the two files \texttt{improveCrossSectionSommerfeld.cpp} and \texttt{BoundStateFormation.cpp} must be added to the \texttt{F3StR/lib/} directory. 
These files need to be customized to the user's need: 
\begin{itemize}
    \item In the file \\ \texttt{improveCrossSection\_Sommerfeld.cpp} the user can turn the Sommerfeld effect on or off with the flag \texttt{int somm\_flag} (off: \texttt{somm\_flag} = 0, on: \texttt{somm\_flag} = 1). In the following nested \texttt{if} block in the function \texttt{SommerfeldFactor\_BSMmodel}, the predefined factor \texttt{kQfac} corresponding to $c^{[\textbf{R}]}_{\ell s}$ in Eq.~\eqref{eq:color_decom_coef_general} for the various processes described in Appendix \ref{app:color_decomp} are given. 
    These coefficients can be altered or new coefficients can be added (e.g. for $\mathbf{8} \otimes \mathbf{8}$) using a new \texttt{if \ldots} block. 
    If the default settings (Eqs.~\eqref{eq:color_decomp_pure8} - \eqref{eq:color_decomp_qq} in Appendix~\ref{app:color_decomp}) shall be used, no further action is required.
    The function \texttt{improveCrossSection} is located in the same file and it is used for our user-defined Sommerfeld factors. 
    This function was originally created by the \texttt{micrOMEGAs} developers to enable the user to modify the tree-level cross section for a given process (e.g. by adding loop corrections through the software package \texttt{DM@NLO}~\cite{Harz:2012fz,Harz:2014tma,Harz:2014gaa,Harz:2016dql,Schmiemann:2019czm,Branahl:2019yot,Harz:2022ipe,Harz:2023llw}) and we have repurposed it for the Sommerfeld effect. 
    In case the user wants to use this original functionality, they can do so by implementing the desired modifications (loop corrections) in a new \texttt{else if (somm\_flag==2)\{\ldots\}} environment. 

    \item In the file \\ \texttt{lib/BoundStateFormation.cpp} the user can define the specifications of how BSF should be implemented in \texttt{micrOMEGAs}. 
    Four lines are necessary to be specified by the user:
    \begin{enumerate}
        \item \texttt{int bsf\_scenario = 0,1,2,3,4} specifies the treatment of BSF, corresponding respectively to: no BSF, the no–transition limit, the efficient–transition limit, ionization equilibrium, or the full solution (see Sec.~\ref{subsec:excited_states_formalism}). Setting this flag to \texttt{0} disables BSF.
        \item \texttt{num\_excited\_states=n} includes bound states with main quantum number up to $n$. Setting \texttt{n=0} turns off BSF.
        \item \texttt{const int num\_of\_mediators} sets the number of colored particles in the dark sector. 
        \item \texttt{int pdg\_nums\_mediators} is an array with length \texttt{[num\_of\_mediators]} containing the PDG numbers for the colored particles in the dark sector.  
    \end{enumerate}
\end{itemize}

Finally, in the \texttt{F3StR/main.cpp} file, one needs to include the two aforementioned files with 
\newline


\begin{lstlisting}[
  frame=trbl,
  framerule=0.4pt,
  framesep=2mm,
  xleftmargin=15pt,  % Critical: makes space for line numbers
  xrightmargin=2mm,
  backgroundcolor=\color{gray!5},
  basicstyle=\small\ttfamily,
  language=Cpp,
  numbers=none,
  numbersep=1pt,
  numberstyle=\tiny\color{gray}\hfill,
  breaklines=true
]
#include"lib/improveCrossSection_Sommerfeld.cpp"
#include"lib/BoundStateFormation.cpp"
\end{lstlisting}

After these steps, the Sommerfeld effect and BSF are included in the calculation of the relic abundance, according to the users specifications defined in the two files in the \texttt{F3StR/lib/} folder and \texttt{micrOMEGAs} can be used as usual.
The relic density is calculated with \\


\begin{lstlisting}[
  frame=trbl,
  framerule=0.4pt,
  framesep=2mm,
  xleftmargin=15pt,  % Critical: makes space for line numbers
  xrightmargin=2mm,
  backgroundcolor=\color{gray!5},
  basicstyle=\small\ttfamily,
  language=Cpp,
  numbers=none,
  numbersep=1pt,
  numberstyle=\tiny\color{gray}\hfill,
  breaklines=true
]
Omega = darkOmegaExt(&Xf, BSF_XS_A, BSF_XS_S);
\end{lstlisting}

instead of the usual \texttt{darkOmega} routine. 
In the current release, our code includes the Sommerfeld effect and BSF in the \textbf{coannihilation regime}, i.e. where $\gdm$ is sufficiently large to maintain chemical and kinetic equilibrium in the dark sector (see the discussion in Sec.~\ref{sec:model}).
We note that there is no check and corresponding warning if a parameter point analyzed does not maintain chemical equilibrium in the dark sector and it remains the task of the user to make sure that this condition is satisfied.
Incorporating also the \textbf{freeze-in} and \textbf{conversion-driven freeze-out} regimes, where excited states have been shown to play a pivotal role \cite{Binder:2023ckj, Beneke:2024nxh}, is straightforward, but requires additional functionalities of \texttt{micrOMEGAs}.
In particular a new function \texttt{darkOmegaNExt}, which is a fusion of \texttt{darkOmegaExt} and \texttt{darkOmegaN}, allowing for a user-defined cross-section for dark matter models in the coscattering regime \cite{Alguero:2023zol}, would be desireable. We are in contact with the \texttt{micrOMEGAs} developers and this will be the focus of a follow-up work.

\bibliographystyle{apsrev4-2}
\bibliography{refs}

@article{Becker:2022iso,
    author = "Becker, Mathias and Copello, Emanuele and Harz, Julia and Mohan, Kirtimaan A. and Sengupta, Dipan",
    title = "{Impact of Sommerfeld effect and bound state formation in simplified t-channel dark matter models}",
    eprint = "2203.04326",
    archivePrefix = "arXiv",
    primaryClass = "hep-ph",
    reportNumber = "ADP-22-6/T1177, MSUHEP-22-002, TUM-HEP-1387-22",
    doi = "10.1007/JHEP08(2022)145",
    journal = "JHEP",
    volume = "08",
    pages = "145",
    year = "2022"
}

@article{Brod:2017bsw,
    author = "Brod, Joachim and Gootjes-Dreesbach, Aaron and Tammaro, Michele and Zupan, Jure",
    title = "{Effective Field Theory for Dark Matter Direct Detection up to Dimension Seven}",
    eprint = "1710.10218",
    archivePrefix = "arXiv",
    primaryClass = "hep-ph",
    reportNumber = "DO-TH-17-21, DO-TH 17/21",
    doi = "10.1007/JHEP10(2018)065",
    journal = "JHEP",
    volume = "10",
    pages = "065",
    year = "2018",
    note = "[Erratum: JHEP 07, 012 (2023)]"
}

@article{Jueid:2024cge,
    author = "Jueid, Adil and Kanemura, Shinya",
    title = "{Dark matter as the trigger of flavor changing neutral current decays of the top quark}",
    eprint = "2402.08652",
    archivePrefix = "arXiv",
    primaryClass = "hep-ph",
    reportNumber = "CTPU-PTC-23-35, CERN-TH-2023-235, OU-HET-1217",
    doi = "10.1103/PhysRevD.110.095009",
    journal = "Phys. Rev. D",
    volume = "110",
    number = "9",
    pages = "095009",
    year = "2024"
}

@article{Cirelli:2005uq,
    author = "Cirelli, Marco and Fornengo, Nicolao and Strumia, Alessandro",
    title = "{Minimal dark matter}",
    eprint = "hep-ph/0512090",
    archivePrefix = "arXiv",
    reportNumber = "DFTT40-2005, IFUP-TH-2005-34",
    doi = "10.1016/j.nuclphysb.2006.07.012",
    journal = "Nucl. Phys. B",
    volume = "753",
    pages = "178--194",
    year = "2006"
}

@article{Cirelli:2009uv,
    author = "Cirelli, Marco and Strumia, Alessandro",
    title = "{Minimal Dark Matter: Model and results}",
    eprint = "0903.3381",
    archivePrefix = "arXiv",
    primaryClass = "hep-ph",
    reportNumber = "IFUP-TH-2009-04, SACLAY-T09-010",
    doi = "10.1088/1367-2630/11/10/105005",
    journal = "New J. Phys.",
    volume = "11",
    pages = "105005",
    year = "2009"
}

@article{Cirelli:2007xd,
    author = "Cirelli, Marco and Strumia, Alessandro and Tamburini, Matteo",
    title = "{Cosmology and Astrophysics of Minimal Dark Matter}",
    eprint = "0706.4071",
    archivePrefix = "arXiv",
    primaryClass = "hep-ph",
    reportNumber = "IFUP-TH-2007-12, SACLAY-T07-052",
    doi = "10.1016/j.nuclphysb.2007.07.023",
    journal = "Nucl. Phys. B",
    volume = "787",
    pages = "152--175",
    year = "2007"
}

@article{Belyaev:2018xpf,
    author = "Belyaev, Alexander and Cacciapaglia, Giacomo and Mckay, James and Marin, Dixon and Zerwekh, Alfonso R.",
    title = "{Minimal Spin-one Isotriplet Dark Matter}",
    eprint = "1808.10464",
    archivePrefix = "arXiv",
    primaryClass = "hep-ph",
    doi = "10.1103/PhysRevD.99.115003",
    journal = "Phys. Rev. D",
    volume = "99",
    number = "11",
    pages = "115003",
    year = "2019"
}

@article{Belyaev:2022qnf,
    author = "Belyaev, Alexander and Cacciapaglia, Giacomo and Locke, Daniel and Pukhov, Alexander",
    title = "{Minimal consistent Dark Matter models for systematic experimental characterisation: fermion Dark Matter}",
    eprint = "2203.03660",
    archivePrefix = "arXiv",
    primaryClass = "hep-ph",
    doi = "10.1007/JHEP10(2022)014",
    journal = "JHEP",
    volume = "10",
    pages = "014",
    year = "2022"
}

@article{Petraki:2015hla,
    author = "Petraki, Kalliopi and Postma, Marieke and Wiechers, Michael",
    title = "{Dark-matter bound states from Feynman diagrams}",
    eprint = "1505.00109",
    archivePrefix = "arXiv",
    primaryClass = "hep-ph",
    reportNumber = "NIKHEF-2015-013",
    doi = "10.1007/JHEP06(2015)128",
    journal = "JHEP",
    volume = "06",
    pages = "128",
    year = "2015"
}

@article{Petraki:2016cnz,
    author = "Petraki, Kalliopi and Postma, Marieke and de Vries, Jordy",
    title = "{Radiative bound-state-formation cross-sections for dark matter interacting via a Yukawa potential}",
    eprint = "1611.01394",
    archivePrefix = "arXiv",
    primaryClass = "hep-ph",
    doi = "10.1007/JHEP04(2017)077",
    journal = "JHEP",
    volume = "04",
    pages = "077",
    year = "2017"
}

@article{Liew:2016hqo,
    author = "Liew, Seng Pei and Luo, Feng",
    title = "{Effects of QCD bound states on dark matter relic abundance}",
    eprint = "1611.08133",
    archivePrefix = "arXiv",
    primaryClass = "hep-ph",
    reportNumber = "UT-16-33, IPMU-16-0176",
    doi = "10.1007/JHEP02(2017)091",
    journal = "JHEP",
    volume = "02",
    pages = "091",
    year = "2017"
}

@book{bethe1957quantum,
  author    = {Bethe, Hans A. and Salpeter, Edwin E.},
  title     = {Quantum Mechanics of One- and Two-Electron Atoms},
  publisher = {Springer},
  year      = {1957},
  address   = {New York},
  note      = {First edition}
}

@article{Yao_2019,
   title={Quarkonium inside the quark-gluon plasma: Diffusion, dissociation, recombination, and energy loss},
   volume={100},
   ISSN={2470-0029},
   url={http://dx.doi.org/10.1103/PhysRevD.100.014008},
   DOI={10.1103/physrevd.100.014008},
   number={1},
   journal={Physical Review D},
   publisher={American Physical Society (APS)},
   author={Yao, Xiaojun and Müller, Berndt},
   year={2019},
   month=jul }

@article{Feng:2003uy,
    author = "Feng, Jonathan L. and Rajaraman, Arvind and Takayama, Fumihiro",
    title = "{SuperWIMP dark matter signals from the early universe}",
    eprint = "hep-ph/0306024",
    archivePrefix = "arXiv",
    reportNumber = "UCI-TR-2003-19",
    doi = "10.1103/PhysRevD.68.063504",
    journal = "Phys. Rev. D",
    volume = "68",
    pages = "063504",
    year = "2003"
}

@article{Binder:2021vfo,
    author = "Binder, Tobias and Filimonova, Anastasiia and Petraki, Kalliopi and White, Graham",
    title = "{Saha equilibrium for metastable bound states and dark matter freeze-out}",
    eprint = "2112.00042",
    archivePrefix = "arXiv",
    primaryClass = "hep-ph",
    doi = "10.1016/j.physletb.2022.137323",
    journal = "Phys. Lett. B",
    volume = "833",
    pages = "137323",
    year = "2022"
}

@article{Covi:1999ty,
    author = "Covi, Laura and Kim, Jihn E. and Roszkowski, Leszek",
    title = "{Axinos as cold dark matter}",
    eprint = "hep-ph/9905212",
    archivePrefix = "arXiv",
    reportNumber = "LANCS-TH-9824",
    doi = "10.1103/PhysRevLett.82.4180",
    journal = "Phys. Rev. Lett.",
    volume = "82",
    pages = "4180--4183",
    year = "1999"
}

@article{Griest:1990kh,
    author = "Griest, Kim and Seckel, David",
    title = "{Three exceptions in the calculation of relic abundances}",
    reportNumber = "CFPA-TH-90-001A, BA-90-79",
    doi = "10.1103/PhysRevD.43.3191",
    journal = "Phys. Rev. D",
    volume = "43",
    pages = "3191--3203",
    year = "1991"
}

@article{Binder:2023ckj,
    author = "Binder, Tobias and Garny, Mathias and Heisig, Jan and Lederer, Stefan and Urban, Kai",
    title = "{Excited bound states and their role in dark matter production}",
    eprint = "2308.01336",
    archivePrefix = "arXiv",
    primaryClass = "hep-ph",
    reportNumber = "TUM-HEP 1469/23, TTK-23-21",
    doi = "10.1103/PhysRevD.108.095030",
    journal = "Phys. Rev. D",
    volume = "108",
    number = "9",
    pages = "095030",
    year = "2023"
}

@article{Garny:2021qsr,
    author = "Garny, Mathias and Heisig, Jan",
    title = "{Bound-state effects on dark matter coannihilation: Pushing the boundaries of conversion-driven freeze-out}",
    eprint = "2112.01499",
    archivePrefix = "arXiv",
    primaryClass = "hep-ph",
    reportNumber = "TUM-HEP 1379/21, TTK-21-52",
    doi = "10.1103/PhysRevD.105.055004",
    journal = "Phys. Rev. D",
    volume = "105",
    number = "5",
    pages = "055004",
    year = "2022"
}

@article{DAgnolo:2017dbv,
    author = "D'Agnolo, Raffaele Tito and Pappadopulo, Duccio and Ruderman, Joshua T.",
    title = "{Fourth Exception in the Calculation of Relic Abundances}",
    eprint = "1705.08450",
    archivePrefix = "arXiv",
    primaryClass = "hep-ph",
    doi = "10.1103/PhysRevLett.119.061102",
    journal = "Phys. Rev. Lett.",
    volume = "119",
    number = "6",
    pages = "061102",
    year = "2017"
}

@article{Hisano:2006nn,
    author = "Hisano, Junji and Matsumoto, Shigeki and Nagai, Minoru and Saito, Osamu and Senami, Masato",
    title = "{Non-perturbative effect on thermal relic abundance of dark matter}",
    eprint = "hep-ph/0610249",
    archivePrefix = "arXiv",
    reportNumber = "KEK-TH-1111",
    doi = "10.1016/j.physletb.2007.01.012",
    journal = "Phys. Lett. B",
    volume = "646",
    pages = "34--38",
    year = "2007"
}

@article{Hisano:2015bma,
    author = "Hisano, Junji and Nagai, Ryo and Nagata, Natsumi",
    title = "{Effective Theories for Dark Matter Nucleon Scattering}",
    eprint = "1502.02244",
    archivePrefix = "arXiv",
    primaryClass = "hep-ph",
    reportNumber = "FTPI-MINN-15-03, IPMU15-0012",
    doi = "10.1007/JHEP05(2015)037",
    journal = "JHEP",
    volume = "05",
    pages = "037",
    year = "2015"
}

@article{Hisano:2010ct,
    author = "Hisano, Junji and Ishiwata, Koji and Nagata, Natsumi",
    title = "{Gluon contribution to the dark matter direct detection}",
    eprint = "1007.2601",
    archivePrefix = "arXiv",
    primaryClass = "hep-ph",
    reportNumber = "IPMU10-0113, ICRR-REPORT-570-2010-3",
    doi = "10.1103/PhysRevD.82.115007",
    journal = "Phys. Rev. D",
    volume = "82",
    pages = "115007",
    year = "2010"
}

@article{Asadi:2016ybp,
    author = "Asadi, Pouya and Baumgart, Matthew and Fitzpatrick, Patrick J. and Krupczak, Emmett and Slatyer, Tracy R.",
    title = "{Capture and Decay of Electroweak WIMPonium}",
    eprint = "1610.07617",
    archivePrefix = "arXiv",
    primaryClass = "hep-ph",
    doi = "10.1088/1475-7516/2017/02/005",
    journal = "JCAP",
    volume = "02",
    pages = "005",
    year = "2017"
}

@misc{becker2025sebsf4dmmicromegaspackage,
      title="{Manual for \texttt{SE+BSF4DM} -- A micrOMEGAs package for Sommerfeld Effect and Bound State Formation in colored Dark Sectors}", 
      author={Mathias Becker and Emanuele Copello and Julia Harz and Martin Napetschnig},
      year={2025},
      eprint={2512.02155},
      archivePrefix={arXiv},
      primaryClass={hep-ph},
      url={https://arxiv.org/abs/2512.02155}, 
}

@article{Harz:2018csl,
    author = "Harz, Julia and Petraki, Kalliopi",
    title = "{Radiative bound-state formation in unbroken perturbative non-Abelian theories and implications for dark matter}",
    eprint = "1805.01200",
    archivePrefix = "arXiv",
    primaryClass = "hep-ph",
    reportNumber = "Nikhef-2018-023",
    doi = "10.1007/JHEP07(2018)096",
    journal = "JHEP",
    volume = "07",
    pages = "096",
    year = "2018"
}

@article{Iengo:2009ni,
    author = "Iengo, Roberto",
    title = "{Sommerfeld enhancement: General results from field theory diagrams}",
    eprint = "0902.0688",
    archivePrefix = "arXiv",
    primaryClass = "hep-ph",
    doi = "10.1088/1126-6708/2009/05/024",
    journal = "JHEP",
    volume = "05",
    pages = "024",
    year = "2009"
}

@article{ElHedri:2017nny,
    author = "El Hedri, Sonia and Kaminska, Anna and de Vries, Maikel and Zurita, Jose",
    title = "{Simplified Phenomenology for Colored Dark Sectors}",
    eprint = "1703.00452",
    archivePrefix = "arXiv",
    primaryClass = "hep-ph",
    reportNumber = "MITP-17-002, TTP17-006",
    doi = "10.1007/JHEP04(2017)118",
    journal = "JHEP",
    volume = "04",
    pages = "118",
    year = "2017"
}

@article{ElHedri:2018atj,
    author = "El Hedri, Sonia and de Vries, Maikel",
    title = "{Cornering Colored Coannihilation}",
    eprint = "1806.03325",
    archivePrefix = "arXiv",
    primaryClass = "hep-ph",
    reportNumber = "MITP-18-048",
    doi = "10.1007/JHEP10(2018)102",
    journal = "JHEP",
    volume = "10",
    pages = "102",
    year = "2018"
}

@article{ElHedri:2016onc,
    author = "El Hedri, Sonia and Kaminska, Anna and de Vries, Maikel",
    title = "{A Sommerfeld Toolbox for Colored Dark Sectors}",
    eprint = "1612.02825",
    archivePrefix = "arXiv",
    primaryClass = "hep-ph",
    reportNumber = "MITP-16-135",
    doi = "10.1140/epjc/s10052-017-5168-z",
    journal = "Eur. Phys. J. C",
    volume = "77",
    number = "9",
    pages = "622",
    year = "2017"
}

@article{Arcadi:2017kky,
    author = "Arcadi, Giorgio and Dutra, Ma\'\i{}ra and Ghosh, Pradipta and Lindner, Manfred and Mambrini, Yann and Pierre, Mathias and Profumo, Stefano and Queiroz, Farinaldo S.",
    title = "{The waning of the WIMP? A review of models, searches, and constraints}",
    eprint = "1703.07364",
    archivePrefix = "arXiv",
    primaryClass = "hep-ph",
    doi = "10.1140/epjc/s10052-018-5662-y",
    journal = "Eur. Phys. J. C",
    volume = "78",
    number = "3",
    pages = "203",
    year = "2018"
}

@article{LHCDarkMatterWorkingGroup:2018ufk,
    author = "Abe, Tomohiro and others",
    collaboration = "LHC Dark Matter Working Group",
    title = "{LHC Dark Matter Working Group: Next-generation spin-0 dark matter models}",
    eprint = "1810.09420",
    archivePrefix = "arXiv",
    primaryClass = "hep-ex",
    reportNumber = "CERN-LPCC-2018-02",
    doi = "10.1016/j.dark.2019.100351",
    journal = "Phys. Dark Univ.",
    volume = "27",
    pages = "100351",
    year = "2020"
}

@article{Planck:2018vyg,
    author = "Aghanim, N. and others",
    collaboration = "Planck",
    title = "{Planck 2018 results. VI. Cosmological parameters}",
    eprint = "1807.06209",
    archivePrefix = "arXiv",
    primaryClass = "astro-ph.CO",
    doi = "10.1051/0004-6361/201833910",
    journal = "Astron. Astrophys.",
    volume = "641",
    pages = "A6",
    year = "2020",
    note = "[Erratum: Astron.Astrophys. 652, C4 (2021)]"
}

@inproceedings{Harris:2022vnx,
    author = "Harris, Philip and Schuster, Philip and Zupan, Jure",
    title = "{Snowmass White Paper: New flavors and rich structures in dark sectors}",
    booktitle = "{Snowmass 2021}",
    eprint = "2207.08990",
    archivePrefix = "arXiv",
    primaryClass = "hep-ph",
    month = "7",
    year = "2022"
}

@article{CMS:2019ykj,
    author = "Sirunyan, Albert M and others",
    collaboration = "CMS",
    title = "{Search for dark matter particles produced in association with a Higgs boson in proton-proton collisions at $ \sqrt{\mathrm{s}} $ = 13 TeV}",
    eprint = "1908.01713",
    archivePrefix = "arXiv",
    primaryClass = "hep-ex",
    reportNumber = "CMS-EXO-18-011, CERN-EP-2019-141",
    doi = "10.1007/JHEP03(2020)025",
    journal = "JHEP",
    volume = "03",
    pages = "025",
    year = "2020"
}

@inproceedings{PerezAdan:2023rsl,
    author = "Perez Adan, Danyer",
    collaboration = "ATLAS, CMS",
    title = "{Dark Matter searches at CMS and ATLAS}",
    booktitle = "{56th Rencontres de Moriond on Electroweak Interactions and Unified Theories}",
    eprint = "2301.10141",
    archivePrefix = "arXiv",
    primaryClass = "hep-ex",
    reportNumber = "CMS-CR-2022-059",
    month = "1",
    year = "2023"
}

@article{ATLAS:2023rvb,
    author = "Aad, Georges and others",
    collaboration = "ATLAS",
    title = "{Combination and summary of ATLAS dark matter searches interpreted in a 2HDM with a pseudo-scalar mediator using 139 fb\ensuremath{-}1 of s=13 TeV pp collision data}",
    eprint = "2306.00641",
    archivePrefix = "arXiv",
    primaryClass = "hep-ex",
    reportNumber = "CERN-EP-2023-088",
    doi = "10.1016/j.scib.2024.06.003",
    journal = "Sci. Bull.",
    volume = "69",
    number = "19",
    pages = "3005--3035",
    year = "2024"
}

@article{LZ:2022lsv,
    author = "Aalbers, J. and others",
    collaboration = "LZ",
    title = "{First Dark Matter Search Results from the LUX-ZEPLIN (LZ) Experiment}",
    eprint = "2207.03764",
    archivePrefix = "arXiv",
    primaryClass = "hep-ex",
    doi = "10.1103/PhysRevLett.131.041002",
    journal = "Phys. Rev. Lett.",
    volume = "131",
    number = "4",
    pages = "041002",
    year = "2023"
}

@misc{binder2025bsffastrapidcomputationboundstate,
      title={BSFfast: Rapid computation of bound-state effects on annihilation in the early Universe}, 
      author={Tobias Binder and Mathias Garny and Jan Heisig and Stefan Lederer},
      year={2025},
      eprint={2512.23812},
      archivePrefix={arXiv},
      primaryClass={hep-ph},
      url={https://arxiv.org/abs/2512.23812}, 
}

@article{LZ:2024zvo,
    author = "Aalbers, J. and others",
    collaboration = "LZ",
    title = "{Dark Matter Search Results from 4.2{\,}{\,}Tonne-Years of Exposure of the LUX-ZEPLIN (LZ) Experiment}",
    eprint = "2410.17036",
    archivePrefix = "arXiv",
    primaryClass = "hep-ex",
    reportNumber = "FERMILAB-PUB-24-0796-V",
    doi = "10.1103/4dyc-z8zf",
    journal = "Phys. Rev. Lett.",
    volume = "135",
    number = "1",
    pages = "011802",
    year = "2025"
}

@article{Drees:1992rr,
    author = "Drees, Manuel and Nojiri, Mihoko M.",
    title = "{New contributions to coherent neutralino - nucleus scattering}",
    eprint = "hep-ph/9210272",
    archivePrefix = "arXiv",
    reportNumber = "MAD-PH-723",
    doi = "10.1103/PhysRevD.47.4226",
    journal = "Phys. Rev. D",
    volume = "47",
    pages = "4226--4232",
    year = "1993"
}

@article{Drees:1993bu,
    author = "Drees, Manuel and Nojiri, Mihoko",
    title = "{Neutralino - nucleon scattering revisited}",
    eprint = "hep-ph/9307208",
    archivePrefix = "arXiv",
    reportNumber = "MAD-PH-768",
    doi = "10.1103/PhysRevD.48.3483",
    journal = "Phys. Rev. D",
    volume = "48",
    pages = "3483--3501",
    year = "1993"
}

@article{Gondolo:2013wwa,
    author = "Gondolo, Paolo and Scopel, Stefano",
    title = "{On the sbottom resonance in dark matter scattering}",
    eprint = "1307.4481",
    archivePrefix = "arXiv",
    primaryClass = "hep-ph",
    reportNumber = "CETUP2013-008",
    doi = "10.1088/1475-7516/2013/10/032",
    journal = "JCAP",
    volume = "10",
    pages = "032",
    year = "2013"
}

@article{Biondini:2023zcz,
    author = "Biondini, Simone and Brambilla, Nora and Qerimi, Gramos and Vairo, Antonio",
    title = "{Effective field theories for dark matter pairs in the early universe: cross sections and widths}",
    eprint = "2304.00113",
    archivePrefix = "arXiv",
    primaryClass = "hep-ph",
    doi = "10.1007/JHEP07(2023)006",
    journal = "JHEP",
    volume = "07",
    pages = "006",
    year = "2023"
}

@article{Heisig_2024,
   title={Probing conversion-driven freeze-out at the LHC},
   volume={110},
   ISSN={2470-0029},
   url={http://dx.doi.org/10.1103/PhysRevD.110.015031},
   DOI={10.1103/physrevd.110.015031},
   number={1},
   journal={Physical Review D},
   publisher={American Physical Society (APS)},
   author={Heisig, Jan and Lessa, Andre and Ramos, Lucas Magno D.},
   year={2024},
   month=jul }

@article{Biondini:2025gpg,
    author = "Biondini, Simone and Tiberi, Lorenzo and Panella, Orlando",
    title = "{Connecting t-channel dark matter models to the Standard Model Effective Field Theory}",
    eprint = "2507.00925",
    archivePrefix = "arXiv",
    primaryClass = "hep-ph",
    doi = "10.1007/JHEP10(2025)060",
    journal = "JHEP",
    volume = "10",
    pages = "060",
    year = "2025"
}

@misc{olgoso2025darkterazfactory,
      title={The Dark Side of a Tera-Z Factory}, 
      author={Pablo Olgoso and Paride Paradisi and Nudzeim Selimovic},
      year={2025},
      eprint={2507.17803},
      archivePrefix={arXiv},
      primaryClass={hep-ph},
      url={https://arxiv.org/abs/2507.17803}, 
}

@misc{arcadi2025veilchartingwimpterritories,
      title={Beyond the Veil: Charting WIMP Territories at the Neutrino Floor}, 
      author={Giorgio Arcadi and Manfred Lindner and Stefano Profumo},
      year={2025},
      eprint={2507.16987},
      archivePrefix={arXiv},
      primaryClass={hep-ph},
      url={https://arxiv.org/abs/2507.16987}, 
}

@article{PICO:2023uff,
    author = "Adams, E. and others",
    collaboration = "PICO",
    title = "{Search for inelastic dark matter-nucleus scattering with the PICO-60 CF3I and C3F8 bubble chambers}",
    eprint = "2301.08993",
    archivePrefix = "arXiv",
    primaryClass = "astro-ph.CO",
    reportNumber = "FERMILAB-PUB-23-059-PPD",
    doi = "10.1103/PhysRevD.108.062003",
    journal = "Phys. Rev. D",
    volume = "108",
    number = "6",
    pages = "062003",
    year = "2023"
}

@article{Beneke:2024nxh,
    author = "Beneke, Martin and Binder, Tobias and de Ros, Lorenzo and Garny, Mathias and Lederer, Stefan",
    title = "{Perturbative unitarity violation in radiative capture transitions to dark matter bound states}",
    eprint = "2411.08737",
    archivePrefix = "arXiv",
    primaryClass = "hep-ph",
    reportNumber = "TUM-HEP-1534/24",
    doi = "10.1007/JHEP02(2025)189",
    journal = "JHEP",
    volume = "02",
    pages = "189",
    year = "2025"
}

@article{XENON:2024wpa,
    author = "Aprile, E. and others",
    collaboration = "XENON",
    title = "{The XENONnT dark matter experiment}",
    eprint = "2402.10446",
    archivePrefix = "arXiv",
    primaryClass = "physics.ins-det",
    doi = "10.1140/epjc/s10052-024-12982-5",
    journal = "Eur. Phys. J. C",
    volume = "84",
    number = "8",
    pages = "784",
    year = "2024"
}

@article{Cassel:2009wt,
    author = "Cassel, S.",
    title = "{Sommerfeld factor for arbitrary partial wave processes}",
    eprint = "0903.5307",
    archivePrefix = "arXiv",
    primaryClass = "hep-ph",
    reportNumber = "OUTP-0910P",
    doi = "10.1088/0954-3899/37/10/105009",
    journal = "J. Phys. G",
    volume = "37",
    pages = "105009",
    year = "2010"
}

@article{Sakharov:1948plh,
    author = "Sakharov, Andrei D.",
    title = "{Interaction of an Electron and Positron in Pair Production}",
    reportNumber = "RT-4471",
    doi = "10.1070/PU1991v034n05ABEH002492",
    journal = "Zh. Eksp. Teor. Fiz.",
    volume = "18",
    pages = "631--635",
    year = "1948"
}

@article{Mitridate:2017izz,
    author = "Mitridate, Andrea and Redi, Michele and Smirnov, Juri and Strumia, Alessandro",
    title = "{Cosmological Implications of Dark Matter Bound States}",
    eprint = "1702.01141",
    archivePrefix = "arXiv",
    primaryClass = "hep-ph",
    reportNumber = "CERN-TH-2017-030, IFUP-TH-2017",
    doi = "10.1088/1475-7516/2017/05/006",
    journal = "JCAP",
    volume = "05",
    pages = "006",
    year = "2017"
}

@article{Gross:2018zha,
    author = "Gross, Christian and Mitridate, Andrea and Redi, Michele and Smirnov, Juri and Strumia, Alessandro",
    title = "{Cosmological Abundance of Colored Relics}",
    eprint = "1811.08418",
    archivePrefix = "arXiv",
    primaryClass = "hep-ph",
    doi = "10.1103/PhysRevD.99.016024",
    journal = "Phys. Rev. D",
    volume = "99",
    number = "1",
    pages = "016024",
    year = "2019"
}

@article{Baker:2015qna,
    author = "Baker, Michael J. and others",
    title = "{The Coannihilation Codex}",
    eprint = "1510.03434",
    archivePrefix = "arXiv",
    primaryClass = "hep-ph",
    reportNumber = "MITP-15-078",
    doi = "10.1007/JHEP12(2015)120",
    journal = "JHEP",
    volume = "12",
    pages = "120",
    year = "2015"
}

@article{Giacchino:2015hvk,
    author = "Giacchino, Federica and Ibarra, Alejandro and Lopez Honorez, Laura and Tytgat, Michel H. G. and Wild, Sebastian",
    title = "{Signatures from Scalar Dark Matter with a Vector-like Quark Mediator}",
    eprint = "1511.04452",
    archivePrefix = "arXiv",
    primaryClass = "hep-ph",
    doi = "10.1088/1475-7516/2016/02/002",
    journal = "JCAP",
    volume = "02",
    pages = "002",
    year = "2016"
}

@article{Colucci:2018vxz,
    author = "Colucci, Stefano and Fuks, Benjamin and Giacchino, Federica and Lopez Honorez, Laura and Tytgat, Michel H. G. and Vandecasteele, J\'er\^ome",
    title = "{Top-philic Vector-Like Portal to Scalar Dark Matter}",
    eprint = "1804.05068",
    archivePrefix = "arXiv",
    primaryClass = "hep-ph",
    reportNumber = "ULB-TH/18-04, ULB-TH-18-04",
    doi = "10.1103/PhysRevD.98.035002",
    journal = "Phys. Rev. D",
    volume = "98",
    pages = "035002",
    year = "2018"
}

@book{KolbTurner1994,
  title={The Early Universe},
  author={Kolb, Edward W. and Turner, Michael S.},
  year={1994},
  publisher={CRC Press},
  isbn={978-0201626748},
  series={Frontiers in Physics},
  address={Boca Raton, FL}
}

@article{DiFranzo:2013vra,
    author = "DiFranzo, Anthony and Nagao, Keiko I. and Rajaraman, Arvind and Tait, Tim M. P.",
    title = "{Simplified Models for Dark Matter Interacting with Quarks}",
    eprint = "1308.2679",
    archivePrefix = "arXiv",
    primaryClass = "hep-ph",
    reportNumber = "UCI-HEP-TR-2013-17, KEK-TH-1659",
    doi = "10.1007/JHEP11(2013)014",
    journal = "JHEP",
    volume = "11",
    pages = "014",
    year = "2013",
    note = "[Erratum: JHEP 01, 162 (2014)]"
}

@article{Beneke:2014hja,
    author = "Beneke, M. and Hellmann, Charlotte and Ruiz-Femenia, P.",
    title = "{Heavy neutralino relic abundance with Sommerfeld enhancements - a study of pMSSM scenarios}",
    eprint = "1411.6930",
    archivePrefix = "arXiv",
    primaryClass = "hep-ph",
    reportNumber = "TUM-HEP-955-14, TTK-14-22, SFB-CPP-14-70, IFIC-14-59",
    doi = "10.1007/JHEP03(2015)162",
    journal = "JHEP",
    volume = "03",
    pages = "162",
    year = "2015"
}

@article{vonHarling:2014kha,
    author = "von Harling, Benedict and Petraki, Kalliopi",
    title = "{Bound-state formation for thermal relic dark matter and unitarity}",
    eprint = "1407.7874",
    archivePrefix = "arXiv",
    primaryClass = "hep-ph",
    reportNumber = "NIKHEF-2014-018",
    doi = "10.1088/1475-7516/2014/12/033",
    journal = "JCAP",
    volume = "12",
    pages = "033",
    year = "2014"
}

@article{Alguero:2023zol,
    author = "Alguero, G. and Belanger, G. and Boudjema, F. and Chakraborti, S. and Goudelis, A. and Kraml, S. and Mjallal, A. and Pukhov, A.",
    title = "{micrOMEGAs 6.0: N-component dark matter}",
    eprint = "2312.14894",
    archivePrefix = "arXiv",
    primaryClass = "hep-ph",
    doi = "10.1016/j.cpc.2024.109133",
    journal = "Comput. Phys. Commun.",
    volume = "299",
    pages = "109133",
    year = "2024"
}

@article{Beneke:2019qaa,
    author = "Beneke, Martin and Szafron, Robert and Urban, Kai",
    title = "{Wino potential and Sommerfeld effect at NLO}",
    eprint = "1909.04584",
    archivePrefix = "arXiv",
    primaryClass = "hep-ph",
    reportNumber = "TUM-HEP-1222/19",
    doi = "10.1016/j.physletb.2019.135112",
    journal = "Phys. Lett. B",
    volume = "800",
    pages = "135112",
    year = "2020"
}

@article{Branahl:2019yot,
    author = "Branahl, J. and Harz, J. and Herrmann, B. and Klasen, M. and Kova\v{r}\'\i{}k, K. and Schmiemann, S.",
    title = "{SUSY-QCD corrected and Sommerfeld enhanced stau annihilation into heavy quarks with scheme and scale uncertainties}",
    eprint = "1909.09527",
    archivePrefix = "arXiv",
    primaryClass = "hep-ph",
    reportNumber = "LAPTH-045/19, MS-TP-19-28, TUM-HEP-1226-19, MS-TP-19-25",
    doi = "10.1103/PhysRevD.100.115003",
    journal = "Phys. Rev. D",
    volume = "100",
    number = "11",
    pages = "115003",
    year = "2019"
}

@article{Harz:2012fz,
    author = "Harz, J. and Herrmann, B. and Klasen, M. and Kovarik, K. and Boulc'h, Q. Le",
    title = "{Neutralino-stop coannihilation into electroweak gauge and Higgs bosons at one loop}",
    eprint = "1212.5241",
    archivePrefix = "arXiv",
    primaryClass = "hep-ph",
    reportNumber = "DESY-12-205, LAPTH-051-12, LPSC-12-341, MS-TP-12-17, KA-TP-42-2012",
    doi = "10.1103/PhysRevD.87.054031",
    journal = "Phys. Rev. D",
    volume = "87",
    number = "5",
    pages = "054031",
    year = "2013"
}

@article{Sommerfeld:1931qaf,
    author = "Sommerfeld, A.",
    title = {{\"Uber die Beugung und Bremsung der Elektronen}},
    doi = "10.1002/andp.19314030302",
    journal = "Annalen Phys.",
    volume = "403",
    number = "3",
    pages = "257--330",
    year = "1931"
}

@article{ATLAS:2017ayi,
    author = "Aaboud, Morad and others",
    collaboration = "ATLAS",
    title = "{Search for new phenomena in high-mass diphoton final states using 37 fb$^{-1}$ of proton--proton collisions collected at $\sqrt{s}=13$ TeV with the ATLAS detector}",
    eprint = "1707.04147",
    archivePrefix = "arXiv",
    primaryClass = "hep-ex",
    reportNumber = "CERN-EP-2017-132",
    doi = "10.1016/j.physletb.2017.10.039",
    journal = "Phys. Lett. B",
    volume = "775",
    pages = "105--125",
    year = "2017"
}

@article{Younkin:2009zn,
    author = "Younkin, James E. and Martin, Stephen P.",
    title = "{QCD Corrections to Stoponium Production at Hadron Colliders}",
    eprint = "0912.4813",
    archivePrefix = "arXiv",
    primaryClass = "hep-ph",
    reportNumber = "FERMILAB-PUB-10-781-T",
    doi = "10.1103/PhysRevD.81.055006",
    journal = "Phys. Rev. D",
    volume = "81",
    pages = "055006",
    year = "2010"
}

@article{Martin:2008sv,
    author = "Martin, Stephen P.",
    title = "{Diphoton decays of stoponium at the Large Hadron Collider}",
    eprint = "0801.0237",
    archivePrefix = "arXiv",
    primaryClass = "hep-ph",
    reportNumber = "FERMILAB-PUB-08-720-T",
    doi = "10.1103/PhysRevD.77.075002",
    journal = "Phys. Rev. D",
    volume = "77",
    pages = "075002",
    year = "2008"
}

@article{Backovic:2015soa,
    author = {Backovi\'c, Mihailo and Kr\"amer, Michael and Maltoni, Fabio and Martini, Antony and Mawatari, Kentarou and Pellen, Mathieu},
    title = "{Higher-order QCD predictions for dark matter production at the LHC in simplified models with s-channel mediators}",
    eprint = "1508.05327",
    archivePrefix = "arXiv",
    primaryClass = "hep-ph",
    reportNumber = "MCNET-15-24, CP3-15-25, TTK-15-19",
    doi = "10.1140/epjc/s10052-015-3700-6",
    journal = "Eur. Phys. J. C",
    volume = "75",
    number = "10",
    pages = "482",
    year = "2015"
}

@article{An:2016gad,
    author = "An, Haipeng and Wise, Mark B. and Zhang, Yue",
    title = "{Effects of Bound States on Dark Matter Annihilation}",
    eprint = "1604.01776",
    archivePrefix = "arXiv",
    primaryClass = "hep-ph",
    reportNumber = "CALT-TH-2016-005",
    doi = "10.1103/PhysRevD.93.115020",
    journal = "Phys. Rev. D",
    volume = "93",
    number = "11",
    pages = "115020",
    year = "2016"
}

@article{Palmiotto:2022rvw,
    author = "Palmiotto, Marco and Arbey, Alexandre and Mahmoudi, Farvah",
    title = "{DarkPACK: A modular software to compute BSM squared amplitudes for particle physics and dark matter observables}",
    eprint = "2211.10376",
    archivePrefix = "arXiv",
    primaryClass = "hep-ph",
    reportNumber = "CERN-TH-2022-197",
    doi = "10.1016/j.cpc.2023.108905",
    journal = "Comput. Phys. Commun.",
    volume = "294",
    pages = "108905",
    year = "2024"
}

@article{Ambrogi:2018jqj,
    author = "Ambrogi, Federico and Arina, Chiara and Backovic, Mihailo and Heisig, Jan and Maltoni, Fabio and Mantani, Luca and Mattelaer, Olivier and Mohlabeng, Gopolang",
    title = "{MadDM v.3.0: a Comprehensive Tool for Dark Matter Studies}",
    eprint = "1804.00044",
    archivePrefix = "arXiv",
    primaryClass = "hep-ph",
    reportNumber = "CP3-18-26, MCnet-18-07, MCNET-18-07",
    doi = "10.1016/j.dark.2018.11.009",
    journal = "Phys. Dark Univ.",
    volume = "24",
    pages = "100249",
    year = "2019"
}

@article{Harz:2023llw,
    author = {Harz, Julia and Herrmann, Bj\"orn and Klasen, Michael and Kova\v{r}\'\i{}k, Karol and Wiggering, Luca Paolo},
    title = "{Precision predictions for dark matter with DM@NLO in the MSSM}",
    eprint = "2312.17206",
    archivePrefix = "arXiv",
    primaryClass = "hep-ph",
    reportNumber = "MITP-23-084, MS-TP-23-52, LAPTH-062/23",
    doi = "10.1140/epjc/s10052-024-12660-6",
    journal = "Eur. Phys. J. C",
    volume = "84",
    number = "4",
    pages = "342",
    year = "2024"
}

@article{Bringmann:2018lay,
    author = {Bringmann, Torsten and Edsj\"o, Joakim and Gondolo, Paolo and Ullio, Piero and Bergstr\"om, Lars},
    title = "{DarkSUSY 6 : An Advanced Tool to Compute Dark Matter Properties Numerically}",
    eprint = "1802.03399",
    archivePrefix = "arXiv",
    primaryClass = "hep-ph",
    doi = "10.1088/1475-7516/2018/07/033",
    journal = "JCAP",
    volume = "07",
    pages = "033",
    year = "2018"
}

@article{Belanger:2018sti,
    author = "B\'elanger, G. and others",
    title = "{LHC-friendly minimal freeze-in models}",
    eprint = "1811.05478",
    archivePrefix = "arXiv",
    primaryClass = "hep-ph",
    doi = "10.1007/JHEP02(2019)186",
    journal = "JHEP",
    volume = "02",
    pages = "186",
    year = "2019"
}

@article{Belanger:2006is,
    author = "Belanger, G. and Boudjema, F. and Pukhov, A. and Semenov, A.",
    title = "{MicrOMEGAs 2.0: A Program to calculate the relic density of dark matter in a generic model}",
    eprint = "hep-ph/0607059",
    archivePrefix = "arXiv",
    reportNumber = "LAPTH-1152-06",
    doi = "10.1016/j.cpc.2006.11.008",
    journal = "Comput. Phys. Commun.",
    volume = "176",
    pages = "367--382",
    year = "2007"
}

@article{Garny:2015wea,
    author = "Garny, Mathias and Ibarra, Alejandro and Vogl, Stefan",
    title = "{Signatures of Majorana dark matter with t-channel mediators}",
    eprint = "1503.01500",
    archivePrefix = "arXiv",
    primaryClass = "hep-ph",
    reportNumber = "CERN-PH-TH-2015-036, TUM-HEP-985-15",
    doi = "10.1142/S0218271815300190",
    journal = "Int. J. Mod. Phys. D",
    volume = "24",
    number = "07",
    pages = "1530019",
    year = "2015"
}

@article{Garny:2017rxs,
    author = {Garny, Mathias and Heisig, Jan and L\"ulf, Benedikt and Vogl, Stefan},
    title = "{Coannihilation without chemical equilibrium}",
    eprint = "1705.09292",
    archivePrefix = "arXiv",
    primaryClass = "hep-ph",
    reportNumber = "TUM-HEP-1085-17, TTK-17-18",
    doi = "10.1103/PhysRevD.96.103521",
    journal = "Phys. Rev. D",
    volume = "96",
    number = "10",
    pages = "103521",
    year = "2017"
}

@article{Garny:2018ali,
    author = "Garny, Mathias and Heisig, Jan",
    title = "{Interplay of super-WIMP and freeze-in production of dark matter}",
    eprint = "1809.10135",
    archivePrefix = "arXiv",
    primaryClass = "hep-ph",
    reportNumber = "TUM-HEP 1166/18, TTK-18-39",
    doi = "10.1103/PhysRevD.98.095031",
    journal = "Phys. Rev. D",
    volume = "98",
    number = "9",
    pages = "095031",
    year = "2018"
}

@article{Biondini:2019int,
    author = "Biondini, Simone and Vogl, Stefan",
    title = "{Scalar dark matter coannihilating with a coloured fermion}",
    eprint = "1907.05766",
    archivePrefix = "arXiv",
    primaryClass = "hep-ph",
    doi = "10.1007/JHEP11(2019)147",
    journal = "JHEP",
    volume = "11",
    pages = "147",
    year = "2019"
}

@article{Ibarra:2015nca,
    author = "Ibarra, A. and Pierce, A. and Shah, N. R. and Vogl, S.",
    editor = "Tecchio, Monica and Levin, Daniel",
    title = "{Anatomy of Coannihilation with a Scalar Top Partner}",
    eprint = "1501.03164",
    archivePrefix = "arXiv",
    primaryClass = "hep-ph",
    reportNumber = "MCTP-15-04, TUM-HEP-974-15",
    doi = "10.1103/PhysRevD.91.095018",
    journal = "Phys. Rev. D",
    volume = "91",
    number = "9",
    pages = "095018",
    year = "2015"
}

@article{Gondolo:1997km,
    author = "Gondolo, Paolo and Edsjo, Joakim",
    editor = "Bottino, A. and Di Credico, Alessandra and Monacelli, Piero",
    title = "{Neutralino relic density including coannihilations}",
    eprint = "hep-ph/9711461",
    archivePrefix = "arXiv",
    reportNumber = "MPI-PHT-97-078, MPI-PHT-97-78",
    doi = "10.1016/S0920-5632(98)00401-0",
    journal = "Nucl. Phys. B Proc. Suppl.",
    volume = "70",
    pages = "120--122",
    year = "1999"
}

@article{Ellis:2014ipa,
    author = "Ellis, John and Olive, Keith A. and Zheng, Jiaming",
    title = "{The Extent of the Stop Coannihilation Strip}",
    eprint = "1404.5571",
    archivePrefix = "arXiv",
    primaryClass = "hep-ph",
    reportNumber = "KCL-PH-TH-2014-17, LCTS-2014-16, CERN-PH-TH-2014-067, FTPI-MINN-14-11, UMN-TH-3333-14",
    doi = "10.1140/epjc/s10052-014-2947-7",
    journal = "Eur. Phys. J. C",
    volume = "74",
    pages = "2947",
    year = "2014"
}

@article{Ellis:2015vaa,
    author = "Ellis, John and Luo, Feng and Olive, Keith A.",
    title = "{Gluino Coannihilation Revisited}",
    eprint = "1503.07142",
    archivePrefix = "arXiv",
    primaryClass = "hep-ph",
    reportNumber = "KCL-PH-TH-2015-12, LCTS-2015-04, CERN-PH-TH-2015-049, UMN-TH-3424-15, FTPI-MINN-15-10, UMN--TH--3424-15, FTPI--MINN--15-10",
    doi = "10.1007/JHEP09(2015)127",
    journal = "JHEP",
    volume = "09",
    pages = "127",
    year = "2015"
}

@article{Saez:2018off,
    author = "S\'aez, Bastian D\'\i{}az and Rojas-Abatte, Felipe and Zerwekh, Alfonso R.",
    title = "{Dark Matter from a Vector Field in the Fundamental Representation of $SU(2)_L$}",
    eprint = "1810.06375",
    archivePrefix = "arXiv",
    primaryClass = "hep-ph",
    doi = "10.1103/PhysRevD.99.075026",
    journal = "Phys. Rev. D",
    volume = "99",
    number = "7",
    pages = "075026",
    year = "2019"
}

@article{Arcadi:2023imv,
    author = "Arcadi, Giorgio and Cabo-Almeida, David and Mescia, Federico and Virto, Javier",
    title = "{Dark Matter Direct Detection in \ensuremath{\mathsf{t}}-channel mediator models}",
    eprint = "2309.07896",
    archivePrefix = "arXiv",
    primaryClass = "hep-ph",
    doi = "10.1088/1475-7516/2024/02/005",
    journal = "JCAP",
    volume = "02",
    pages = "005",
    year = "2024"
}

@article{Mohan:2019zrk,
    author = "Mohan, Kirtimaan A. and Sengupta, Dipan and Tait, Tim M. P. and Yan, Bin and Yuan, C. P.",
    title = "{Direct detection and LHC constraints on a $t$-channel simplified model of Majorana dark matter at one loop}",
    eprint = "1903.05650",
    archivePrefix = "arXiv",
    primaryClass = "hep-ph",
    reportNumber = "UCI-HEP-TR-2019-02, MSUHEP-19-00",
    doi = "10.1007/JHEP05(2019)115",
    journal = "JHEP",
    volume = "05",
    pages = "115",
    year = "2019",
    note = "[Erratum: JHEP 05, 232 (2023)]"
}

@article{Arina:2020tuw,
    author = "Arina, Chiara and Fuks, Benjamin and Mantani, Luca and Mies, Hanna and Panizzi, Luca and Salko, Jakub",
    title = "{Closing in on $t$-channel simplified dark matter models}",
    eprint = "2010.07559",
    archivePrefix = "arXiv",
    primaryClass = "hep-ph",
    reportNumber = "P3H-20-058, TTK-20-35, CP3-20-47",
    doi = "10.1016/j.physletb.2020.136038",
    journal = "Phys. Lett. B",
    volume = "813",
    pages = "136038",
    year = "2021"
}

@article{Arina:2020udz,
    author = "Arina, Chiara and Fuks, Benjamin and Mantani, Luca",
    title = "{A universal framework for t-channel dark matter models}",
    eprint = "2001.05024",
    archivePrefix = "arXiv",
    primaryClass = "hep-ph",
    reportNumber = "CP3-20-01, MCNET-20-01",
    doi = "10.1140/epjc/s10052-020-7933-7",
    journal = "Eur. Phys. J. C",
    volume = "80",
    number = "5",
    pages = "409",
    year = "2020"
}

@book{Del_Nobile_2022,
   title={The Theory of Direct Dark Matter Detection: A Guide to Computations},
   ISBN={9783030952280},
   ISSN={1616-6361},
   url={http://dx.doi.org/10.1007/978-3-030-95228-0},
   DOI={10.1007/978-3-030-95228-0},
   journal={Lecture Notes in Physics},
   publisher={Springer International Publishing},
   author={Del Nobile, Eugenio},
   year={2022} }

@article{Arina:2023msd,
    author = {Arina, Chiara and Fuks, Benjamin and Heisig, Jan and Kr\"amer, Michael and Mantani, Luca and Panizzi, Luca},
    title = "{Comprehensive exploration of t-channel simplified models of dark matter}",
    eprint = "2307.10367",
    archivePrefix = "arXiv",
    primaryClass = "hep-ph",
    reportNumber = "TTK-23-19",
    doi = "10.1103/PhysRevD.108.115007",
    journal = "Phys. Rev. D",
    volume = "108",
    number = "11",
    pages = "115007",
    year = "2023"
}

@inproceedings{Ghosh:2022zef,
    author = "Ghosh, Tathagata and Kelso, Chris and Kumar, Jason and Sandick, Pearl and Stengel, Patrick",
    title = "{Simplified dark matter models with charged mediators}",
    booktitle = "{Snowmass 2021}",
    eprint = "2203.08107",
    archivePrefix = "arXiv",
    primaryClass = "hep-ph",
    reportNumber = "HRI-RECAPP-2022-005",
    month = "3",
    year = "2022"
}

@article{Morgante:2018tiq,
    author = "Morgante, Enrico",
    title = "{Simplified Dark Matter Models}",
    eprint = "1804.01245",
    archivePrefix = "arXiv",
    primaryClass = "hep-ph",
    reportNumber = "DESY-18-047, DESY 18-047",
    doi = "10.1155/2018/5012043",
    journal = "Adv. High Energy Phys.",
    volume = "2018",
    pages = "5012043",
    year = "2018"
}

@article{Abdallah:2015ter,
    author = "Abdallah, Jalal and others",
    title = "{Simplified Models for Dark Matter Searches at the LHC}",
    eprint = "1506.03116",
    archivePrefix = "arXiv",
    primaryClass = "hep-ph",
    reportNumber = "FERMILAB-PUB-15-283-CD, CERN-PH-TH-2015-139",
    doi = "10.1016/j.dark.2015.08.001",
    journal = "Phys. Dark Univ.",
    volume = "9-10",
    pages = "8--23",
    year = "2015"
}

@article{Ibarra:2022nzm,
    author = "Ibarra, Alejandro and Reichard, Merlin and Nagai, Ryo",
    title = "{Anapole moment of Majorana fermions and implications for direct detection of neutralino dark matter}",
    eprint = "2207.01014",
    archivePrefix = "arXiv",
    primaryClass = "hep-ph",
    doi = "10.1007/JHEP01(2023)086",
    journal = "JHEP",
    volume = "01",
    pages = "086",
    year = "2023"
}

@article{Arina:2025zpi,
    author = "Arina, Chiara and others",
    title = "{t-channel dark matter models {\textendash} a whitepaper}",
    eprint = "2504.10597",
    archivePrefix = "arXiv",
    primaryClass = "hep-ph",
    reportNumber = "CERN-LPCC-2025-001, IRMP-CP3-25-07, TTK-25-07",
    doi = "10.1140/epjc/s10052-025-14635-7",
    journal = "Eur. Phys. J. C",
    volume = "85",
    pages = "975",
    year = "2025",
    note = "[Erratum: Eur.Phys.J.C 85, 1105 (2025)]"
}

@article{Hall:2009bx,
    author = "Hall, Lawrence J. and Jedamzik, Karsten and March-Russell, John and West, Stephen M.",
    title = "{Freeze-In Production of FIMP Dark Matter}",
    eprint = "0911.1120",
    archivePrefix = "arXiv",
    primaryClass = "hep-ph",
    reportNumber = "OUTP-09-18-P, UCB-PTH-09-32",
    doi = "10.1007/JHEP03(2010)080",
    journal = "JHEP",
    volume = "03",
    pages = "080",
    year = "2010"
}

@article{Cahill-Rowley:2015aea,
    author = "Cahill-Rowley, Matthew and El Hedri, Sonia and Shepherd, William and Walker, Devin G. E.",
    title = "{Perturbative Unitarity Constraints on Charged/Colored Portals}",
    eprint = "1501.03153",
    archivePrefix = "arXiv",
    primaryClass = "hep-ph",
    reportNumber = "SLAC-PUB-16190, MITP-14-104, SCIPP-15-01",
    doi = "10.1016/j.dark.2018.04.003",
    journal = "Phys. Dark Univ.",
    volume = "22",
    pages = "48--59",
    year = "2018"
}

@inproceedings{Schuessler:2007av,
    author = "Schuessler, Alexander and Zeppenfeld, Dieter",
    title = "{Unitarity constraints on MSSM trilinear couplings}",
    booktitle = "{15th International Conference on Supersymmetry and the Unification of Fundamental Interactions (SUSY07)}",
    eprint = "0710.5175",
    archivePrefix = "arXiv",
    primaryClass = "hep-ph",
    pages = "236--239",
    month = "10",
    year = "2007"
}

@article{Harz:2014tma,
    author = "Harz, J. and Herrmann, B. and Klasen, M. and Kovarik, K.",
    title = "{One-loop corrections to neutralino-stop coannihilation revisited}",
    eprint = "1409.2898",
    archivePrefix = "arXiv",
    primaryClass = "hep-ph",
    reportNumber = "LAPTH-049-14, LCTS-2014-33, MS-TP-14-24",
    doi = "10.1103/PhysRevD.91.034028",
    journal = "Phys. Rev. D",
    volume = "91",
    number = "3",
    pages = "034028",
    year = "2015"
}

@article{Harz:2014gaa,
    author = "Harz, J. and Herrmann, B. and Klasen, M. and Kova{\v{r}}{\'\i}k, K. and Meinecke, M.",
    title = "{SUSY-QCD corrections to stop annihilation into electroweak final states including Coulomb enhancement effects}",
    eprint = "1410.8063",
    archivePrefix = "arXiv",
    primaryClass = "hep-ph",
    doi = "10.1103/PhysRevD.91.034012",
    journal = "Phys. Rev. D",
    volume = "91",
    number = "3",
    pages = "034012",
    year = "2015"
}

@article{Harz:2016dql,
    author = "Harz, J. and Herrmann, B. and Klasen, M. and Kovarik, K. and Steppeler, P.",
    title = "{Theoretical uncertainty of the supersymmetric dark matter relic density from scheme and scale variations}",
    eprint = "1602.08103",
    archivePrefix = "arXiv",
    primaryClass = "hep-ph",
    reportNumber = "LAPTH-035-15, MS-TP-16-03",
    doi = "10.1103/PhysRevD.93.114023",
    journal = "Phys. Rev. D",
    volume = "93",
    number = "11",
    pages = "114023",
    year = "2016"
}

@article{Schmiemann:2019czm,
    author = "Schmiemann, S. and Harz, J. and Herrmann, B. and Klasen, M. and Kova{\v{r}}{\'\i}k, K.",
    title = "{Squark-pair annihilation into quarks at next-to-leading order}",
    eprint = "1903.10998",
    archivePrefix = "arXiv",
    primaryClass = "hep-ph",
    reportNumber = "LAPTH-008-19, LAPTH-008/19, MS-TP-19-07, TUM-HEP-1194-19",
    doi = "10.1103/PhysRevD.99.095015",
    journal = "Phys. Rev. D",
    volume = "99",
    number = "9",
    pages = "095015",
    year = "2019"
}

@article{Harz:2022ipe,
    author = "Harz, Julia and Klasen, Michael and Sassi, Mohamed Younes and Wiggering, Luca Paolo",
    title = "{Dipole formalism for massive initial-state particles and its application to dark matter calculations}",
    eprint = "2210.03409",
    archivePrefix = "arXiv",
    primaryClass = "hep-ph",
    reportNumber = "MS-TP-22-09, MITP-22-078",
    doi = "10.1103/PhysRevD.107.056020",
    journal = "Phys. Rev. D",
    volume = "107",
    number = "5",
    pages = "056020",
    year = "2023"
}

@article{Harz:2019rro,
    author = "Harz, Julia and Petraki, Kalliopi",
    title = "{Higgs-mediated bound states in dark-matter models}",
    eprint = "1901.10030",
    archivePrefix = "arXiv",
    primaryClass = "hep-ph",
    reportNumber = "TUM-HEP-1186-19; Nikhef-2019-004",
    doi = "10.1007/JHEP04(2019)130",
    journal = "JHEP",
    volume = "04",
    pages = "130",
    year = "2019"
}

@article{Harz:2017dlj,
    author = "Harz, Julia and Petraki, Kalliopi",
    title = "{Higgs Enhancement for the Dark Matter Relic Density}",
    eprint = "1711.03552",
    archivePrefix = "arXiv",
    primaryClass = "hep-ph",
    reportNumber = "NIKHEF-2017-058",
    doi = "10.1103/PhysRevD.97.075041",
    journal = "Phys. Rev. D",
    volume = "97",
    number = "7",
    pages = "075041",
    year = "2018"
}

\end{document}